\documentclass[aps,prb,twocolumn,groupedaddress,floatfix,letterpaper,nofootinbib]{revtex4-2}

\newcommand{\id}{\mathrm{id}}

\newcommand{\RZ}{{\mathbb{R}/\mathbb{Z}}}

\newcommand\toZ[1]{\lfloor #1 \rfloor}
\newcommand{\cRep}{\cR\mathrm{ep}}
\newcommand{\scRep}{\mathrm{s}\cR\mathrm{ep}}
\newcommand{\eRep}{\eR\mathrm{ep}}
\newcommand{\cVec}{\cV\mathrm{ec}}
\newcommand{\eVec}{\eV\mathrm{ec}}

\newcommand{\eGau}{\eG\mathrm{au}}

\newcommand{\catsymm}{\textsf{categorical symmetry}}

\newcommand{\map}[1]{\overset{#1}{\rightarrow}}

\usepackage{amsthm}
\theoremstyle{theorem}

\newtheorem{Proposition}{Proposition}

\newtheorem{Definition}{Definition}
\newtheorem{DefinitionPH}[Definition]{Definition$^\text{ph}$}
\theoremstyle{definition}

\usepackage{tikzit}

\tikzstyle{cyan rect}=[fill={rgb,255: red,225; green,255; blue,255}, draw=black, shape=rectangle]
\tikzstyle{white circle}=[fill=white, draw=black, shape=circle, minimum width=3pt]
\tikzstyle{basic circle}=[fill=white, draw=black, shape=circle]
\tikzstyle{wide rect}=[fill=white, draw=black, shape=rectangle, minimum width=25pt]
\tikzstyle{red hazy circle}=[fill={rgb,255: red,255; green,159; blue,152}, draw=none, shape=circle, fill opacity=0.4]
\tikzstyle{small cyan circle}=[fill={rgb,255: red,194; green,236; blue,255}, draw=black, shape=circle, minimum size=2pt]
\tikzstyle{red circle}=[fill={rgb,255: red,255; green,156; blue,156}, draw=black, shape=circle]
\tikzstyle{transparent blue circle}=[fill={rgb,255: red,194; green,236; blue,255}, draw=black, shape=circle, fill opacity=0.3]

\tikzstyle{blue dashed}=[draw=blue, -, dashed]
\tikzstyle{thicc}=[-, line width=2pt]
\tikzstyle{thicc blue}=[-, fill=none, line width=2pt, draw={rgb,255: red,25; green,28; blue,255}]
\tikzstyle{black dots}=[-, draw=black, dashed, dash pattern=on 2pt off 2pt]
\tikzstyle{arrow-to-source}=[<-]
\tikzstyle{arrow-to-dest}=[->]

\usepackage{braket}

\usepackage{array}
\makeatletter
\newcommand{\thickhline}{%
	\noalign {\ifnum 0=`}\fi \hrule height 1pt
	\futurelet \reserved@a \@xhline
}
\newcolumntype{"}{@{\hskip\tabcolsep\vrule width 1pt\hskip\tabcolsep}}
\makeatother

\newcommand{\ml}[1]{\( #1 \)}

\begin{document}

\begin{titlepage}

\title{
Symmetry as a shadow of topological order\\
and a derivation of topological holographic principle
}

\author{Arkya Chatterjee} 
\affiliation{Department of Physics, Massachusetts Institute of Technology,
Cambridge, Massachusetts 02139, USA}

\author{Xiao-Gang Wen} 
\affiliation{Department of Physics, Massachusetts Institute of Technology,
Cambridge, Massachusetts 02139, USA}

\begin{abstract} 

Symmetry is usually defined via transformations described by a (higher) group.
But a symmetry really corresponds to an algebra of local symmetric operators,
which directly constrains the properties of the system.  In this paper, we
point out that the algebra of local symmetric operators contains a special
class of extended operators -- transparent patch operators, which reveal the
selection sectors and hence the corresponding symmetry. The algebra of those
transparent patch operators in $n$-dimensional space gives rise to a
non-degenerate braided fusion $n$-category, which happens to describe a
topological order in one higher dimension (for finite symmetry).  Such a
holographic theory not only describes (higher) symmetries, it also describes
anomalous (higher) symmetries, non-invertible (higher) symmetries (also known
as algebraic higher symmetries), and non-invertible gravitational anomalies.
Thus, topological order in one higher dimension, replacing group, provides a
unified and systematic description of the above generalized symmetries.  This
is referred to symmetry/topological-order (Symm/TO) correspondence.  Our
approach also leads to a derivation of topological holographic principle:
\emph{boundary uniquely determines the bulk}, or more precisely, the algebra of
local boundary operators uniquely determines the bulk topological order.  
As an application of the Symm/TO correspondence, we show the equivalence
between $\mathbb{Z}_2\times \mathbb{Z}_2$ symmetry with mixed anomaly and
$\mathbb{Z}_4$ symmetry, as well as between many other symmetries, in
1-dimensional space.

\end{abstract}

\maketitle

\end{titlepage}

\setcounter{tocdepth}{1} 
{\small \tableofcontents }

\section{Introduction}

It is well known that symmetry, higher symmetry,\cite{NOc0605316, NOc0702377,
KT13094721, GW14125148} gravitational anomaly,\cite{AW8469, W8597} and
anomalous (higher) symmetry\cite{H8035} can all constrain the properties of
quantum many-body systems or quantum field theory.\cite{HW0541, BSm0511710,
GP07083051, BH10034485, KT13094721, GW14125148, S150804770, TK151102929,
BM160606639, CT171104186, T171209542, BT170402330, P180201139, DT180210104, L180207747,
CI180204790, HI180209512, BT180300529, BH180309336, NW180400677, ZW180809394,
HO181005338, WW181211968, WW181211967, GW181211959, WW181211955, W181202517}
Recently, motivated by anomaly in-flow\cite{CH8527, W8951, W9125, KT14043230,
WY190908775} as well as the equivalence\cite{JW190513279} between
non-invertible gravitational anomalies\cite{W1313, KW1458, KZ150201690,
FV14095723, M14107442, JW190513279} and symmetries, it was proposed that
non-invertible gravitational anomalies, (higher) symmetries, anomalous
symmetries,\cite{H8035} algebraic higher symmetries,\cite{KZ200308898,
KZ200514178} etc, can be unified by viewing all of them as shadow of
topological order\cite{W8987, W9039} in one higher dimension.\cite{W1313,
KW1458, F14047224, KZ150201690, KZ170200673, FT180600008, TW191202817,
JW190513279, JW191209391, JW191213492, KZ200514178} A comprehensive theory was
developed along this line.\cite{JW191213492, KZ200308898, KZ200514178} More
specifically, the properties of quantum many-body systems constrained by a
non-invertible gravitational anomaly or a finite (anomalous and/or higher
and/or algebraic) symmetry are the same as the boundary properties constrained
by a bulk topological order in one higher dimension.  Thus gravitational
anomaly and/or finite symmetry can be fully replaced and is equivalent to
topological order in one higher dimension.  Such a point of view is called the
\emph{holographic point view of symmetry}.

To place the above holographic point view on a firmer foundation, we note that
even though we use transformations described by groups or higher groups to
define symmetries, in fact, a symmetry is not about transformations. What a
symmetry really does is to select a set of local symmetric operators which form
an algebra.  The algebra of all local symmetric operators determines the
possible quantum phases and phase transitions, as well as all other properties
allowed by the symmetry. However, the algebra of local symmetric operators does
not contain symmetry transformations and it is hard to identify the
corresponding symmetry group from such an algebra (but see Refs.
\onlinecite{MM210810324} and \onlinecite{MM220903370} where symmetry is
reconstructed using commutant algebras).

In this paper, we show that the algebra generated by local symmetric operators
includes not only point-like local operators, but it also includes extended
operators algebraically generated by local symmetric operators, such as string-like operators,
membrane-like operators, \etc.  We find that a subclass of the extended operators
-- transparent patch operators -- are important.  These transparent patch
operators reveal the symmetry selection sectors hidden in the algebra of local
symmetric operators, and thus reveal the selection rules and the corresponding
symmetry.  Thus, isomorphic algebras of transparent patch operators give rise
to equivalent symmetries.\footnote{Such equivalent symmetries were call
holo-equivalent symmetries in \Rf{KZ200514178}.}  Those isomorphic classes of
algebra were referred to as \textsf{categorical symmetries} in \Rf{JW191213492,
KZ200514178}, which, by definition, describe all known and unknown types of
symmetries.  However, the term ``categorical symmetry'' has also been used to refer
to algebraic higher symmetry (\ie non-invertible symmetry) by some authors.  So
here, we use \catsymm\ to stress that the term is used in the sense of
\Rf{JW191213492, KZ200514178}.  

We find that, for a finite symmetry in $n$-dimensional space, such an algebra
of transparent patch operators determines a braided fusion $n$-category.  If
the algebra include all local symmetric operators, the braided fusion
$n$-category will be non-degenerate.  Further more, isomorphic algebras of
transparent patch operators give rise to the same non-degenerate braided fusion
$n$-category.  Thus \textsf{categorical symmetries} are described by
non-degenerate braided fusion $n$-categories, which happen to correspond to
topological orders in one higher dimension.\cite{KZ200308898, KZ200514178}  In
other words, we suggest that group is not a proper description of symmetry,
since (higher) symmetries and anomalous (higher) symmetries described by
different (higher) groups can be equivalent.  Finite symmetries are really
described by non-degenerate braided fusion $n$-categories (\ie topological
orders in one higher dimension). 

The calculation in this paper is based on operator algebra\footnote{\ie algebra
generated by local symmetric operators (LSOs), which we will refer to as LSO
algebra for short throughout the rest of the paper}.  A similar picture was
obtained in \Rf{KZ210808835} based on ground state and their excitations.  The
operator algebra discussed in this paper may be related to the nets of local
observable algebras in \Rf{HK6448} and topological net of extended defects in
\Rf{KZ220105726}. See also \Rf{FT180600008} for related discussion on some of
the examples discussed in this paper. 

The holographic theory of symmetry allows us to identify equivalent (higher
and/or anomalous) symmetries, that can look quite different.  For example, two
(higher and/or anomalous) symmetries can be realized at boundaries of two
symmetry protected topological (SPT) states with those symmetries in one higher
dimension. If after gauging the respected symmetries in the SPT states, we
obtain the same topological order, then the two corresponding symmetries have
the same \catsymm\ and are equivalent.  This is a systematic way to identify
equivalent symmetries and their \catsymm.

In \Rf{KZ200514178} it was conjectured that \emph{if two anomaly-free
(invertible or non-invertible) symmetries described by local fusion higher
categories, $\cR$ and $\cR'$, are equivalent (\ie have the equivalent monoidal
center $\eZ(\cR) \simeq \eZ(\cR')$), then the two symmetries provide the same
constraint on the physical properties.} This leads to the following conjecture:
for any pair of equivalent symmetries, there is a lattice duality map, that
maps a lattice model with one symmetry $\cR$ to a lattice model with another
symmetry $\cR'$.  More specifically, the sets of local symmetric operators
selected by the two symmetries, $\{ O_\cR\}$ and $\{ O_{\cR'}\}$, have an
one-to-one correspondence and generate the same algebra, under such a  duality
map.  The duality map also maps the lattice Hamiltonians (as sums of local
symmetric operators) of the two lattice models into each other.  The two
lattice models have identical dynamical properties, \eg they have identical
energy spectrum in symmetric sub Hilbert space.\cite{{JW191213492}} This can be
viewed as the physical meaning of ``equivalent symmetry''.

This conjecture is motivated and supported by the studies of some explicit
examples of well known and new dualities.  The notion of dual symmetry was
introduced in \Rf{T171209542, BT170402330} via gauging.  \Rf{JW191213492} used
Kramers–Wannier duality and its generalization to study the equivalence and its
holographic understanding of 1d $\cRep_{\Z_2}$-symmetry (the $\Z_2$ 0-symmetry)
and $\cVec_{\Z_2}$-symmetry (the dual $\Z_2$ 0-symmetry), as well as 2d
$2\cRep_{\Z_2}$-symmetry (the $\Z_2$ 0-symmetry) and $2\cVec_{\Z_2}$-symmetry
(the $\Z_2^{(1)}$ 1-symmetry). 
\Rf{KZ200514178}  used a lattice duality map to
study the equivalence and its holographic understanding of $n$d
$n\cRep_G$-symmetry (the 0-symmetry described by a finite group $G$) and
$n\cVec_G$-symmetry (the dual non-invertible $(n-1)$-symmetry).
\Rf{KZ210808835} studied the duality maps and holographic equivalence of 1d
$\cRep_{\Z_2}$-symmetry, $\cVec_{\Z_2}$-symmetry, and $\scRep_{\Z_2}$-symmetry
(the 1d $\Z_2^f$ fermionic symmetry).  In the above examples, the duality map
can be viewed as gauging process.  In \Rf{BA10065823,LV211209091}, a more
general duality map between lattice systems is discussed via category theory
and tensor network.  

In this paper, we studied a duality between anomaly-free symmetry and anomalous
symmetry.  We obtain new duality maps between many pairs of equivalent
symmetries, such as 1d $\Z_2\times \Z_2$ symmetry with the mixed anomaly and
anomaly-free $\Z_4$ symmetry (see Section \ref{eqsymm} for many more examples).

Viewing symmetry as topological order in one higher dimension generalizes the
fundamental concept of symmetry.  It allows us to describe new type of
non-invertible symmetries (also called algebraic (higher)
symmetries)\cite{FSh0607247,DR10044725,DR11070495,CY180204445,TW191202817,KZ200514178}
that are beyond group and higher group,  as well as new type of symmetries that
are neither anomalous nor anomaly-free.  But why do we want a more general
notion of symmetry?

We know that symmetry can emerge at low energies.  So we hope our notion of
symmetry can include all the possible emergent symmetry.  It turns out that the
low energy emergent symmetries can be the usual higher and/or anomalous
symmetries.  They can also be non-invertible symmetries.  They can even be
symmetries that are neither anomalous nor anomaly-free.  Therefore, we need a
most general and unified view of higher and/or anomalous symmetries and beyond,
if we want to use emergent symmetry as a guide to systematically understand or
even classify gapless states of matter.  

For example, using this generalized notion of symmetry, we gain a deeper
understanding of quantum critical points.  We find that the symmetry breaking
quantum critical point for a symmetry described by a finite group $G$ in
$n$-dimensional space is the same as the symmetry breaking quantum critical
point for an algebraic higher symmetry described by fusion $n$-category
$n\cVec_G$.\cite{JW191213492, KZ200514178}  In fact, both the ordinary symmetry
described by group $G$ and the algebraic higher symmetry (a non-invertible
symmetry) described by fusion $n$-category $n\cVec_G$ are present and are not
spontaneously broken at this critical point.  The $G$-symmetry and the
algebraic higher symmetry $n\cVec_G$ may give us a more comprehensive
understanding of the symmetry breaking quantum critical point.

Symmetry can constrain the properties of a physical system.  On the other hand,
when certain excitations in a system have a large energy gap, below that energy
gap, the system can have emergent symmetry, which can be anomalous and/or
non-invertible.\cite{MT170707686, JW191213492, KZ200514178}  In this case, we
can use the emergent symmetry to reflect and to characterize the special low
energy properties of the system below the gap.  Here we make a preparation to
go one step further.  We intend to propose that the low energy properties and
the emergent symmetries are the same thing.  In other words, we intend to
propose that the full emergent symmetry may fully characterize the low energy
effective theory.  We may be able to study and to classify all possible low
energy effective theories by studying and classifying all possible emergent
symmetries.  

Such an idea cannot be correct if the above symmetries are still considered as
being described by groups and higher groups. This is because the symmetries
described by groups and higher groups are quite limited, and they cannot
capture the much richer varieties of possible low energy effective theories.
However, after we greatly generalize the notion of symmetry to algebraic higher
symmetry, and even further to topological order in one higher dimensions --
which includes (anomalous and/or higher) symmetries, (invertible and
non-invertible) gravitational anomalies, and beyond -- then it may be possible
that those generalized symmetries can largely capture the low energy properties
of quantum many-body systems.  This may be a promising new direction to study
low energy properties of quantum many-body systems.

The above proposal is supported by the recent study of 1d gapless conformal
field theory where a topological skeleton was identified for each conformal
field theory.\cite{KZ170501087, KZ190504924, KZ191201760} Such a topological
skeleton is a non-degenerate braided fusion category corresponding to a 2d
topological order, where the involved conformal field theory is one of the
gapless boundary.

The low energy properties of quantum many-body systems are described by quantum
field theories.  A systematic understanding and classification of low energy
properties is equivalent to a systematic understanding and classification of
quantum field theories.  Thus the holographic view of symmetry can have an
impact on our general understanding of quantum field theories.  Using this
holographic point of view of symmetry, one can also obtain a classification of
topological order and symmetry protected topological orders, with those
generalized symmetry, for bosonic and fermionic systems, and in any
dimensions.\cite{KZ200308898, KZ200514178}  

The holographic point view of symmetry has a close relation to AdS/CFT duality,
where a boundary CFT and a bulk quantum gravity in AdS space determine each
other.  In the holographic point view of symmetry, there is a topological
holographic principle: \emph{boundary determines bulk, while bulk does not
determine boundary}.  In this paper, we give the above statement a more precise
meaning which allows us to derive the topological holographic principle.  We
regard \emph{boundary} as an algebra of local boundary operators.  From the
algebra of local boundary operators, we can obtain the sub-algebra of a special
class of extended operators -- transparent patch operators, which in turn encodes a non-degenerate braided fusion (higher) category.  This category describes a topological order in one higher
dimension, which is the \ml{bulk}.  We see that \emph{boundary uniquely
determines bulk}.

\section{Notations and terminology}

In this paper, we will use $n+1$D to represent spacetime dimensions, and $n$d
to represent spatial dimensions.  We will use mathcal font $\cA,\cB,\cC$ to
describe fusion categories, and euscript font $\eA, \eB, \eC$ to describe
braided fusion categories. We will use the theorem style
\textbf{Definition$^\text{ph}$} to provide "physical definitions", which serve
the purpose of introducing concepts without delving into mathematical rigor.

Let us also remark on some terminology.  In this paper we use \catsymm\ to mean
the combination of symmetry and dual symmetry \cite{JW191213492}.  If a
\catsymm\ is finite, it turns out that the \catsymm\ corresponds to a
topological order \cite{CGW1038, ZW1490} in one higher
dimension.\cite{KZ200308898, KZ200514178} Such a topological order in one
higher dimension has also been referred to as symmetry topological field theory
(symmetry TFT) in the field theory literature.\cite{AS211202092} In this context, one describes the topological operators corresponding to the (finite) symmetries of a quantum field theory in $ d $ dimensions in terms of the topological excitations of a corresponding TFT in $ d+1 $ dimensions. The symmetry data are encoded in the global topological properties of this TFT, which may be described in the form of some action. However, such an action is not necessarily unique. So one should keep in mind that "symmetry TFT" really refers to the topological data of the theory which is independent of the fields one uses to describe it. The topological data encoded by such a TFT may also be captured by a lattice model exhibiting topologically ordered ground states. In this limit, the two notions of symmetry TFT and \catsymm\ coincide. This concept was also explored under the name of "topological symmetry" in \Rf{FT220907471}. 

Let us note that in physics contexts, one usually interprets
topological field theory as a particular kind of \emph{field theory}, \ie a theory in
terms of smoothly varying fields. If we have a lattice regularization in mind for the field theory, we must first take the limit where the lattice spacing vanishes. Under such an interpretation, topological
field theory describes a topological order near a critical point, where the
smoothly varying field describes the long wavelength fluctuations
near the critical point. Since a topologically ordered phase can have many
different phase boundaries described by different critical points, it is common
that different topological field theories can describe the same topological order. 
Moreover, for continuous or infinite symmetry, \catsymm\ does not correspond to
topological order or symmetry TFT in one higher dimension. We need some generalization of fusion categories with an infinite number of objects to describe such symmetries. Whatever this generalization is, it is clear that the notion of \catsymm\ for such symmetries is
more general than topological order/TFT in one higher dimension. 

In this paper, we conjecture that \catsymm\ (as the combination of symmetry and
dual symmetry) corresponds to equivalence class of isomorphic algebra of
transparent patch operators.  So we will use this as a more precise definition
of \catsymm.  We conjecture that, in $n$-dimensional space, \catsymm\ (as
equivalence class of isomorphic algebras of transparent patch operators) is
described by non-degenerate braided fusion $n$-category.  For continuous or
infinite symmetry, the corresponding braided fusion $n$-category will have
infinite objects/morphisms.  We will discuss some simple examples to support
our conjecture.

In this paper, we also interpret \emph{quantum field theory} as an algebra of
local operators, along with a Hamiltonian.  Under such an interpretation, the
algebra of local operators may have an energy dependence: we may exclude some
local operators that generate high energy excitations.  Then, the remaining
local operators may generate a different algebra.  This low energy operator
algebra gives rise to emergent \catsymm.  We propose that the full low energy
emergent \catsymm\ may largely characterize gapless liquid states.

Similarly, we also interpret \emph{boundary} of a topological order as an
algebra of local boundary operators along with a boundary Hamiltonian.  Here, the
local boundary operators only create excitations with energy less then the bulk
energy gap which is assumed to be infinite.  Under such an interpretation, we
see a close connection between quantum field theory and boundary of topological
order.

We would like to point out that \catsymm\ (as non-degenerate braided fusion
$n$-category $\eM$) is not algebraic higher symmetry \cite{KZ200308898,
KZ200514178} nor fusion category symmetry \cite{TW191202817}.  The latter are
described by local fusion higher category $\cR$.  In fact, the \catsymm\ $\eM$
is given by the center of $\cR$ \cite{KZ200308898, KZ200514178} \begin{align}
\eM =\eZ(\cR).  \end{align}

\section{Bosonic quantum system and its algebra of local operators} 
\label{bosonsys}

\subsection{Total Hilbert space, local operator algebra, and local Hamiltonian} 

\begin{figure}[t]
\begin{center}
 \includegraphics[scale=0.6]{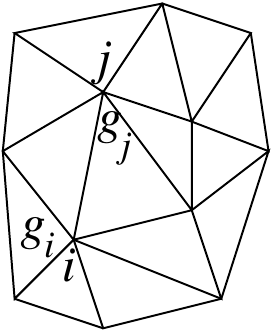}
\end{center}
\caption{A 2d lattice bosonic model, whose degrees of freedom live on the vertices and are labeled by the elements in a set: $g_i\in G$} \label{TriLattG} 
\end{figure}

A lattice bosonic quantum system is defined by four components:
\begin{enumerate}
\item
A triangulation of space (see Fig. \ref{TriLattG}).

\item
A total Hilbert space 
\begin{align}
\cV = \bigotimes_i \cV_i,
\end{align}
 where \ml{\cV_i = \mathrm{span}\{ |g\>\ \big|\ g \in G\}} is the local Hilbert
space on vertex-\ml{i}.  The basis vectors of $\cV_i$ are labeled by the
elements in a finite set $G$.

\item An \textbf{algebra  of local operators} formed by all the local
operators, \ml{\cA=\{O_i\}}.  Here \textbf{local operator} is defined as an
operator $O_i$ that acts within the tensor product of a few nearby local
Hilbert spaces, say near a vertex-$i$.

\item A {local Hamiltonian} \ml{H= - \sum_i O_i } which is a sum of hermitian
local operators.

\end{enumerate}

\subsection{Transparent patch operators} 

\begin{figure}[t]
\begin{center}
 \includegraphics[scale=0.8]{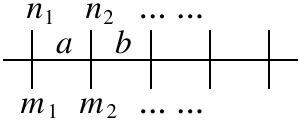} 
\end{center}
\caption{
The matrix elements of a string-like tensor network operator,
$O_{m_1,m_2,\cdots;n_1,n_2,\cdots}$, can be given by a contraction of rank-4
tensors $T_{n_2,m_2,a,b}$, \etc.  Each tensor is represented by a vertex, where
the legs of the vertex correspond to the indices of the tensor.  The connected
legs have the same index and is summed over (which correspond to the tensor
contraction).  This is just one representation of tensor network
operator. 
} \label{TNOp} 
\end{figure}

The algebra of local operators will play a central role in this paper.  This algebra, which is generated by the local operators, does not only contain local operators but, beyond 0-dimensional space, also contains the products of local operators. These products can
generate extended operators that can be string-like, membrane-like, \etc. Thus
the closure of the algebra of local operators must contain those extended
operators.  An algebra of local operators may have many different extensions.
Since we are going to use the algebra of local operators to describe
symmetries, we will consider a particular extension. We would like to organize those
local and extended operators into point operators, string operators, disk
operators, \etc, with a special transparency property. We refer to these operators
generally as transparent patch operators. More precisely, 
\begin{DefinitionPH}
a \textbf{patch operator} is a \textbf{tensor network
operator} (see Fig. \ref{TNOp}). It also has the following form
\begin{align}
\label{patchop}
 O_\mathrm{patch} = \sum_{\{a_i\}} \Phi(\{a_i\}) 
\prod_{i \in \mathrm{patch}} O_i^{a_i}
\end{align}
where ``patch''  has a topology of $n$-dimensional disk, $n=0,1,2,\cdots$.  A
\textbf{transparent patch (t-patch) operator} is a patch operator that
satisfies the following \textbf{transparency} condition (or
\textbf{invisible-bulk} condition):
\begin{align}
\label{pcomLSO}
O_\mathrm{patch} 
O_\mathrm{LSO}
=
O_\mathrm{LSO}
O_\mathrm{patch}, 
\end{align}
if the boundaries of the patch, $\prt\mathrm{patch}$
is \textbf{far away} from the LSO $O_\mathrm{LSO}$.
The above condition is also equivalent to
\begin{align}
\label{pcom}
O_\mathrm{patch} 
O_\mathrm{patch'}
=
O_\mathrm{patch'}
O_\mathrm{patch}, 
\end{align}
if the boundaries of two patches, $\prt\mathrm{patch}$ and
$\prt\mathrm{patch'}$, are not linked and are \textbf{far away} from each other.
\end{DefinitionPH}
In the above definition, $O_i^{a_i}$'s are local operators acting near
vertex-$i$.  For each vertex, there can be several different local operators
(including the trivial identity operator) which are labeled by $a_i$.
$\prod_{i \in \mathrm{patch}} O_i^{a_i}$ is a product of those local  operators
over all the vertices $i$ in the patch.  Different choices of $\{a_i\}$ give
rise different operator pattern.  $\sum_{\{a_i\}} \Phi(\{a_i\})$ is the sum of
all operator pattern with complex weight $\Phi(\{a_i\})$.  One may think
$O_i^{a_i}$'s create different types of particles labeled by $a_i$ at
vertex-$i$.  Then $ O_\mathrm{patch} = \sum_{\{a_i\}} \Phi(\{a_i\}) \prod_{i
\in \mathrm{patch}} O_i^{a_i} $ creates a quantum liquid state of those
particles on the patch.  The quantum liquid state is described by the many-body
wave function $ \Phi(\{a_i\})$.  

In the above definition, we also used a notion of \emph{far away} which is not
rigorously defined.  To define such a notion, we first introduce a notion of
\textbf{small local} operators as operators acting on vertices whose
separations are less then a number $L_\text{op}$.  (The separations between two
vertices is defined as the minimal number of links connecting the two
vertices.) In the rest of this paper, the terms ``local operator'' and
``0-dimensional patch operator'' will refer to this kind of small local
operators.  

However, the algebra of small local operators contains \textbf{big local}
operators, acting on vertices whose separations are larger then the number
$L_\text{op}$.  ``$n$-dimensional patch operator'' for $n>0$ refer to those big
local operators.  The notion of far away means further than the distance
$L_\text{op}$.  When we take the large system size limit: $L_\text{sys}\to
\infty$, we also assume $L_\text{op}\to \infty$ and $L_\text{op}/L_\text{sys}
\to 0$. We will see in this paper that it is this particular way to take the
large system size limit that ensures the algebra of small local operators to
contain large extended operators.  Such an algebra of small local operators and
large extended operators in $n$-dimensional space have a structure of
\emph{non-degenerate braided fusion $n$-category}. This emergent phenomenon is
the key point of this paper.

There is another important motivation to introduce transparent patch operators.
The bulk of transparent patch operators is invisible.  Thus a transparent open
string operator can be viewed as two point-like particles, one for each string
end.  A transparent disk operator can be viewed as a closed string at the
boundary of the disk.  In general, a transparent patch operator gives rise to
an extended excitation in one lower dimension, corresponding to the boundary of
the patch.  Later we will see that those point-like, string-like, \etc\
excitations can fuse and braid, forming a braided fusion category that describe
the operator algebra.

The boundaries of transparent patch operators can be viewed as charged
particles, although the  patch operators are fromed by LSO's that carry no
symmetry charge.  The boundaries of transparent patch operators can also be
viewed as fractionalized particles, which may carries fractionalized degrees of
freedom and/or fractionalized quantum numbers.  So the boundaries of
transparent patch operators reveal the selection sectors of a symmetry. Such
selection sectors are hidden in the algebra generated by the LSOs.

\subsection{Patch symmetry and patch charge operators}

Symmetry transformation operators and symmetry-charge creation operators play
important roles in our theory about symmetry (including higher symmetry and
algebraic higher symmetry).  Those operators also appear in our setup of local
operator algebra after we include the extended operators.
\begin{DefinitionPH}
\label{def:chargesymm}
A t-patch operator is said to have an \textbf{empty bulk} if
$O_i^{a_i} = \id_i$ for all $i$'s far away from the boundary of the patch.  A
t-patch operator with an empty bulk is also referred to as a \textbf{patch
charge operator}.  A t-patch operator with non-empty bulk is referred to as a
\textbf{patch symmetry operator} (see Section \ref{psymm} for a concrete example).
\end{DefinitionPH}
We would like to remark that due to the transparency condition \eqn{pcom}, a charge
patch operator always commutes with symmetry transformation operator (acting on
the whole space for 0-symmetry, or closed sub-manifold for higher symmetries).
Thus the patch charge operator always carry zero total charge.  So the patch
charge operators are not charged operators, since charged operators do not
commute with symmetry transformations.  The patch charge operators defined
above are something like operators that create a pair of charge and
anti-charge, which correspond to a charge fluctuations with vanishing total net
charge.

We want to point out that the definition \ref{def:chargesymm} is not that
important physically, since the notations of \emph{charge} and \emph{symmetry
transformation} are not the notions of algebra of local operators.  They are
the notions of a representation of an operator algebra.  For different
representations of the same operator algebra, the same operator in the algebra
can some times be patch charge operator and other times be patch symmetry
operator.  

In next section, we will discuss a concrete simple example: a bosonic system in
1-dimensional space with $\Z_2$ symmetry, to illustrate the above abstract
definition. We will give the explicit form of t-patch operators, to show how
they reveal a braided fusion category in the algebra of local operators.  In
Appendix \ref{example3d}, we will discuss an example of bosonic system in
3-dimensional space without symmetry.  We will illustrate how they give rise to
a non-degenerate braided fusion $3$-category $3\eVec$.  

\subsection{Algebra of t-patch operators and Categorical symmetry} 

The symmetric Hamiltonian is a sum of local symmetric operator \ml{H=\sum_i
O_i^\mathrm{symm}}.  If our measurement equipments do not break the symmetry,
then the measurement results are correlations of local symmetric operators.  We
see that a symmetry is actually described by the algebra of local symmetric
operators, rather than by the symmetry transformations.  Or more precisely,
symmetry is defined by the commutant algebra of local symmetric operators.
Here a commutant algebra of a local operator algebra is formed by all the
operators (local or non-local) that commute with all the operators in the
local operator algebra.  In particular
\begin{align}
\text{Isomorphic commutant algebras} \leftrightarrow \text{Equivalent symmetry}.
\end{align}
In this paper, we will view symmetry from this operator algebra point of view:
\begin{DefinitionPH}
A \textbf{\catsymm} is an equivalence class of isomorphic commutant algebra.
\end{DefinitionPH}
We remark that if the  operator algebra contains all the local operators in a
lattice model, then the \textbf{\catsymm} is trivial, describing a trivial
symmetry (\ie no symmetry).  If the local operator algebra contain only a
subset of local operators (such as containing only symmetric local operators),
then the \textbf{\catsymm} is non-trivial.  Also note that \catsymm\ is
different from the usual symmetry defined via the symmetry transformations.
Two symmetries defined by different symmetry transformations may have
isomorphic algebra of local symmetric operators.  In that case, the two
symmetries correspond to the same \catsymm, and are said to be equivalent.

\section{A 1\texorpdfstring{\MakeLowercase{d}}{d} bosonic quantum system with $\Z_2$ symmetry} 
\label{Z2symm1d}

In this section, we consider the simplest symmetry -- $\Z_2$ symmetry in one
spatial dimension.  A bosonic system with $\Z_2$ symmetry is obtained by
modifying the algebra of the local operators.  For convenience, let we assume
the degrees of freedom live on vertices, which are labeled by elements in the
$\Z_2$ group $+1$ and $-1$.

\subsection{$\Z_2$ symmetry and its algebra of local symmetric operators} 

In the standard approach, a symmetry is described by a symmetry transformation,
which has the following form for our example:
\begin{align}
W = \bigotimes_{i\in \text{whole space}} X_i, \ \ \ \
X = 
\begin{pmatrix}
 0&1\\
 1&0\\
\end{pmatrix}
. 
\end{align}
Since \ml{W^2=1} which generates a \ml{\Z_2} group, we call the symmetry a $\Z_2$
symmetry.  We can use the $\Z_2$ transformation $W$ to define an algebra of
local operators:
\begin{align}
\cA=\{O_i^\mathrm{symm} \ \big|\  O_i^\mathrm{symm}W = W O_i^\mathrm{symm}\}
\end{align}
The local operator $O_i^\mathrm{symm}$, satisfying $O_i^\mathrm{symm}W = W
O_i^\mathrm{symm}$, is called \emph{local symmetric operator}.

To see the connections between operator algebra and braided fusion category, we use the t-patch operators
introduced in last section to organize the local symmetric operators:
\begin{enumerate}
\item
0-dimensional t-patch operators: $X_i,\ Z_iZ_{i+1}$, where $Z=\begin{pmatrix}
 1&0\\
 0&-1\\
\end{pmatrix}$.

\item
1-dimensional t-patch operators -- string operators: 
for $i<j$
\begin{align}
Z_{\mathrm{str}_{ij}} = Z_iZ_j,\ \ 
Z_{\mathrm{str}_{ji}} \equiv Z_{\mathrm{str}_{ij}}^\dag,
\end{align}
where the string$_{ij}$ connects the vertex-$i$ and vertex-$j$.  The above
string operator has an empty bulk and is called as patch charge
operator. We have another string operator: for $i<j$
\begin{align}
X_{\mathrm{str}_{ij}} &= 
X_{i+1} X_{i+2} \cdots
X_{j}, \ \ 
X_{\mathrm{str}_{ji}} \equiv X_{\mathrm{str}_{ij}}^\dag
.
\end{align}
Note that the boundaries of $X$-strings actually live on the links $\<i,i+1\>$
and $\<j,j+1\>$.  We labeled those links by $i,j$.  This leads to the special
choice of the boundary of the string operator.  The second string operator has
a non trivial bulk, which generates our $\Z_2$ symmetry.

\end{enumerate}
We remark that, in general, the operators in the string may not commute and the
order of the operator product will be important in that case.  Here we adopted
a convention that in string operator $O_{\mathrm{str}_{ij}}$, the operators
near $i$ appear on the left side of the operators near $j$.

\begin{figure}[t]
\begin{center}
\includegraphics[scale=0.35]{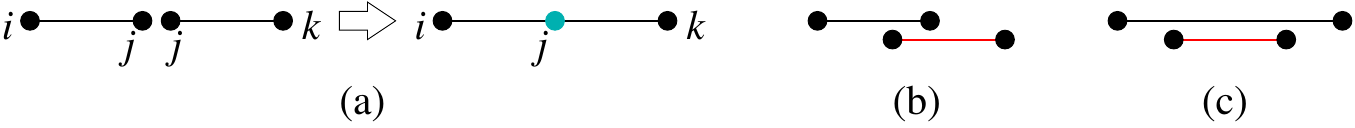} 
\end{center}
\caption{
(a) ``Fusion'' of two string operators.
(b) Non-trivial ``braiding'' between two string operators.
(c) Trivial ``braiding'' between two string operators.
} \label{OAstring} 
\end{figure}

In terms of t-patch operators, algebra of local symmetric operators takes the
following form (only important operator relations are listed)
\begin{align}
\label{OA1dZ2a}
Z_{\mathrm{str}_{ij}} Z_{\mathrm{str}_{jk}} &= Z_{\mathrm{str}_{ik}},
\\ 
\label{OA1dZ2b}
X_{\mathrm{str}_{ij}} X_{\mathrm{str}_{jk}} &= X_{\mathrm{str}_{ik}},
\\
\label{OA1dZ2c}
Z_{\mathrm{str}_{ij}} X_{\mathrm{str}_{kl}} &= - X_{\mathrm{str}_{kl}} Z_{\mathrm{str}_{ij}}, \ \ (i<k<j<l)
\\
\label{OA1dZ2d}
Z_{\mathrm{str}_{ij}} X_{\mathrm{str}_{kl}} &= + X_{\mathrm{str}_{kl}} Z_{\mathrm{str}_{ij}}, \ \ \text{(else)}
\end{align}
Eqn. (\ref{OA1dZ2a}) and (\ref{OA1dZ2b}) describe the fusion of string operators (see Fig.
\ref{OAstring}a).  
The commutator between the two kinds of string operators depends on their
relative positions.  If one string straddles the boundary of the other string,
such as $i< k < j< l$ as in Fig.  \ref{OAstring}b, commutator has a non-trivial
phase.  Otherwise (see Fig. \ref{OAstring}c), the string operators commute,
which ensure the string operators are indeed \emph{transparent} patch (t-patch) operators. All such "non-straddling" orderings of $ i,j,k,l $ are understood to be captured in \eqn{OA1dZ2d}. In Section
\ref{z2all}, we will discuss the full algebra of extended t-patch operators in more
detail.

\subsection{Patch symmetry transformation}
\label{psymm}

We note that $Z_i$ operator transforms as the non-trivial representation of
$\Z_2$ group:
\begin{align}
WZ_i W^{-1} = -Z_i.
\end{align}
Thus we say $Z_i$ carries a non-trivial representation, or more commonly, a
non-trivial $\Z_2$ charge.  The string operator $Z_\mathrm{str}$ is formed by
two $\Z_2$ charges and carry a trivial total $\Z_2$ charge.  In fact, by
definition, all local symmetric operators carry trivial $\Z_2$ charge (see later
discussion).

We have stressed that a symmetry is fully characterized by its algebra of local
symmetric operators.  But all those local symmetric operators carry no symmetry
charge.  It appears that a key component of symmetry, the symmetry charge ( \ie
the symmetry representation) is missing in our description.

In fact, the symmetry representation can be recovered.  As pointed out in
\Rf{JW191213492}, there is a better way to describe symmetry transformations
using t-patch operators.  We notice that the only use of the symmetry
transformations is to select local symmetric operators.  After that we no
longer need the symmetry transformations.  Since local symmetric operators are
local, we do not need the symmetry transformations that act on the whole space.
We only need symmetry transformations that act on patches to select local
symmetric operators.  This motivates us to introduce patch symmetry
transformation
\begin{align}
W_\mathrm{patch} =
\bigotimes_{i\in \text{patch}} X_i. 
\end{align}
We can use the patch symmetry  transformation $W_\mathrm{patch}$ to define 
the local symmetric operators:
\begin{align}
\cA=\{O_i^\mathrm{symm} \ \big|\ & O_i^\mathrm{symm}W_\mathrm{patch} 
= W_\mathrm{patch} O_i^\mathrm{symm}, 
\nonumber\\
& i \text{ far away from } \prt \mathrm{patch}\} .
\end{align}
So a symmetry can also be defined via the patch symmetry transformations.  

For the $\Z_2$ symmetry in 1-dimensional space, the patch symmetry
transformations happen to be generated by one of the string operators with
non-empty bulk, $X_\mathrm{str}$, and this is why we call them patch symmetry
operators.  In this example, we also see that the string operator
$Z_{\text{str}_{ij}}$ with empty bulk corresponds to a charge-anti-charge pair
operator.  This is the why we call t-patch operators with empty bulk as patch
charge operators.  

\begin{figure}[t]
\begin{center}
\includegraphics[scale=0.5]{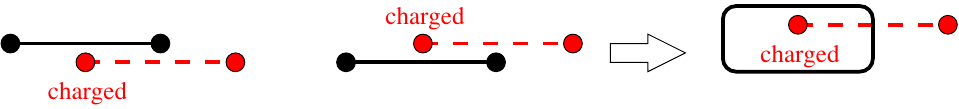}
\end{center}
\caption{ Non-trivial ``braiding'' between two string operators, the patch
symmetry operator (the solid line) and the patch charge operator (the
dashed-line), measures the symmetry charge carried by boundary of
patch charge operator, if the patch symmetry operator generates the symmetry.
} \label{measurecharge1d} 
\end{figure}

The patch symmetry transformations have an advantage that they can detect the
symmetry charge hidden in the patch charge operators (which have zero total
charge): when the patch charge operator $Z_\mathrm{str}$ straddle the boundary
of the patch symmetry transformation $W_\mathrm{patch}$, the two operators have
a non-trivial commutation relation:
\begin{align}
\label{ZWWZ1}
Z_\mathrm{str} W_\mathrm{patch} = - W_\mathrm{patch} Z_\mathrm{str}.
\end{align}
This non-trivial commutation relation measures the charge carried by one end of
the string operator.  

If we view the order of the operator product as the order in time, and view the
string as world line of a particle in spacetime (see Fig.
\ref{measurecharge1d}), then the commutation relation \eqn{ZWWZ1} can be viewed
as a braiding of the charged particle around the boundary of the patch symmetry
operator.  The boundary of the patch symmetry operator can be viewed as a
``symmetry twist flux''.  The charge is measured by a braiding of symmetry
charge around symmetry twist flux.  This is why we refer to \eqn{OA1dZ2c} and
\eqn{OA1dZ2d} as ``braiding'' relations in Fig.  \ref{OAstring}.

\subsection{The algebra of patch charge operators and its braided fusion
category}

\label{ncpatch}

Let us concentrate on patch charge operators.  The properties of the charges of
a symmetry can be systematically and fully described by a braided fusion
category.  To connect the $\Z_2$ symmetry charges to fusion category, we view
the local symmetric operators \ml{O_i^\mathrm{symm}} as the morphisms, and the
ends of string operator $Z_{\mathrm{str}_{ij}}$ (\ie the point-like
\ml{\Z_2}-charge) as objects \ml{ e_i} and \ml{\bar e_j} in a fusion category.
In other words, we write the string operator as 
\begin{align}
Z_{\mathrm{str}_{ij}} = T_e(i\to j) .
\end{align}
The notation $T_e(i\to j)$ is more precise and carries several meanings.  (1) We
view $T_e(i\to j)$ as a world-line of a particle labeled by $e$ that travels
from $i$ to $j$.  $T_{e}(i\to j)$ can also be viewed as a hopping operator of
$e$ from $i$ to $j$.  Here, we have adopted a convention that the arrow
indicate the direction of the hopping.  (2) The notation of string operator
$T_e(i\to j)$ also specify the ordering of operators: the operators near left
index $i$ appears to the left of the operators near the right index $j$.

Since the local symmetric operators $Z_{\mathrm{str}_{ii'}}$ (the morphisms)
can move the string ends (the \ml{\Z_2}-charges): 
\begin{align}
e_i \stackrel{O^\mathrm{symm}}{\to} e_{i'},\ \ e_{i'} \stackrel{O^\mathrm{symm}}{\to} e_i,  
\end{align}
the \ml{\Z_2}-charges (at the string ends) at different places are isomorphic
\ml{e_i\cong e_{i'}}, \ie they belong to the same type of excitations.  More
generally, two excitations that can be connected by local symmetric operators
are regarded as the same type of excitations.


From the above expression of t-patch operators, we can compute the fusion ring
\begin{align}
 a\otimes b = \bigoplus_c N^{ab}_c c
\end{align}
of the braided fusion category.  Notice that $T_e(-\infty \to i)$ creates an
$e$ particle at $i$ (and creates another particle at $-\infty$ which we
ignore).  Creating two $e$ particles, we obtain
\begin{align}
T_e(-\infty \to i) T_e(-\infty \to i) =\id.\label{e2=1}
\end{align}
In other words, we get a trivial particle $\one$.  This allows us to obtain the
fusion rule
\begin{align} \label{eifusion}
e_i\otimes e_i  =  \one.
\end{align}
The  isomorphic relation is an equivalence relation.  After quotienting out the
equivalence relation, \ml{e_i\cong e_j}, we find that the fusion category has
only two objects: \ml{\one,e}.  The morphism of the fusion category  is given
by local symmetric operators $O_i^\mathrm{symm}$ near a vertex-$i$. Also, with this equivalence relation, we can interpret \eqn{eifusion} as 
\begin{align} \label{efusion}
	e\otimes e  =  \one,
\end{align}
which tells us that the $ e $ particle is its own anti-particle. 

\begin{figure}
\centering
\includegraphics[scale=0.6]{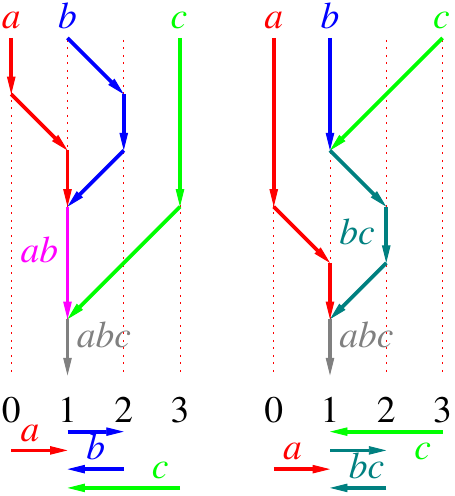}
\caption{Two ways to fuse three particles $a,b,c$ into $abc$, as operator
product.  The phase difference of the two resulting operators is $ F(a,b,c) $.
The horizontal lines and the corresponding 45$^\circ$ lines correspond to
hopping operators. For example $1\xrightarrow{b} 2 \sim T_b(1\to 2)$.  The
hopping operators with higher location are applied first.  Thus we have the
relation $ T_c(3\to 1) T_b(2\to 1) T_a(0\to 1) = F(a,b,c)  T_a(0\to 1) T_c(3\to
1) T_b(2\to 1) $.  } 
\label{Fsym-KL}
\end{figure}

However, the fusion rule $N^{ab}_c$ fails to completely determines the fusion
category, because it is possible for two different fusion categories to have the same
fusion ring.  To complete the description of the fusion category, we also need
to compute the $F$-symbol, which is defined as the relative phases of different
ways to fuse three particles $a,b,c$ together, $a\otimes b \otimes c \to
(ab)\otimes c \to (ab)c$ and $a\otimes b \otimes c \to a\otimes (bc) \to a(bc)$
(see Fig. \ref{Fsym-KL}), if we treat the result of fusion, as quantum state or as
operator:
\begin{align}
 |(ab)c\> &= F(a,b,c)|a(bc)\>,
\nonumber\\
 O((ab)c) &= F(a,b,c) O(a(bc)). 
\end{align}

Following \Rf{Kawagoe}, the $ F $-symbol is computed from the relative phase
of the two ways to compute operator products in Fig \ref{Fsym-KL}. It is trivial to check that
\begin{align}
	\label{Fsyme}
	&\ \ \ \
	T_e(1\to 2) 
	T_e(0\to 1) 
	T_e(2\to 1) 
	T_e(3\to 1)
	\nonumber\\
	&\equiv 
	Z_{\mathrm{str}_{12}} 
	Z_{\mathrm{str}_{01}} 
	Z_{\mathrm{str}_{12}}^\dag 
	Z_{\mathrm{str}_{13}}^\dag
	\nonumber\\
	&=
	Z_{\mathrm{str}_{13}}^\dag 
	Z_{\mathrm{str}_{01}} 
	\nonumber\\
	&=
	T_e(3\to 1) 
	T_\one(1\to 2) 
	T_e(0\to 1) 
	T_\one(2\to 1) 
\end{align}
therefore $ F(e,e,e) =1$.  Similarly, we can show that $ F(\one,\one,\one)=
F(e,\one,\one)= F(\one,e,\one)= F(\one,\one,e)= F(e,e,\one)= F(e,\one,e)=
F(\one,e,e)=1$, since the hopping operators of $e$ and $\one$ particles all
commute.  This implies that the category formed by $\one,e$ and described by
data $N^{ab}_c,\ F(a,b,c)$ is a fusion category $\cRep_{\Z_2}$ -- the fusion
category of the representations of $\Z_2$ group.

\begin{figure}
\centering
\includegraphics[scale=0.6]{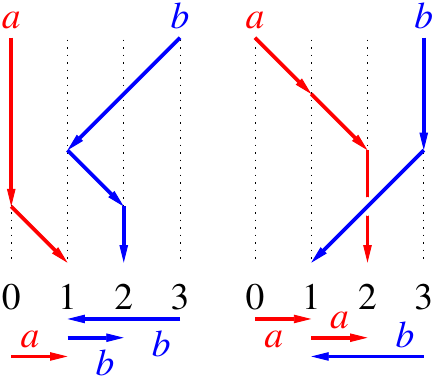}
\caption{The two ways of $a,b$ particle hopping give rise to two configurations
which exchange their positions.  When $a=b$, the phase difference of the two
resulting operators is $ \ee^{\ii\th_a}$, which is the self statistics of the
$a$-particle.  Thus we have a relation $ T_b(3\to 1) T_b(1\to 2) T_a(0\to 1) =
\ee^{\ii \th_a}  T_a(0\to 1) T_b(1\to 2) T_b(3\to 1) $.  } \label{exchange}
\end{figure}

\begin{figure}
\centering
\includegraphics[scale=0.6]{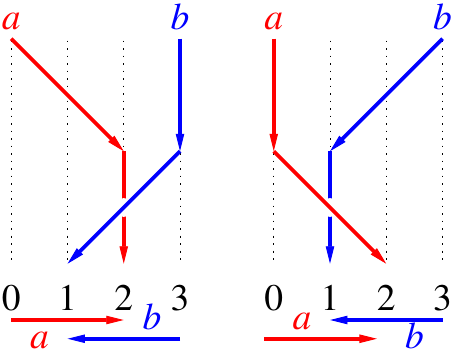}
\caption{The two ways of $a,b$ particles hopping give rise to the same final
configuration but via different braiding paths.  The phase difference of two
hopping processes is $ \ee^{\ii\th_{ab}}$, which is the mutual statistics of
the $a$- and $b$-particles.  Thus we have a relation $ T_b(3\to 1) T_a(0\to 2)
= \ee^{\ii \th_{ab}}  T_a(0\to 2) T_b(3\to 1) $.  } \label{mutual}
\end{figure}

In fact, the $\one,e$ particles not only form a fusion category, they actually
form a braided fusion category.  To calculate the braiding properties, we first
calculate the self statistics of $e $ particle using the statistical hopping
algebra prescription introduced in \Rf{LW0316} and depicted in Fig
\ref{exchange},
\begin{equation}\label{hopalge}
\begin{split}
&\ \ \ \ T_e(0\rightarrow 1) T_e(1\rightarrow 2)
T_e(3\to 1) 
\\
&=Z_{\mathrm{str}_{01}} Z_{\mathrm{str}_{12}} Z_{\mathrm{str}_{13}}^\dag \\
&=\ee^{\ii \th_e} Z_{\mathrm{str}_{13}}^\dag Z_{\mathrm{str}_{12}} Z_{\mathrm{str}_{01}}\\
&=\ee^{\ii \th_e} 
T_e(3\to 1) 
T_e(1\rightarrow 2) T_e(0\rightarrow 1)
\end{split}
\end{equation}
from which we can read off the self-statistical angle $ \ee^{\ii \th_e} =1 $.
This shows that $ e $ particles have \textit{bosonic} self-statistics.  We can
also use Fig \ref{mutual} to compute mutual statistics of $\one,e$ particles.
We find that $\one,e$ particles are bosons with trivial mutual statistics.
This implies that the category formed by $\one,e$ is a braided fusion category
$\eRep_{\Z_2}$.  It is actually a special braided fusion category called
symmetric fusion category, since all the mutual statistics are trivial.

According Tannaka duality, the symmetric fusion category \ml{\eRep_{\Z_2}} can
fully describe the symmetry group \ml{G=\Z_2}.  So, instead of using a group $G$
(formed by the transformations), we can also use a symmetric fusion category of
patch charge operators (\ie formed by charged objects or the representations)
to fully describe a symmetry.

\subsection{Representation category and symmetry}
\label{repcat}

Above picture also works for generic finite group $G$: a symmetry $G$ can also
be described by a symmetric fusion category $\eRep_G$ (formed by the
representations of $G$).  This is the categorical point of view of symmetry,
which was used in \Rf{LW160205936, KZ200308898} and will be used in this paper.
The symmetric fusion category generated by patch charge operators is nothing
but the mathematical framework that describes the properties of symmetry
charges (such as their fusion and braiding).
\begin{DefinitionPH}
\label{def:repcat1}
We will call the  symmetric fusion category $\eR$ formed by patch charge
operators as \textbf{representation category}.\cite{W181202517}  
\end{DefinitionPH}
In fact, there is another definition of representation category.  We may ignore
the braiding structure and consider the  fusion category $\cR$ formed by patch
charge operators. Instead of the braiding structure, we consider a
symmetry-breaking structure, \ie a faithful functor $\bt: \cR \to \cVec$ (which
is also called fiber functor) that describe the process of ignoring the
symmetry:
\begin{DefinitionPH}
\label{def:repcat2}
If a fusion category $\cR$ has a fiber functor $\bt$, then the pair $(\cR,\bt)$
will be called a \textbf{local fusion category}.  Such a local fusion category
can also be viewed as the \textbf{representation category} of the symmetry.
\cite{KZ200514178}.
\end{DefinitionPH}
In higher dimension, the notion of symmetric fusion higher category used in
\Rf{KZ200308898} may be hard to define.  In this case, the second Definition
\ref{def:repcat2} is can be used as in \Rf{KZ200514178}.

Thus we can say that an anomaly-free symmetry in 1-dimensional space described
by a finite group $G$ is fully described by its representation category, a
symmetric fusion category $\eRep_G$ or a local fusion category $(\cRep_G,\bt)$.
This point of view can be generalized to described anomaly-free symmetries
beyond group and higher group.  In \Rf{KZ200514178}, it is proposed that the
most general anomaly-free symmetries in $n$-dimensional space are fully
described by \textbf{local fusion $n$-categories} $(\cR,\bt)$.  Such a
description includes non-invertible symmetries (\ie algebraic higher
symmetries).

In the above we have used a notion of \textbf{anomaly-free symmetry}.  For
symmetry described by group and/or higher group, an anomaly-free symmetry is
defined as a symmetry that can be gauged.  But such a definition does not apply
to non-invertible symmetries, for which we do not how to gauge them.  To solve
this problem, \Rf{KZ200514178} proposed the following macroscopic definition
without using gauging
\begin{DefinitionPH}
\label{def:anofree1}
\textbf{Anomaly-free symmetry} is the symmetry that allows non-degenerate
symmetric gapped states for any closed space manifolds.
\end{DefinitionPH}
A microscopic definition was also proposed
\begin{DefinitionPH}
\label{def:anofree2}
\textbf{Anomaly-free symmetry} is the symmetry that allows symmetric state of
form $|\Psi\>=\bigotimes_i |\psi_i\>$, where $|\psi_i\>$ is a symmetric state
on site-$i$.
\end{DefinitionPH}

We would like to remark that representation categories (\ie symmetric fusion
$n$-categories or local fusion $n$-categories) only fully describe anomaly-free
symmetries, but fail to fully describe anomalous symmetries.  This is because
different anomalous symmetries can have the same representation category.  In
fact, an anomalous symmetry $G$ can be described by symmetry transformations
$W_g, \ g \in G$: $W_g W_h = W_{gh}$ that may not be on-site.  The non-invariant
local operators that form representations of of the symmetry group $G$. Thus
\begin{Proposition}
all the different anomalous symmetries of the same group $G$ have the same
representation category $\eRep_G$.  
\end{Proposition}

Later in Section \ref{z2z2beyond}, we will give a 1d example of emergent
$\Z_2\times \Z_2$ symmetry, whose representation category formed by the
$\Z_2\times \Z_2$ charge is not a local fusion category.  This implies that the
emergent $\Z_2\times \Z_2$ symmetry is not an anomaly-free symmetry, since the
representation categories of all anomaly-free symmetries are local fusion
categories.  This also implies that the  emergent $\Z_2\times \Z_2$ symmetry is
not an anomalous symmetry (in the usual sense), since the representation
categories of all anomalous symmetries are also described by local fusion
categories.  Here we have an example of an emergent symmetry that is neither
anomalous nor anomaly-free.

\subsection{The algebra of patch symmetry operators and its braided fusion category -- transformation category}
\label{transcat}

In the above, we show that the operator algebra of a class of string operators,
the patch charge operators $Z_\text{str}$, gives rise to a symmetric fusion
category $\eRep_{\Z_2}$.  In this section, we are going to consider the operator
algebra of another class of t-patch operators $X_\text{str}$, the patch
symmetry operators, and show that they give rise to a fusion category
$\eVec_{\Z_2}$ which happens to be isomorphic to $\eRep_{\Z_2}$.

Patch symmetry operators are defined by restricting the global symmetry to
finite patches as discussed in Sec \ref{psymm} (assume $i<j$)
\begin{equation}\label{patchOpDef}
W_{\text{patch}_{ij}}=X_{\text{str}_{ij}}=
X_{i+1}
\cdots
X_{j-1}
X_{j}
.
\end{equation} 
Since the bulk of a patch symmetry operator is invisible, it is completely
legitimate to think of the boundary of the 1d  patch symmetry operator as
particles.  We can define a fusion operation of those particles, which is
called $ m $.\footnote{The reason for this name will become clear in next
subsection.} Analogous to the discussion in the previous subsection, we may
construct a braided fusion category corresponding to these $ m $ particles.

To do so, we can think of these patch symmetry operators as operators that
transport $ m $ particles from one point to another on the one-dimensional
space 
\begin{equation}\label{Tm}
W_{\text{patch}_{ij}}=T_m(i\to j),\ \
\end{equation}
The above can also be viewed as a world-line of $ m $ particle from $i$ to $j$.
In fact, the $m$ particle live on the link, such as $\<i,i+1\>$. In the above,
we view such a $m$ particle as located at $i$.  


We can work out the fusion of the $ m $ particles as we did for the $ e $
particles in Sec \ref{ncpatch}.  From
\begin{align}
T_m(-\infty \to i)
T_m(-\infty \to i) = \id
\end{align}
we find the fusion rule $ m\otimes m=\one $. It tells us that the $ m $ particles are their own antiparticles.

Next, we work out $F$-symbol from Fig. \ref{Fsym-KL}.  We find that $F(a,b,c)=1$
for $a,b,c = \one,m$.   This is not surprising because the patch operators all commute with each other since they are just products of Pauli $X$ operators
and identity operators.
Thus $\one,m$ form a fusion category $\cVec_{\Z_2}$,
which is isomorphic to $\cRep_{\Z_2}$.

$\one,m$ also have a braiding structure and form a braided fusion category.
Using Fig. \ref{exchange}, we find that $ m $ particles have trivial
self-statistics.  Using Fig. \ref{mutual}, we find that $\one$ and $ m $
particles have trivial mutual statistics.  This allows us to show that $\one,m$
form a symmetric fusion category $\eVec_{\Z_2}$.

\begin{DefinitionPH}
\label{def:trancat1}
We will call the symmetric fusion category $\eT$ formed by patch symmetry
operators as \textbf{transformation category}.\cite{W181202517}  
\end{DefinitionPH}
Similar to representation category, we believe that the transformation category
in $n$-dimension space can also be described by local fusion $n$-categories.
We ignore the braiding structure and consider the fusion category $\cT$ formed
by patch symmetry operators.  We replace braiding structure with a faithful
functor $\bt: \cT \to \cVec$.\cite{KZ200514178} The local fusion category
$(\cT,\bt)$ can also be viewed as the transformation category of the symmetry.

\subsection{The algebra of all string operators and its non-degenerate braided
fusion category}\label{z2all}

In this subsection, we are going to consider the operator algebra of all string
operators, \ie the patch charge operators  $Z_\text{str}$ and the patch
symmetry operators $X_\text{str}$.  The isomorphic class of such a complete
operator algebra is called a \catsymm.

We have seen that the algebra of $Z_\text{str}$ corresponds to a  symmetric
fusion category $\eRep_{\Z_2}$, and the algebra of $X_\text{str}$ corresponds to
a symmetric fusion category $\eVec_{\Z_2} $.  The algebra of $Z_\text{str}$ and
$X_\text{str}$ corresponds to a braided fusion category formed by $\eRep_{\Z_2}$
and  $\eVec_{\Z_2} $.  Here, we would like to show that such a  braided fusion
category describes the topological excitations in $\Z_2$-topological order with
topological excitations $\one,e,m,f$ in 2-dimensional space.  We will denote
such a  braided fusion category as $\eGau_{\Z_2}$. 

The  algebra  of $Z_\text{str}$ and $X_\text{str}$ also contain their product
\begin{equation}\label{Wf}
X_{\text{str}_{ij}}Z_{\text{str}_{ij}} =T_f(i\to j)
\end{equation}
$T_f(i\to j)$ is the world-line
of a new particle $f$.
We see that a $ f $ particle at $ i $ is the bound
state of an $ e $ particle at $ i $ and an $ m $ particle on the link
$\<i, i+1\> $.
$T_f(i\to j)$ satisfies the following algebra
\begin{align*}
T_f(i\to j) T_f(j\to k)
&= X_{i+1}\cdots X_j Z_i Z_j  X_{j+1}\cdots X_k Z_j Z_k
\nonumber\\
& =X_{i+1}\cdots X_k Z_i Z_k= T_f(i\to k)
\end{align*}
Next, we compute the fusion rules for the $ f $ particles:
\begin{align}
T_f(-\infty\to i)
T_f(-\infty\to i) = -\id
\end{align}
This implies the fusion rule $f\otimes f = \one$, upto an overall phase factor. This phase factor does not carry any meaning for the fusion rule.

We can also compute the $ F $ symbol for the $ f $ particle, using the prescription in Fig \ref{Fsym-KL}. It is easy to check that all the components of $ F(a,b,c) $ with $ a,b,c = \one, f $ are equal to 1. Let us compute two cases explicitly, $ F(f,f,f) $ and $ F(f,1,f) $. The first, $ F(f,f,f) $, is obtained as follows:
\begin{align}
	\label{Ffff}
	&\ \ \ \
	T_f(1\to 2) 
	T_f(0\to 1) 
	T_f(2\to 1) 
	T_f(3\to 1)
	\nonumber\\
	&\equiv 
	X_{\mathrm{str}_{12}} Z_{\mathrm{str}_{12}} 
	X_{\mathrm{str}_{01}} Z_{\mathrm{str}_{01}} 
	Z_{\mathrm{str}_{12}}^\dag X_{\mathrm{str}_{12}}^\dag 
	Z_{\mathrm{str}_{13}}^\dag X_{\mathrm{str}_{13}}^\dag
	\nonumber\\
	&=
	Z_{\mathrm{str}_{13}}^\dag X_{\mathrm{str}_{13}}^\dag 
	 X_{\mathrm{str}_{01}} Z_{\mathrm{str}_{01}}
	\nonumber\\
	&=
	T_f(3\to 1) 
	T_\one(1\to 2) 
	T_f(0\to 1) 
	T_\one(2\to 1) 
\end{align}
\ie $ F(f,f,f)=1 $
and the second, $ F(f,1,f) $, is obtained from:
\begin{align}
	\label{Ff1f}
	&\ \ \ \
	T_\one(1\to 2) 
	T_f(0\to 1) 
	T_\one(2\to 1) 
	T_f(3\to 1)
	\nonumber\\
	&\equiv 
	X_{\mathrm{str}_{01}} Z_{\mathrm{str}_{01}} 
	Z_{\mathrm{str}_{13}}^\dag X_{\mathrm{str}_{13}}^\dag
	\nonumber\\
	&=
	Z_{\mathrm{str}_{13}}^\dag X_{\mathrm{str}_{13}}^\dag 
	X_{\mathrm{str}_{12}} Z_{\mathrm{str}_{12}} 
	X_{\mathrm{str}_{01}}  Z_{\mathrm{str}_{01}}
	Z_{\mathrm{str}_{12}}^\dag X_{\mathrm{str}_{12}}^\dag 
	\nonumber\\
	&=
	T_f(3\to 1) 
	T_f(1\to 2) 
	T_f(0\to 1) 
	T_f(2\to 1) 
\end{align}
\ie $ F(f,1,f)=1 $. The product $ F(f,f,f) F(f,1,f) = 1$ is gauge invariant; in fact, (the sign of) this product encodes the Frobenius-Schur indicator of $ f $.

Further, we can calculate the self-statistics of the $ f $ particle using the hopping
algebra method used in previous subsections,
\begin{align}
\label{hopalgf}
&\ \ \ \
T_f(3\to 1) 
T_f(1\to 2)
T_f(0\to 1) 
\nonumber\\
&=
(X_2X_3Z_1Z_3)^\dag
(X_2Z_1Z_2)
(X_1Z_0Z_1)
\nonumber\\
&= -
(X_1Z_0Z_1)
(X_2Z_1Z_2)
(X_2X_3Z_1Z_3)^\dag
\nonumber\\
&=  - 
T_f(0\to 1) 
T_f(1\to 2) 
T_f(3\to 1), 
\end{align}
from which we find that $ f $ particles have fermionic self-statistics. 

Mutual statistics of $ e $, $ m $, and $f$ particles can be obtained by the use of the patch operators. For example, when $i<k<j<l$, we have
\begin{equation}
Z_{\text{str}_{ij}} X_{\mathrm{str}_{kl}}  = - X_{\mathrm{str}_{kl}}  Z_{\text{str}_{ij}}. 
\end{equation}
Thus the $ e $ and $ m $ particles have $ \pi $ mutual statistics.  In fact,
the $ e $, $m$, and $ f $ particles all have $ \pi $ mutual statistics respect
to each other. 

Since every non-trivial topological excitations (\ie $e,m,f$) can be detected
remotely via  mutual statistics, the particles $\one,e,m,f$ form a
non-degenerate braided fusion category $\eGau_{\Z_2}$.  We believe that such a
non-degenerate braided fusion category fully characterized the isomorphic class
of the algebras of local symmetric operators.  Thus \catsymm\ is
fully characterized by non-degenerate braided fusion category.  Since the
non-degenerate braided fusion category describes a topological order in
2-dimensional space, we can also say that \catsymm\ is fully
characterized by topological order in one higher dimension.  This connection
between  algebra of local symmetric operators and non-degenerate braided fusion
category, as well as topological order in one higher dimension is the key
result of this paper.

\subsection{A holographic way to compute \catsymm}

In the above, we have computed the \catsymm\ of $\Z_2$ symmetry
directly from the definition of \catsymm, \ie from the algebra of
local symmetry operators and their string extensions.  We find that the
\catsymm\ of $\Z_2$ symmetry is a topological order in one higher
dimension.  In fact, we can compute this topological order in one higher
dimension directly.

We know that a system with $\Z_2$ symmetry can be realized as a boundary of a
trivial product state with $\Z_2$ symmetry in one higher dimension.  If we gauge
the bulk symmetric product state, we will obtain a $\Z_2$ topological order
$\eGau_{\Z_2}$ described by $\Z_2$ gauge theory.  Such a $\Z_2$ topological order
in one higher dimension happen to be the \catsymm\ of $\Z_2$ symmetry.

This result can be generalized.  An anomaly-free (higher) symmetry described by
(higher) group $G$ can be realized as a boundary of a trivial product state
with $G$ (higher) symmetry in one higher dimension.  If we gauge the bulk
symmetric product state, we will obtain a topological order $\eGau_G$ described
by $G$ (higher) gauge theory.  Such a topological order in one higher dimension
is the \catsymm\ of the $G$ (higher) symmetry.

We note that in \Rf{AAX211112096}, the authors consider various $ G $-symmetric 1+1D models as realized on the edge of 2+1D $ G $-gauge theory (\ie $ G $ quantum double). This is one particular instance of the general argument we present in this paper.

\section{A 1\texorpdfstring{\MakeLowercase{d}}{d} bosonic quantum system with anomalous $ \Z_2 $ symmetry}

Now we discuss the next simplest example: a bosonic system 
\begin{align}
\label{aZ2}
 H_{\text{a}\Z_2} 
&=
- B  \sum_{i=1}^L Z_iZ_{i+1} 
- J_1 \sum_{i=1}^{L} (X_i - Z_{i-1}X_i Z_{i+1} ) 
\nonumber\\
&
+ J_2  \sum_{i=1}^{L} Z_{i-1}( X_i + Z_{i-1}X_i Z_{i+1}),
\end{align}
in 1-dimensional space with an anomalous $\Z_2$ symmetry (\ie a non-on-site
symmetry).  Our discussions here follow closely the discussions in the last
section.

The non-on-site $ \Z_2 $ symmetry
\cite{CLW1141, CGL1314, WW170506728, JW191213492}) is described by the symmetry
operator
\begin{equation}\label{Uz2anom}
W=\prod_i X_i \prod_i s_{i,i+1}=\prod_i X_i \prod_i \ii^{\frac{-Z_i+Z_{i+1}+Z_i Z_{i+1}-1}{2}}
\end{equation}
which we represent pictorially as
\begin{equation*}
\begin{tikzpicture}
	\begin{pgfonlayer}{nodelayer}
		\node [style=none] (1) at (7, 0) {};
		\node [style=none] (2) at (-7, 0) {};
		\node [style=none] (3) at (-6, 0.5) {};
		\node [style=none] (4) at (-6, -0.5) {};
		\node [style=none] (7) at (-4, 0.5) {};
		\node [style=none] (8) at (-2, 0.5) {};
		\node [style=none] (9) at (0, 0.5) {};
		\node [style=none] (10) at (0, -0.5) {};
		\node [style=none] (11) at (-2, -0.5) {};
		\node [style=none] (12) at (-4, -0.5) {};
		\node [style=none] (13) at (2, 0.5) {};
		\node [style=none] (14) at (4, 0.5) {};
		\node [style=none] (15) at (6, 0.5) {};
		\node [style=none] (17) at (2, -0.5) {};
		\node [style=none] (18) at (4, -0.5) {};
		\node [style=none] (19) at (6, -0.5) {};
		\node [style=none] (21) at (-6.5, -2) {};
		\node [style=none] (22) at (-5, -2) {};
		\node [style=none] (29) at (-2, -1.25) {$X$};
		\node [style=none] (30) at (0, -1.25) {$X$};
		\node [style=none] (31) at (2, -1.25) {$X$};
		\node [style=none] (32) at (4, -1.25) {$X$};
		\node [style=none] (34) at (-1.75, -2.25) {};
		\node [style=none] (35) at (-0.25, -2.25) {};
		\node [style=none] (37) at (0.25, -2.25) {};
		\node [style=none] (38) at (1.75, -2.25) {};
		\node [style=none] (39) at (2.25, -2.25) {};
		\node [style=none] (40) at (3.75, -2.25) {};
		\node [style=none] (43) at (-2.25, -2.25) {};
		\node [style=none] (44) at (-3.75, -2.25) {};
		\node [style=cyan rect] (45) at (-3, -2.25) {$s$};
		\node [style=cyan rect] (46) at (-1, -2.25) {$s$};
		\node [style=cyan rect] (47) at (1, -2.25) {$s$};
		\node [style=cyan rect] (48) at (3, -2.25) {$s$};
		\node [style=none] (50) at (-4, -1.25) {$X$};
		\node [style=none] (58) at (6.5, -2) {};
		\node [style=none] (59) at (5, -2) {};
	\end{pgfonlayer}
	\begin{pgfonlayer}{edgelayer}
		\draw (2.center) to (1.center);
		\draw (3.center) to (4.center);
		\draw (7.center) to (12.center);
		\draw (8.center) to (11.center);
		\draw (9.center) to (10.center);
		\draw (13.center) to (17.center);
		\draw (14.center) to (18.center);
		\draw (15.center) to (19.center);
		\draw [style=thicc blue, in=180, out=0] (44.center) to (43.center);
		\draw [style=thicc blue] (34.center) to (35.center);
		\draw [style=thicc blue] (37.center) to (38.center);
		\draw [style=thicc blue] (39.center) to (40.center);
		\draw [style=black dots] (22.center) to (21.center);
		\draw [style=black dots] (59.center) to (58.center);
	\end{pgfonlayer}
\end{tikzpicture}
\end{equation*}
where the operators on top act first. The phase factor $ s_{i,i+1} $ is real despite appearances as can be checked by substituting $ \{+1,-1\} $ for $ Z_i $ and $ Z_{i+1} $ (\ie we work in the Z basis). It's easy to see that it evaluates to +1 when there is no domain wall between $ i $ and $ i+1 $. Moreover it evaluates to $ -1 $ for only one kind of domain walls, the $ +1 \to -1 $ kind; it evaluates to $ +1 $ for the $ -1 \to +1 $ kind.

\subsection{Braided fusion category of patch symmetry transformation operator}

In order to identify the braided fusion category (\ie the \catsymm)
corresponding to this anomalous symmetry, we will work out the patch symmetry
operators corresponding to the above global symmetry operation. Legitimate
patch symmetry operators must satisfy the transparent and fusion
properties \ie
\begin{enumerate}
\item $ W_{\text{patch}_{ij}} W_{\text{patch}_{jk}}=W_{\text{patch}_{ik}} $ for $ i<j<k $
\item $ W_{\text{patch}_{ij}} W_{\text{patch}_{kl}} W_{\text{patch}_{ij}}^\dag= W_{\text{patch}_{kl}} $ for $ i<k<l<j $
\end{enumerate}
In order to ensure these properties are satisfied, we need to choose appropriate boundary operators for the $  W_{\text{patch}_{ij}} $. To that end, we propose the following definition:
\begin{equation}\label{anomZ2patch}
W_{\text{patch}_{ij}}=O_i^\dag \prod_{k=i+1}^j X_k \ \ (-Z_j O_j) \prod_{k=i+1}^j s_{k,k+1}
\end{equation}
where $ O_j=(1-\ii Z_j)/\sqrt2 $. We may write this operator pictorially as
\begin{equation*}
\begin{tikzpicture}
	\begin{pgfonlayer}{nodelayer}
		\node [style=none] (1) at (8.5, 0) {};
		\node [style=none] (2) at (-6.5, 0) {};
		\node [style=none] (3) at (-6, 0.5) {};
		\node [style=none] (4) at (-6, -0.5) {};
		\node [style=none] (7) at (-4, 0.5) {};
		\node [style=none] (8) at (-2, 0.5) {};
		\node [style=none] (9) at (0, 0.5) {};
		\node [style=none] (10) at (0, -0.5) {};
		\node [style=none] (11) at (-2, -0.5) {};
		\node [style=none] (12) at (-4, -0.5) {};
		\node [style=none] (13) at (2, 0.5) {};
		\node [style=none] (14) at (4, 0.5) {};
		\node [style=none] (15) at (6, 0.5) {};
		\node [style=none] (17) at (2, -0.5) {};
		\node [style=none] (18) at (4, -0.5) {};
		\node [style=none] (19) at (6, -0.5) {};
		\node [style=none] (21) at (-6, 1.25) {};
		\node [style=none] (22) at (-4, 1.25) {$i$};
		\node [style=none] (23) at (-2, 1.25) {};
		\node [style=none] (24) at (0, 1.25) {};
		\node [style=none] (25) at (2, 1.25) {};
		\node [style=none] (26) at (4, 1.25) {$j$};
		\node [style=none] (27) at (6, 1.25) {};
		\node [style=none] (29) at (-2, -1.25) {$X$};
		\node [style=none] (30) at (0, -1.25) {$X$};
		\node [style=none] (31) at (2, -1.25) {$X$};
		\node [style=none] (34) at (-1.75, -2.25) {};
		\node [style=none] (35) at (-0.25, -2.25) {};
		\node [style=none] (37) at (0.25, -2.25) {};
		\node [style=none] (38) at (1.75, -2.25) {};
		\node [style=none] (39) at (2.25, -2.25) {};
		\node [style=none] (40) at (3.75, -2.25) {};
		\node [style=cyan rect] (46) at (-1, -2.25) {$s$};
		\node [style=cyan rect] (47) at (1, -2.25) {$s$};
		\node [style=cyan rect] (48) at (3, -2.25) {$s$};
		\node [style=none] (49) at (-4, -1.25) {$O^\dagger$};
		\node [style=none] (50) at (4.25, -2.25) {};
		\node [style=none] (51) at (5.75, -2.25) {};
		\node [style=cyan rect] (52) at (5, -2.25) {$s$};
		\node [style=none] (54) at (8, 0.5) {};
		\node [style=none] (55) at (8, -0.5) {};
		\node [style=none] (58) at (4, -1.25) {$ZX$};
		\node [style=none] (59) at (4, -3.25) {$O$};
	\end{pgfonlayer}
	\begin{pgfonlayer}{edgelayer}
		\draw (2.center) to (1.center);
		\draw (3.center) to (4.center);
		\draw (7.center) to (12.center);
		\draw (8.center) to (11.center);
		\draw (9.center) to (10.center);
		\draw (13.center) to (17.center);
		\draw (14.center) to (18.center);
		\draw (15.center) to (19.center);
		\draw [style=thicc blue] (34.center) to (35.center);
		\draw [style=thicc blue] (37.center) to (38.center);
		\draw [style=thicc blue] (39.center) to (40.center);
		\draw [style=thicc blue] (50.center) to (51.center);
		\draw (54.center) to (55.center);
	\end{pgfonlayer}
\end{tikzpicture}
\end{equation*}
It is straightforward to check that this satisfies the properties mentioned above. Let us label the particles at the boundaries of this patch operator as $ s $. The patch symmetry operator $ W_{\text{patch}_{ij}} $ can also be understood as an operator transporting an $ s $ particle from $ i $ to $ j $, \ie 
\begin{equation}\label{Ts}
T_s(i\to j)=W_{\text{patch}_{ij}}
\end{equation}

The fusion of $ s $ particles turns out to be identical to that of the $ e $
particles discussed above: they are their own antiparticles so that $ s\otimes
s=\one $.  Here $\one$ is the trivial particle, an end of trivial string formed
by product of identity operators.  To see this fact, we consider the product of
two semi-infinite strings as in \eqn{e2=1}.
\begin{equation}\label{ss}
T_s(-\infty \to i) T_s(-\infty \to i) = O_{-\infty} (-Z_{i}Z_{i+1})
\end{equation}
where we use $ O_{-\infty} $ to represent a local symmetric
operator at $ -\infty $ of the type $ Z_j Z_{j+1} $ (see section \ref{aZ2ch}). Note that such a local
symmetric operator represents a trivial particle $\one$, so we can ignore it.  A graphical representation of this is shown below.
\begin{equation*}
\begin{tikzpicture}
	\begin{pgfonlayer}{nodelayer}
		\node [style=none] (1) at (8.5, 0) {};
		\node [style=none] (2) at (-6.5, 0) {};
		\node [style=none] (7) at (-4, 0.5) {};
		\node [style=none] (8) at (-2, 0.5) {};
		\node [style=none] (9) at (0, 0.5) {};
		\node [style=none] (10) at (0, -0.5) {};
		\node [style=none] (11) at (-2, -0.5) {};
		\node [style=none] (12) at (-4, -0.5) {};
		\node [style=none] (13) at (2, 0.5) {};
		\node [style=none] (14) at (4, 0.5) {};
		\node [style=none] (15) at (6, 0.5) {};
		\node [style=none] (17) at (2, -0.5) {};
		\node [style=none] (18) at (4, -0.5) {};
		\node [style=none] (19) at (6, -0.5) {};
		\node [style=none] (21) at (-5, 1) {};
		\node [style=none] (22) at (-4, 1.25) {};
		\node [style=none] (23) at (-2, 1.25) {};
		\node [style=none] (24) at (0, 1.25) {};
		\node [style=none] (25) at (2, 1.25) {};
		\node [style=none] (26) at (4, 1.25) {};
		\node [style=none] (27) at (6, 1.25) {};
		\node [style=none] (29) at (-2, -1.5) {$X$};
		\node [style=none] (30) at (0, -1.5) {$X$};
		\node [style=none] (31) at (2, -1.5) {$X$};
		\node [style=none] (32) at (4, -3.5) {$O$};
		\node [style=none] (34) at (-1.75, -2.5) {};
		\node [style=none] (35) at (-0.25, -2.5) {};
		\node [style=none] (37) at (0.25, -2.5) {};
		\node [style=none] (38) at (1.75, -2.5) {};
		\node [style=none] (39) at (2.25, -2.5) {};
		\node [style=none] (40) at (3.75, -2.5) {};
		\node [style=none] (43) at (-2.25, -2.5) {};
		\node [style=none] (44) at (-3.75, -2.5) {};
		\node [style=cyan rect] (45) at (-3, -2.5) {$s$};
		\node [style=cyan rect] (46) at (-1, -2.5) {$s$};
		\node [style=cyan rect] (47) at (1, -2.5) {$s$};
		\node [style=cyan rect] (48) at (3, -2.5) {$s$};
		\node [style=none] (50) at (4.25, -2.5) {};
		\node [style=none] (51) at (5.75, -2.5) {};
		\node [style=cyan rect] (52) at (5, -2.5) {$s$};
		\node [style=none] (54) at (8, 0.5) {};
		\node [style=none] (55) at (8, -0.5) {};
		\node [style=none] (58) at (4, -1.5) {$ZX$};
		\node [style=none] (59) at (-6, -2.5) {};
		\node [style=none] (60) at (-4.75, -2.5) {};
		\node [style=none] (61) at (-4, -1.5) {$X$};
		\node [style=none] (62) at (-6.5, 1) {};
		\node [style=none] (63) at (-5.75, 1.75) {$-\infty$};
		\node [style=none] (64) at (-2, -4.5) {$X$};
		\node [style=none] (65) at (0, -4.5) {$X$};
		\node [style=none] (66) at (2, -4.5) {$X$};
		\node [style=none] (67) at (4, -6.5) {$O$};
		\node [style=none] (68) at (-1.75, -5.5) {};
		\node [style=none] (69) at (-0.25, -5.5) {};
		\node [style=none] (70) at (0.25, -5.5) {};
		\node [style=none] (71) at (1.75, -5.5) {};
		\node [style=none] (72) at (2.25, -5.5) {};
		\node [style=none] (73) at (3.75, -5.5) {};
		\node [style=none] (74) at (-2.25, -5.5) {};
		\node [style=none] (75) at (-3.75, -5.5) {};
		\node [style=cyan rect] (76) at (-3, -5.5) {$s$};
		\node [style=cyan rect] (77) at (-1, -5.5) {$s$};
		\node [style=cyan rect] (78) at (1, -5.5) {$s$};
		\node [style=cyan rect] (79) at (3, -5.5) {$s$};
		\node [style=none] (80) at (4.25, -5.5) {};
		\node [style=none] (81) at (5.75, -5.5) {};
		\node [style=cyan rect] (82) at (5, -5.5) {$s$};
		\node [style=none] (83) at (4, -4.5) {$ZX$};
		\node [style=none] (84) at (-6, -5.5) {};
		\node [style=none] (85) at (-4.75, -5.5) {};
		\node [style=none] (86) at (-4, -4.5) {$X$};
		\node [style=none] (87) at (0, -7.25) {};
		\node [style=none] (88) at (0, -7.25) {=};
		\node [style=none] (104) at (-4, -8.25) {};
		\node [style=none] (105) at (-2, -8.25) {};
		\node [style=none] (106) at (0, -8.25) {};
		\node [style=none] (107) at (2, -8.25) {};
		\node [style=none] (108) at (4, -8.25) {};
		\node [style=none] (109) at (6, -8.25) {};
		\node [style=none] (114) at (6, -8.25) {$Z$};
		\node [style=none] (115) at (4, -8.25) {$-Z$};
		\node [style=none] (116) at (-7.5, -8.25) {$O_{-\infty}$};
		\node [style=none] (117) at (-6, -8.25) {};
		\node [style=none] (118) at (-4.75, -8.25) {};
	\end{pgfonlayer}
	\begin{pgfonlayer}{edgelayer}
		\draw (2.center) to (1.center);
		\draw (7.center) to (12.center);
		\draw (8.center) to (11.center);
		\draw (9.center) to (10.center);
		\draw (13.center) to (17.center);
		\draw (14.center) to (18.center);
		\draw (15.center) to (19.center);
		\draw [style=thicc blue, in=180, out=0] (44.center) to (43.center);
		\draw [style=thicc blue] (34.center) to (35.center);
		\draw [style=thicc blue] (37.center) to (38.center);
		\draw [style=thicc blue] (39.center) to (40.center);
		\draw [style=thicc blue] (50.center) to (51.center);
		\draw (54.center) to (55.center);
		\draw [style=black dots] (60.center) to (59.center);
		\draw [style=arrow-to-source] (62.center) to (21.center);
		\draw [style=thicc blue, in=180, out=0] (75.center) to (74.center);
		\draw [style=thicc blue] (68.center) to (69.center);
		\draw [style=thicc blue] (70.center) to (71.center);
		\draw [style=thicc blue] (72.center) to (73.center);
		\draw [style=thicc blue] (80.center) to (81.center);
		\draw [style=black dots] (85.center) to (84.center);
		\draw [style=black dots] (118.center) to (117.center);
	\end{pgfonlayer}
\end{tikzpicture}
\end{equation*}
The product is identical to a $ \one $-patch operator modulo the LSOs at $ -\infty $ and near $ i $. So we can conclude that this corresponds to the fusion $ s\otimes s = \one $. The complete fusion ring is given by
\begin{align}
 s\otimes s= \one,\ \ \
 s\otimes \one= s,\ \ \
 \one \otimes \one= \one,
\end{align}
or equivalently,
\begin{align}
 N^{\one\one}_\one =
 N^{ss}_\one =
 N^{s\one}_s =
 N^{\one s}_s =1,\ \
\text{ others } = 0.
\end{align}

Fusion ring $N^{ab}_c$ is only a part of data that describe the braided fusion
category.  We need to supply the $F$-symbol, $F(a,b,c)$, to promote the fusion
ring to a fusion category.  Similar to the $e$ and $ f $ particles, we have $F(s,\one,s)=1$. 
However, the $ F $-symbol $ F(s,s,s) $ is different (again, referring to Fig
\ref{Fsym-KL}):
\begin{multline}\label{Fsss}
W_{\text{patch}_{jk}} W_{\text{patch}_{ij}} W_{\text{patch}_{jk}}^\dag W_{\text{patch}_{jl}}^\dag\\
=F(s,s,s) W_{\text{patch}_{jl}}^\dag \one_{\text{patch}_{jk}} W_{\text{patch}_{ij}} \one_{\text{patch}_{jk}}^\dag
\end{multline}
Working out the algebra (see Appendix \ref{Fsymbdeets}) gives us $ F(s,s,s)=-1 $. The guage-invariant product $ F(s,s,s) F(s,\one,s) =-1$ gives us a non-trivial Frobenius-Schur indicator, unlike in the cases of $ e $ and $ f $ discussed in the previous sections.
This distinguishes the fusion category encoded by the anomalous $ \Z_2 $ patch
symmetry operators from that of the anomaly-free $ \Z_2 $ symmetry without even
considering the braiding structure.

Similarly, the fusion category data, $(N^{ab}_c, F(a,b,c))$, is only a part of
data to describe a braided fusion category.  To obtain the full data to
describe a braided fusion category, we need to supply the data that describes
mutual and self statistics.  The mutual statistics between $s$ and $\one$ is
trivial $\th_{s\one}=0$.  We can calculate the self statistics of $s$ by
calculating the statistical hopping algebra of the particle-like endpoints of
the patch symmetry operator, as outlined above in Fig.  \ref{exchange}. In this
case, we find (see Appendix  \ref{ssstat}) $ \ee^{\ii \th_s}=\ii$, \ie a
statistical phase of $ \th_s= \pi/2 $. This shows that the endpoints are
\emph{semions}.  Thus unlike the anomaly-free $\Z_2$ symmetry, the
transformation category of anomalous $ \Z_2 $ symmetry is not a symmetric fusion
category.  The transformation category happens to be non-degenerate, and
correspond to the single-semion topological order in 2d, which will be denoted
as $\eM_\text{single-semion}$. Note that, in general, a transformation category may be
degenerate, in which case it does not correspond to a topological order in one higher
dimension.

\subsection{Braided fusion category of patch charge operator}\label{aZ2ch}

We can also define patch charge operators for anomalous $ \Z_2 $ symmetry, which
have empty bulk and a pair of $ \Z_2 $ charges at the endpoints,
\begin{equation}\label{dualpatch}
Z_{\text{string}_{ij}}=Z_i Z_j
\end{equation}
Let us label the particles at the ends of this operator as $ b $.  This
operator is identical to the patch charge operator in the case of anomaly-free
$ \Z_2 $ symmetry discussed in the previous section. All the results discussed
there carry forward to this case. In particular, these patch charge operators
produce the representation category, which is a symmetric fusion category
$\eRep_{\Z_2}$.  We see that the representation category cannot distinguish
anomalous and anomaly-free symmetries, but the transformation category can.

\subsection{Braided fusion category of all t-patch operators} 

To consider all t-patch operators, we must consider fusion of the semion and
the boson. The $ b $ particles fuse with $ s $ to give another semion, let's
call it $ \t s $.  Along with the trivial one, we thus end up with four
particles. We can easily check that $ s $ and $ b $ have $ \pi $ mutual
statistics,
\begin{equation}\label{sbstat}
Z_{\text{string}_{ij}} W_{\text{patch}_{kl}}  = -W_{\text{patch}_{kl}}  Z_{\text{string}_{ij}} 
\end{equation}
Combining this with the fact that $ s $ has semionic self-statistics, we see
that $ s $ and $ \t s \equiv s\otimes b $ have trivial mutual statistics.

Putting the transformation category $\eM_\text{single-semion}$ and the
representation category $\eRep_{\Z_2}$ together, the above set of anyons and
their braiding and fusion data corresponds to the double-semion topological
order $\eM_\text{double-semion}$.  Double-semion is an Abelian topological
order which are classified $K$-matrix.\cite{WZ9290, FS9333}  The K-matrix for
the double-semion topological order is given by
\begin{equation}\label{DSK}
K_\text{DS}=\begin{pmatrix}
-2 & 0\\
0 & 2
\end{pmatrix}
\end{equation}
The topological quasiparticles are described by integer vectors $l$, and there
det$(K)=16$ is them.  The trivial particle $\one$ is described by $\one \sim
(0,0)^\top $, semion $s\sim (0,1)^\top $, semion $\t s\sim (1,0)^\top $, and
boson $b\sim (1,1)^\top $. The self statistics of anyon $l$ is given by
$\th_l=\pi l^\top K_\text{DS}^{-1} l$, the mutual statistics between anyon
$l_1$ and $l_2$ is given by $\th_{l_1l_2}=2\pi l_1^\top K_\text{DS}^{-1} l_2$.
The above $K$-matrix reproduces the self/mutual statistics of $s,\t s, b$.
Thus, the \catsymm\ for the anomalous $ \Z_2 $ symmetry in
1-dimensional space is the double-semion topological order
$\eM_\text{double-semion}$ in 2-dimensional space.

\subsection{A holographic way to compute \catsymm}

We can also compute \catsymm\ of anomalous symmetry directly by
computing the corresponding topological order in one higher dimension.  We know
that a system with (certain) anomalous $G$ (higher) symmetry can be realized as
a boundary of a $G$-symmetry protected topological (SPT) state in one higher
dimension.  If we gauge the $G$-symmetry in the bulk SPT state, we will obtain
a topological order described by a twisted $G$ (higher) gauge theory.  Such a
topological order in one higher dimension is the \catsymm\ of the $G$
(higher) symmetry.

Applying this method to 1d anomalous $\Z_2$ symmetry, we find the corresponding
\catsymm\ to be the 2d double-semion topological order.  The
connection between 1d anomalous $\Z_2$ symmetry and 2d double-semion topological
order was first observed in \Rf{LG1209}.

\section{A 1\texorpdfstring{\MakeLowercase{d}}{d} bosonic quantum system with $
\Z_2\times \Z_2 $ symmetry with a mixed anomaly} \label{Z2Z2mixed}

In this section, we calculate the \catsymm\ (\ie the non-degenerate
braided fusion category) for bosonic $\Z_2\times \Z_2$ symmetry with the mixed
anomaly in 1-dimensional space.  Following \Rf{WW170506728}, (see Appendix
\ref{z2z2global} for details) we have two qubits on each site and
 two symmetry generators of $ \Z_2\times \Z_2 $,
\begin{align}
W&=\prod_i X_i \label{z2z2W1}\\
\t W&=\prod_i \t X_i\prod_i s_{i,i+1} \label{z2z2W2}
\end{align}
where $ s_{i,i+1} =\ii^{\frac{1}{2}(Z_{i+1}-Z_i)(\t Z_{i+1}+1)}$ is the
non-on-site phase factor that encodes the mixed anomaly.  $X_i, Z_i$ act on one
qubit and $\t X_i,\t Z_i$ on the other qubit.

\subsection{Braided fusion category of patch operators}

The operators $ W $ and $ \t W $ above are global symmetry transformations,
which have corresponding t-patch symmetry operators as discussed in the
previous sections. 
\begin{align}
W_{\text{patch}_{ij}} &= \t O_{i} \left (\prod_{k=i}^{j-1} X_k\right ) \t O_{j}^\dag\label{wz2z2}\\ 
\t W_{\text{patch}_{ij}}&= \prod_{k=i+1}^j \t X_k \prod_{k=i}^{j-1} s_{k,k+1}
\end{align}
To satisfy the transparency condition and the composition algebra of the
t-patch operators (see Fig.  \ref{OAstring}), $\t O_j$ in \eqn{wz2z2} needs to
be chosen carefully: $ \t O_j=(1-\ii \t Z_j)/\sqrt2 $.  Pictorially, we can
represent $ W_{\text{patch}_{ij}} $ as
\begin{equation*}
\begin{tikzpicture}
	\begin{pgfonlayer}{nodelayer}
		\node [style=none] (1) at (7, 0) {};
		\node [style=none] (2) at (-7, 0) {};
		\node [style=none] (3) at (-6, 0.5) {};
		\node [style=none] (4) at (-6, -0.5) {};
		\node [style=none] (7) at (-4, 0.5) {};
		\node [style=none] (8) at (-2, 0.5) {};
		\node [style=none] (9) at (0, 0.5) {};
		\node [style=none] (10) at (0, -0.5) {};
		\node [style=none] (11) at (-2, -0.5) {};
		\node [style=none] (12) at (-4, -0.5) {};
		\node [style=none] (13) at (2, 0.5) {};
		\node [style=none] (14) at (4, 0.5) {};
		\node [style=none] (15) at (6, 0.5) {};
		\node [style=none] (17) at (2, -0.5) {};
		\node [style=none] (18) at (4, -0.5) {};
		\node [style=none] (19) at (6, -0.5) {};
		\node [style=none] (29) at (-2, -2.75) {$X$};
		\node [style=none] (30) at (0, -2.75) {$X$};
		\node [style=none] (31) at (2, -2.75) {$X$};
		\node [style=none] (32) at (-4, -2.75) {$X$};
		\node [style=none] (50) at (-4, -1.75) {$\t O$};
		\node [style=none] (60) at (-4, 1.25) {$i$};
		\node [style=none] (61) at (4, 1.25) {$j$};
		\node [style=none] (62) at (4, -1.75) {$\t O^\dagger$};
	\end{pgfonlayer}
	\begin{pgfonlayer}{edgelayer}
		\draw (2.center) to (1.center);
		\draw (3.center) to (4.center);
		\draw (7.center) to (12.center);
		\draw (8.center) to (11.center);
		\draw (9.center) to (10.center);
		\draw (13.center) to (17.center);
		\draw (14.center) to (18.center);
		\draw (15.center) to (19.center);
	\end{pgfonlayer}
\end{tikzpicture}
\end{equation*}
and $ \t W_{\text{patch}_{ij}}  $ as
\begin{equation*}
\begin{tikzpicture}
	\begin{pgfonlayer}{nodelayer}
		\node [style=none] (1) at (7, 0) {};
		\node [style=none] (2) at (-7, 0) {};
		\node [style=none] (3) at (-6, 0.5) {};
		\node [style=none] (4) at (-6, -0.5) {};
		\node [style=none] (7) at (-4, 0.5) {};
		\node [style=none] (8) at (-2, 0.5) {};
		\node [style=none] (9) at (0, 0.5) {};
		\node [style=none] (10) at (0, -0.5) {};
		\node [style=none] (11) at (-2, -0.5) {};
		\node [style=none] (12) at (-4, -0.5) {};
		\node [style=none] (13) at (2, 0.5) {};
		\node [style=none] (14) at (4, 0.5) {};
		\node [style=none] (15) at (6, 0.5) {};
		\node [style=none] (17) at (2, -0.5) {};
		\node [style=none] (18) at (4, -0.5) {};
		\node [style=none] (19) at (6, -0.5) {};
		\node [style=none] (29) at (-2, -1.75) {$\t X$};
		\node [style=none] (34) at (-1.75, -2.75) {};
		\node [style=none] (35) at (-0.25, -2.75) {};
		\node [style=none] (37) at (0.25, -2.75) {};
		\node [style=none] (38) at (1.75, -2.75) {};
		\node [style=none] (39) at (2.25, -2.75) {};
		\node [style=none] (40) at (3.75, -2.75) {};
		\node [style=none] (43) at (-2.25, -2.75) {};
		\node [style=none] (44) at (-3.75, -2.75) {};
		\node [style=cyan rect] (45) at (-3, -2.75) {$s$};
		\node [style=cyan rect] (46) at (-1, -2.75) {$s$};
		\node [style=cyan rect] (47) at (1, -2.75) {$s$};
		\node [style=cyan rect] (48) at (3, -2.75) {$s$};
		\node [style=none] (60) at (-4, 1.5) {$i$};
		\node [style=none] (61) at (4, 1.5) {$j$};
		\node [style=none] (65) at (0, -1.75) {$\t X$};
		\node [style=none] (66) at (2, -1.75) {$\t X$};
		\node [style=none] (67) at (4, -1.75) {$\t X$};
	\end{pgfonlayer}
	\begin{pgfonlayer}{edgelayer}
		\draw (2.center) to (1.center);
		\draw (3.center) to (4.center);
		\draw (7.center) to (12.center);
		\draw (8.center) to (11.center);
		\draw (9.center) to (10.center);
		\draw (13.center) to (17.center);
		\draw (14.center) to (18.center);
		\draw (15.center) to (19.center);
		\draw [style=thicc blue, in=180, out=0] (44.center) to (43.center);
		\draw [style=thicc blue] (34.center) to (35.center);
		\draw [style=thicc blue] (37.center) to (38.center);
		\draw [style=thicc blue] (39.center) to (40.center);
	\end{pgfonlayer}
\end{tikzpicture}
\end{equation*}
We label the endpoints of these patch operators $ m $ and $ \t m $,
respectively. More carefully, we should choose one end of the string to be $ m
$ (or $ \t m $) while the other end is its antiparticle $ \bar m $ ($ \bar{\t
m} $ respectively). The patch charge operators are generated by
\begin{align}
Z_{\text{string}_{ij}}&=Z_i Z_j\\
\t Z_{\text{string}_{ij}}&=\t Z_i \t Z_j
\end{align}
Let us name the charge operators at the ends of these as $ e $ and $ \t e $. We
note here that $ m,\t m $ are order 4 whereas $ e,\t e $ are order 2. We can
see this from the fact that $ W_{\text{patch}_{ij}}^4=\one=\t
W_{\text{patch}_{ij}}^4 $ while $ W_{\text{patch}_{ij}}^2 \neq \one, \t
W_{\text{patch}_{ij}}^2 \neq \one$. On the other hand, $
Z_{\text{string}_{ij}}^2=\one =\t Z_{\text{string}_{ij}}^2  $. The fusion of $
m $ and $ \t m $ gives $ s $ (say). The self-statistics of $ e $ and $ \t e $
are trivial, by the same logic as in the anomaly-free $ \Z_2 $ symmetry
discussed in Sec \ref{ncpatch}. We can also check that $ m $ and $ \t m $ have
trivial self-statistics. However, $ s $ particles have semionic
self-statistics, as can be seen from the hopping algebra calculation. This is
closely related to the fact that $ m $ and $  \t m $ have $ \pi/2 $ mutual
statistics; we find (cf. Fig \ref{mutual})
\begin{equation}\label{mutualmm}
W_{\text{patch}_{02}} \t W_{\text{patch}_{13}}=\ii \t W_{\text{patch}_{13}} W_{\text{patch}_{02}}
\end{equation}
Further details may be found in Appendix \ref{ZtZdeets}. We also note that the
$ m $ and $ e $ particles have $ \pi $ mutual statistics, and so do $ \t m $
and $ \t e $. 

The particles $m,\t m, e,\t e$ generate a non-degenerate braided fusion
category that correspond to a 2d Abelian topological order.  By comparing the
self/mutual statistics of those topological excitations, we find that the 2d
Abelian topological order is described by the $K$-matrix
\begin{equation}
\label{KmatSPT}
K=\begin{pmatrix}
0 & 2 & -1 & 0\\
2 & 0 & 0 & 0\\
-1 & 0 & 0 &2\\
0 & 0 &2 &0
\end{pmatrix}
\end{equation}
This 2d topological order is the \catsymm\ for the $\Z_2\times \Z_2$
symmetry with the mixed anomaly in 1-dimensional space.  The topological
excitations in such an Abelian topological order are labeled by integer vectors
$l$.  The $m,\t m, e,\t e$ correspond to the following integer vectors:
\begin{align}
 e   \sim (1,0,0,0)^\top,\ \ \ 
 m   \sim (0,1,0,0)^\top,
\nonumber\\
\t e \sim (0,0,1,0)^\top,\ \ \ 
\t m \sim (0,0,0,1)^\top.
\end{align}
The self statistics of particle $l$ and mutual statistics between particles
$l_1$ and $l_2$ can be calculated via
\begin{align}
 \th_l = \pi l^\top K^{-1} l,\ \ \ \
 \th_{l_1,l_2} = 2\pi l^\top_1 K^{-1} l_2,
\end{align}
where
\begin{align}
K^{-1}=\begin{pmatrix}
0 & \frac12 & 0 & 0\\
\frac12 & 0 & 0 & \frac14\\
0 & 0 & 0 &\frac12\\
0 & \frac14 &\frac12 &0
\end{pmatrix}.
\end{align}
The entry $\frac14$ in $K^{-1}$ gives rise to the $ \pi/2 $ mutual statistics
between $ m $ and $  \t m $.

\subsection{A holographic calculation of \catsymm}

The above 2d Abelian topological order (\ie the \catsymm) can be
obtained via another approach.  We know that the $\Z_2 \times \Z_2$ symmetry with
the mixed anomaly is realized by the boundary of a 2d $\Z_2\times \Z_2$ SPT
state.  If we gauge the 2d $\Z_2\times \Z_2$ symmetry, we will turn the  2d
$\Z_2\times \Z_2$ SPT state into a  2d topological order.  Such a 2d topological
order is the Abelian topological order described above.  Such an Abelian
topological order was given by the $K$-matrix in equations (64) and (67) in
\Rf{W1447}.  For our case, we need to substitute the values $ n_1=n_2=2$, and
$m_0=m_3=0, m_2=1 $, which gives us the $K$-matrix in \eqn{KmatSPT}.  This
Abelian topological order is the \catsymm\ for the 1d $\Z_2 \times
Z_2$ symmetry with the mixed anomaly.  The holographic calculation gives rise
to the same result as the operator algebra calculation.

\subsection{The equivalence between 1d $\Z_2 \times \Z_2$ symmetry with mixed
anomaly and 1d $\Z_4$ symmetry}
\label{Z2Z2aZ4}

Generalizing our $\Z_2$ result, we know that the \catsymm\ of 1d
anomaly-free $\Z_4$ symmetry is the 2d $ \Z_4 $ topological order ($ \Z_4 $
gauge theory), denoted as $\eGau_{\Z_4}$ and described by the $K$-matrix,
\begin{equation}\label{Z4K}
K_{\Z_4}=\begin{pmatrix}
0 & 4\\
4 & 0
\end{pmatrix}
\end{equation}
The set of topological quasiparticle is described by integer vectors $
\{(a,b)^\top |a,b\in \Z_4 \}$, and there also $ \left|\det K_{\Z_4}\right|=16 $
of them. Their self and mutual statistics can be read off from the inverse of
the $2\times 2$ $K$-matrix, which are the same as those for the $4\times 4$
$K$-matrix in \eqn{KmatSPT}.  This allows us to make the following
identification
\begin{equation}\label{corresp}
\begin{split}
(0,1)^\top & \leftrightarrow m,\ \ (1,0)^\top \leftrightarrow \t m, \ \ (1,1)^\top  \leftrightarrow s  \\
(2,0)^\top & \leftrightarrow e, \ \ (0,2)^\top  \leftrightarrow \t e 
\end{split}
\end{equation}
For example, note that $ (0,1)^\top $ and $ (1,0)^\top $ have trivial self statistics,
\begin{align}
\pi \ \ (0,1)\cdot K^{-1} \cdot (0,1)^\top &=0\\
\pi \ \ (1,0)\cdot K^{-1} \cdot (1,0)^\top &=0
\end{align}
but have $\frac\pi 2$ mutual statistics,
\begin{equation}\label{Z4mm}
2\pi \ \ (0,1)\cdot K^{-1}\cdot (1,0)^\top =\frac\pi 2
\end{equation} 
so these must correspond to the $ m,\t m $ particles. 
Note also that these are order 4 quasiparticle vectors, \ie 4 of them will fuse to a trivial quasiparticle. On the other hand, the quasiparticle vectors $ (2,0)^\top $ and $ (0,2)^\top $ correspond to $ e,\t e $ particles because not only do they have trivial self statistics, 
\begin{align}
\pi \ \ (0,2)\cdot K^{-1} \cdot (0,2)^\top &=0\\
\pi \ \ (2,0)\cdot K^{-1} \cdot (2,0)^\top &=0
\end{align}
but they also have trivial mutual statistics,
\begin{equation}\label{Z4ee}
2\pi \ \ (0,2)\cdot K^{-1}\cdot (2,0)^\top =2\pi
\end{equation}
Similar calculations show that $ (0,2)^\top $ and $ (1,0)^\top $ have $ \pi $ mutual statistics, and so do $ (2,0)^\top $ and $ (0,1)^\top $. 

In fact, 2d Abelian topological orders described by \eq{KmatSPT} and \eq{Z4K}
are actually the same topological order \cite{EW211211394}.  
It turns out, this K-matrix in
\eq{KmatSPT} can be transformed $ K\to W K W^\top $ by an integer matrix $ W $
with det$(W) = \pm 1$ into a $ \Z_4 $ K-matrix, direct summed with a trivial
block.
\begin{equation}\label{WKWT}
W=\begin{pmatrix}
1 & 0 & 0 & 0\\
0 & 0 & 1 & 0\\
0 & 1 & 2 & 0\\
2 & 0 & 0 & 1\end{pmatrix}
\implies  W K W^\top =\begin{pmatrix}
0 & -1 & 0 & 0\\
-1 & 0 & 0 & 0\\
0 & 0 & 0 &4\\
0 & 0 &4 &0\end{pmatrix}
\end{equation}

To summarize, 1d $\Z_2 \times \Z_2$ symmetry with the mixed anomaly is realized
by the boundary of a 2d $\Z_2\times \Z_2$ SPT state. 1d anomaly-free $\Z_4$
symmetry is realized by the boundary of a 2d trivial $\Z_4$ SPT state.  The
\catsymm\ of the 1d mixed-anomalous $\Z_2 \times \Z_2$ symmetry is the
$\Z_2 \times \Z_2$ gauging of the 2d $\Z_2\times \Z_2$ SPT state.  The
\catsymm\ of the 1d $\Z_4$ symmetry is the $\Z_4$ gauging of the 2d
trivial $\Z_4$ SPT state.  The two symmetries give rise to the same 2d
topological order. Thus 1d $\Z_2 \times \Z_2$ symmetry with the mixed anomaly and
1d anomaly-free $\Z_4$ symmetry have the same \catsymm\ and are
equivalent.

\subsection{A new duality mapping}
\label{newdual}

\begin{table}
\caption{Group "multiplication" table of $ \Z_4 \equiv \Z_2 \leftthreetimes_{e_2} \Z_2 $. The entries left blank are redundant since the group is Abelian. }
\label{table:multZ4}
\centering
\begin{tabular}{|c|c|c|c|c|}
\hline
$ \Z_2 \leftthreetimes_{e_2} \Z_2 $& (0,0) & (0,1) & (1,0) & (1,1) \\ 
\hline  
(0,0)& (0,0) &  &  &  \\ 
\hline 
(0,1)& (0,1) & (0,0) &  &  \\ 
\hline 
(1,0)& (1,0) & (1,1) & (0,1) &  \\ 
\hline 
(1,1)& (1,1) & (1,0) & (0,0) & (0,1) \\ 
\hline 
\end{tabular} 
\end{table}
By comparing with the corresponding table for $ \Z_4 $ in the additive presentation $ \{0,1,2,3\} $, we can make the following (non-unique) one-to-one mapping between these two representations of $ \Z_4 $.
\begin{align}\label{map}
(0,0) \leftrightarrow 0 \qquad & (0,1) \leftrightarrow  2\\
(1,0) \leftrightarrow  3 \qquad & (1,1) \leftrightarrow  1
\end{align}

\begin{table*}
	\centering
	\caption{Patch operators of $ \Z_2\times \Z_2 $ with mixed anomaly and their dual $ \Z_4 $ patch operators. ($ O_i'= \frac{1-\ii Z'_i}{\sqrt2}$)}
	\label{table:dualpatch}
	\begin{tabular}{|c|c|}
		\hline
		$ \Z_2 \times \Z_2$ with mixed anomaly & Anomaly-free $ \Z_4 $ \\
		\thickhline
		$ W_{\text{patch}_{ij}}= \t O_{i} \left (\prod_{k=i}^{j-1} X_k\right ) \t O_{j}^\dag $ & $ O'_{i} \bar Z_{i-\frac12} \left (O'_{j}\right )^\dag \bar Z_{j-\frac12} $ \\ 
		\hline 
		$ \t W_{\text{patch}_{ij}}= \prod_{k=i+1}^j \t X_k \prod_{k=i}^{j-1} s_{k,k+1} $ & $ \prod_{k=i}^j X'_k CX(g'_k, \bar{g}_{k-\frac12}) \equiv \prod_{k=i}^j L_{+3}\vert_k $ \\ 
		\hline 
		$ Z_{\text{string}_{ij}}=Z_i Z_j $ & $ \prod_i^{j-1} \bar X_{k+\frac12} \equiv \prod_{k=i+1}^j L_{+2}\vert_k $\\ 
		\hline 
		$ \t Z_{\text{string}_{ij}}=\t Z_i \t Z_j $ & $ Z'_i Z'_j $   \\ 
		\hline 
	\end{tabular} 
\end{table*}

The above holographic equivalence of 1d mixed-anomalous $\Z_2 \times \Z_2$
symmetry and 1d anomaly-free $\Z_4$ symmetry suggests the existence of a new
duality mapping, between a model with $\Z_2\times \Z_2$ non-on-site symmetry and
model with $\Z_4$ on-site symmetry.  Such an exact duality maps between the $
\Z_4 $ patch symmetry/charge operators and the patch symmetry/charge operators
of the mixed-anomalous $ \Z_2\times \Z_2  $ symmetry we have been outlining in
this section. This duality is a Kramers-Wannier-like transformation that
transforms one set of $ \Z_2 $ variables from order to disorder (or site to
link) variables, followed by an on-site (local) unitary transformation. To
state the duality mapping, we first re-write the group $ \Z_4 $ as a
cocycle-twisted product of two $ \Z_2 $ groups, as described in Appendix N of
\Rf{LW180901112}.  With $ G=\Z_4 $, and $ A=\Z_2\leq G $, we extend $ A $ by $
H=\Z_2 $ with $ \alpha=\id $ and $ e_2(h_1,h_2)=\toZ{h_1.h_2}_{\text{mod 2}}
$.\footnote{The multiplication of elements of $ H $ in the definition of $ e_2
$ is understood to be done in $ \Z $ and then mapped back to $ \Z_2 $.} The
group operation with these choices can be expressed as 
\begin{equation}\label{hxop}
(h_1,x_1) * (h_2,x_2)= \left(h_1+h_2,x_1+x_2+e_2(h_1,h_2) \right )
\end{equation}
where the additions are to be understood modulo 2. Using this, we may write elements of $ \Z_4 $ using two $ \Z_2 $ labels as $ g\equiv (h,x) $ where $ x\in A $ and $ h\in H $. There are four $ \Z_4 $ symmetry transformations: one trivial and three non-trivial. Taking $ \Z_4 $ to be represented as $ \{0,1,2,3\} $, with the group operation being addition modulo 4, we have two generators $ L_{+1} $ and $ L_{+3} $ of the symmetry group,
\begin{equation}\label{gen}
\begin{split}
L_{+1}\ket{g}&=\ket{g+1 \text{ mod 4}}\\
L_{+3}\ket{g}&=\ket{g+3 \text{ mod 4}}
\end{split}
\end{equation}
In the $ (h,x) $ representation, what do these generators look like? We can work this out by looking at the group "multiplication" table of $ \Z_4 $ in this representation: see Table \ref{table:multZ4}.

Using this mapping, we re-write \eqn{gen} as follows.
\begin{equation}
\begin{split}
L_{+1}\ket{(h,x)}&=\ket{(h,x)*(1,1)}\\
L_{+3}\ket{(h,x)}&=\ket{(h,x)*(1,0)}
\end{split}
\end{equation}
Inspecting this case-by-case, one observes that the generator $ L_{+3} $ is nothing but the operator $ X_1 CX_{1,0} $, acting on kets $ \ket{(h,x)} $. Here $ h $ and $ x $ are labeled as qubits 1 and 0 respectively, and $ CX_{1,0} $ denotes the controlled NOT gate with qubit 1 as the control. 

Now we apply a duality transformation on the t-patch operators of the $ \Z_2 \times \Z_2 $ symmetry with mixed anomaly in order to show that we recover the t-patch operators of anomaly-free $ \Z_4 $ symmetry. 
The reader who is interested in the explicit form of the duality instead of the steps leading up to it is invited to skip to the end of this subsection. 

On the $ \Z_2\times \Z_2 $ side, our states are defined by a pair of $ \Z_2 $ variables on each site $ i $, denoted $ (g_i,\t g_i) $. The definitions $ g_i=\frac{Z_i-1}{2}, \t g_i= \frac{\t Z_i-1}{2}$ map the Z-basis $ \{\pm 1\} $ to the additive $ \Z_2 $ basis $ \{0,1\} $. 

Step 1 of duality transformation $ \cD $: We transform $ (g_i,\t g_i) $ to $ (g'_i,\bar g_{i-1/2}) $ by defining $ \bar g_{i-\frac{1}{2}} = g_i -g_{i-1}\text{ mod 2}$ and $ g'_i=\t g_i $. The Pauli operators transform as \begin{equation}\label{xzopD1}
X_i \to \bar X_{i-\frac{1}{2}} \bar X_{i+\frac{1}{2}}, \ \ Z_i Z_{i+1}\to \bar Z_{i+\frac12}
\end{equation}
The new degrees of freedom may be shown pictorially as
\begin{equation*}
\begin{tikzpicture}
	\begin{pgfonlayer}{nodelayer}
		\node [style=none] (1) at (8, 0) {};
		\node [style=none] (2) at (-8, 0) {};
		\node [style=wide rect] (45) at (-2, 0) {$\bar g$};
		\node [style=white circle] (50) at (-4, 0) {$g'$};
		\node [style=white circle] (60) at (0, 0) {$g'$};
		\node [style=white circle] (61) at (4, 0) {$g'$};
		\node [style=wide rect] (62) at (2, 0) {$\bar g$};
		\node [style=wide rect] (63) at (6, 0) {$\bar g$};
		\node [style=none] (65) at (-4, -1.5) {$i-1$};
		\node [style=none] (66) at (0, -1.5) {$i$};
		\node [style=none] (67) at (4, -1.5) {$i+1$};
		\node [style=none] (68) at (7, -1.5) {$\cdots$};
		\node [style=wide rect] (69) at (-6, 0) {$\bar g$};
		\node [style=none] (70) at (-7, -1.5) {$\cdots$};
	\end{pgfonlayer}
	\begin{pgfonlayer}{edgelayer}
		\draw (2.center) to (1.center);
	\end{pgfonlayer}
\end{tikzpicture}
\end{equation*}
For each site $ i $, let us define $ g''_i=\bar g_{i-\frac{1}{2}} $. Then we have a two-qubit Hilbert space labeled as $(g'_i,g''_i) $ associated with site $ i $. Let us choose $ g'_i $ as qubit-1 and $ g''_i $ as qubit-2.

Step 2 of duality transformation $ \cD $: Now we perform a Hadamard transformation on qubit-2 of each site. The states transform as
\begin{equation}\label{Htransf}
\ket{g'_i}\otimes \ket{g''_i}  \to  \ket{g'}\otimes \left(H \ket{g''}\right )
\end{equation}
where $ H $ is the Hadamard operator. We will instead work in the Heisenberg picture, where the Hadamard transformation acts on the operators and interchanges $ \bar X  $ and $ \bar Z $. Then the states on which these transformed operators act are labeled by $ \Z_4 $ elements in the $ (h,x) $ representation with $ h_i=g'_i $ and $ x_i=g''_i= \bar g_{i-\frac{1}{2}}=g_i -g_{i-1}\text{ mod 2}$.

Summarizing the mapping of the basis states,
\begin{equation}\label{basismap}
(g_i,\t g_i)\to (g'_i=\t g_i, g''_i=g_i -g_{i-1}\text{ mod 2})
\end{equation}
with $ (g_i,\t g_i)\in \Z_2\times \t \Z_2 $ and $ (g'_i,g''_i)\in \Z_2\leftthreetimes_{e_2} \Z_2 \cong \Z_4 $. On the other hand, under the combined effect of steps 1 and 2 of $ \cD $, we have the operator maps.
\begin{equation}\label{xzopD2}
	X_i \to Z''_{i} Z''_{i+1}, \ \ Z_i Z_{i+1}\to X''_{i+1}
\end{equation}
Using this, one finds that the operator $ s_{i-1,i} $ becomes $ CX(g'_i, g''_i) $. We can also denote this as $ CX_{1,0}\vert_{i} $ with the qubit labels 1 and 0 as described above. 
In fact, we can check that the patch operators in the left column of Table \ref{table:dualpatch} are transformed to those in the right column, under the transformation $ \cD $.

In particular, we find the dual of $ \t W_{\text{patch}_{ij}} $ to be the patch
symmetry operator corresponding to the $ L_{+3} $ transformation discussed above. This operator then generates all the $ \Z_4 $
patch symmetry operators in the $ \Z_2 \leftthreetimes_{e_2} \Z_2 $
representation. On the other hand, the dual of $ W_{\text{patch}_{ij}} $ is a
t-patch operator with empty bulk that has order 4. This operator may be
identified with one of the charge patch operators of anomaly-free $ \Z_4 $
symmetry. This completes the mapping between patch operators on both sides of
our duality $ \cD: \left (\Z_2\times \Z_2\right )^{\om_{12}}\leftrightarrow \Z_4
$.  This exact duality mapping allows us to show that the 1d $ \Z_2 \times
\Z_2$ symmetry with mixed anomaly and anomaly-free $ \Z_4 $ symmetry have
\emph{isomorphic local symmetric operator algebra} \ie they have the \emph{same
\catsymm}.

\emph{A comment on gauging:} The duality we described above can also be understood as coupling the degrees of freedom of $ \Z_4 $ symmetric system to a $ \Z_2 $ gauge field. The Kramers-Wannier-like transformation in the first step of $ \cD $ essentially amounts to such a gauging procedure. In the case of $ \Z_2 $ symmetry in 1d, the Kramers-Wannier duality transformation allows one to relate $ \Z_2 $ order and disorder operators, where the latter can be obtained from the former by gauging the local $ \Z_2 $ symmetry and then restricting to the $ \Z_2 $ charge even sector of the Ising gauge theory. Our duality transformation above involves an on-site unitary (Hadamard) transformation in addition to this gauging procedure.

\section{A 1\texorpdfstring{\MakeLowercase{d}}{d} bosonic quantum system with
an emergent $ \Z_2\times \Z_2 $ symmetry which is ``beyond
anomaly''}\label{z2z2beyond}

In this section, we are going to study a case of emergent symmetry.  We find
that the  emergent symmetry is neither anomaly-free nor anomalous.  It
illustrates that \catsymm\ (\ie topological order in one higher
dimension) is a better way to view symmetry. We get a simpler, more uniform,
and more systematic picture.

Let us briefly recall the model from section II.C of \Rf{JW191213492}. This
model describes a 1+1D bosonic quantum system with spin-$ 1/2 $ degrees of
freedom on each site and each link. The Hamiltonian describing the model is:
\begin{equation}
\label{Hztz}
\begin{split}
H=-\sum_i \left (B \t X_{i-\frac{1}{2}} X_i \t X_{i+\frac{1}{2}}+J \t Z_{i+\frac{1}{2}}\right )\\
+U\sum_i\left (1-Z_i \t Z_{i+\frac{1}{2}} Z_{i+1}\right )
\end{split}
\end{equation} 
This Hamiltonian has two on-site (\ie anomaly-free) $ \Z_2 $ symmetries,
generated by
\begin{equation}\label{z2symms}
W = \prod_k X_k, \quad \t W =\prod_k \t Z_{k+\frac12}
\end{equation}
Let us denote the corresponding symmetries as $ \Z_2 $ and $ \t \Z_2 $.
The algebra of local operators is constrained by these symmetries. We add an additional constraint on this algebra: the \emph{low-energy constraint}. This constraint is imposed by taking the limit of $ U\to \infty $. Low energy sector of the Hilbert space must then satisfy
\begin{equation}\label{lowE}
1-Z_i \t Z_{i+\frac{1}{2}} Z_{i+1}=0, \ \ \forall i
\end{equation}
In operator language, we demand that the allowed local operators commute with
the operator appearing in \eqn{lowE}.  The algebra of the allowed local
operators will give rise to emergent low energy symmetry.

The question is then, how does this additional constraint\footnote{The
experienced reader may note that this is sometimes colloquially referred to as
"gauging" in the literature. We are particular about not calling it by this
name since we don't introduce any extra unphysical, or \emph{gauge}, degrees of
freedom in this discussion. Instead we are restricting to a subspace of the
full Hilbert space to focus on the effective theory.} 
change the algebra of
t-patch operators? It turns out that this modified algebra
involves a non-trivial relationship between the $ \Z_2 $ and $ \t \Z_2 $ symmetries. To be clear, this is not a case
of mixed anomaly of two $ \Z_2 $ symmetries like the case discussed in the
previous section. Nor is this a case of an anomaly-free symmetry: the patch symmetry
operators form a non-symmetric fusion category. This is thus an example of a
symmetry that is, in some sense, beyond the usual notion of ``anomalous
symmetry''. The \catsymm\ of the low-energy sector of this model is
not $ \eGau_{\Z_2\times \Z_2} $ (\ie $ \Z_2\times \Z_2 $ gauge theory coupled to matter), as would be the case for a anomaly-free global $
\Z_2\times \Z_2 $ symmetry. Instead it has the \catsymm\ $ \eGau_{\Z_2}
$, same as that of anomaly-free global $ \Z_2 $ symmetry. Let us now expand on
this using the language we have been developing in the previous sections.

The algebra generated by the LSOs can be organized in terms of the t-patch operators, which serve as a particular convenient choice of generators:
\begin{enumerate}
\item 0-dimensional t-patch operators are the local symmetric operators that act within the low-energy sector: 
\begin{align}
 \t X_{i-\frac12}X_i\t X_{i+\frac12},\ \ \  \t Z_{i+\frac12} .
\end{align}
\item 1-dimensional t-patch operators -- string operators: 
\begin{align}
 Z_{\text{str}_{ij}}  &=\prod_{k=i}^{j-1} \t Z_{k+\frac12} = Z_i Z_j , 
\nonumber\\
X_{\text{str}_{ij}}&=\t X_{i-\frac12} \prod_{k=i}^j X_k \t X_{j+\frac12}.
\end{align}
 
\end{enumerate}  
One may note that the new constraint, \eqn{lowE} has the effect of restricting
the set of allowed t-patch operators.  For example, the two string operators
$\prod_{k=i}^{j-1} \t Z_{k+\frac12}$ and $Z_i Z_j$ become identical within the
low energy subspace.  Also two string operators $\prod_{k=i}^{j} X_{k}$ and $\t
X_{i+1/2} \t X_{j+1/2}$ must appear together.  Without this constraint, the
list of t-patch operators would be a bigger one -- one that would encode
anomaly-free $ \Z_2\times \Z_2 $ symmetry. 

The algebra of the \emph{extended} t-patch operators takes the form:
\begin{align}
Z_{\text{str}_{ij}} X_{\text{str}_{kl}} &=\pm X_{\text{str}_{kl}} Z_{\text{str}_{ij}}\label{zxmutual}\\
Z_{\text{str}_{ij}} Z_{\text{str}_{jk}} &= Z_{\text{str}_{ik}}\label{zzfusion}\\
X_{\text{str}_{ij}} X_{\text{str}_{jk}} &= X_{\text{str}_{ik}}\label{xxfusion}
\end{align}
where the sign in \eqn{zxmutual} is $ - $ if $ i< k<j<l $, and $ + $ otherwise.
We see here that the algebra of the patch operators above mirrors that of
anomaly-free $ \Z_2 $ symmetry, as discussed in Section \ref{z2all}.
Specifically, note that \eqn{zzfusion} and \eqn{xxfusion} are identical to
\eqn{OA1dZ2a} and \eqn{OA1dZ2b} respectively. These represent the fusion of the
endpoints of these t-patch operators. The mutual statistics of these endpoints
are also identical in the two cases as can be seen by comparing \eqn{zxmutual}
with \eqn{OA1dZ2c} and \eqn{OA1dZ2d}.

Therefore, the exact 1d $\Z_2\times \Z_2$ on-site symmetry in the model
\eq{Hztz} becomes a different $\Z_2\times \Z_2$ symmetry at low energies.  The
new $\Z_2\times \Z_2$ symmetry has the \catsymm\ $\eGau_{\Z_2}$, while the
original $\Z_2\times \Z_2$ on-site symmetry has the \catsymm\
$\eGau_{\Z_2\times \Z_2}$.  The new  $\Z_2\times \Z_2$ symmetry has a special
property: a gapped state must spontaneously break one of the $\Z_2$ symmetry. A
state with both $\Z_2$ symmetry must be gapless.  There is no state that can
spontaneously break both the $\Z_2$ symmetries.\cite{L190309028, JW191213492}
Those properties have some similarities to $\Z_2\times \Z_2$ symmetry with the
mixed anomaly.  But the  $\Z_2\times \Z_2$ symmetry with the mixed anomaly has
the \catsymm\ $\eGau_{\Z_4}$.  Since the \catsymm\ $\eGau_{\Z_2}$ for the new
$\Z_2\times \Z_2$ symmetry is different from both the \catsymm\
$\eGau_{\Z_2\times \Z_2}$ for anomaly-free $\Z_2\times \Z_2$ symmetry and the
\catsymm\ $\eGau_{\Z_4}$ for $\Z_2\times \Z_2$ symmetry with the mixed anomaly,
the new $\Z_2\times \Z_2$ symmetry is beyond anomaly.

\section{2\texorpdfstring{\MakeLowercase{d}}{d} $\Z_2$ symmetry and its dual} 
\label{2dZ2}

In the above, we have discussed symmetries and \textsf{categorical symmetries} in
1-dimensional space.  In this section, we will start to consider symmetries in
higher dimensions, where we will encounter higher symmetries.

First, we consider the simplest symmetry -- $\Z_2$ symmetry, in 2-dimensional
space.  For convenience, let we assume the degrees of freedom on each vertex
(labeled by $i$) are labeled by elements in the $\Z_2$ group.

\subsection{$\Z_2$ symmetry transformation and t-patch operators as
local symmetric operators} 

\label{Z2symm2d}

The $\Z_2$ symmetry is described by a symmetry transformation:
\begin{align}
W = \prod_{i\in \text{whole space}} X_i,\ \ \ W^2=\id. 
\end{align}
The $\Z_2$ transformation $W$ select a set of local symmetric operators which
form an algebra:
\begin{align}
\cA=\{O_i^\mathrm{symm} \ \big|\  O_i^\mathrm{symm}W = W O_i^\mathrm{symm}\}
\end{align}

As before, we can use the t-patch operators to organize the
local symmetric operators:
\begin{enumerate}
\item
0-dimensional t-patch operators, $X_i, Z_iZ_{i+\v \mu}$, where $\v \mu$ connects
vertex-$i$ to its neighbors.

\item
1-dimensional t-patch operators -- string operators, 
\begin{align}
Z_{\mathrm{str}_{ij}} = Z_iZ_j,
\end{align}
where the string$_{ij}$ connects the vertex-$i$ and vertex-$j$.
We see that the string operator has an empty bulk.

\item
2-dimensional t-patch operators -- disk operators, 
\begin{align}
X_{\mathrm{disk}} &= \prod_{i\in \mathrm{disk} }X_i.
\end{align}
The disk operator has a non trivial bulk, which generate our $\Z_2$ symmetry.

\end{enumerate}

In terms of t-patch operators, algebra of local symmetric operators
takes the following form
\begin{align}
\label{OAZ2a}
Z_{\mathrm{str}_{ij}} Z_{\mathrm{str}_{jk}} &= Z_{\mathrm{str}_{ik}},
\\ 
\label{OAZ2b}
X_{\mathrm{disk}_1} X_{\mathrm{disk}_2} &= X_{\mathrm{disk}_{1+2}}
\\
\label{OAZ2c}
Z_\mathrm{str} X_\mathrm{disk} &= - X_\mathrm{disk} Z_\mathrm{str}, 
\\
\label{OAZ2d}
Z_\mathrm{str} X_\mathrm{disk} &= + X_\mathrm{disk} Z_\mathrm{str} 
\end{align}
Eqn. (\ref{OAZ2a}) describes the fusion of string operators (see Fig.
\ref{OApatch}a).  Eqn. (\ref{OAZ2b}) describes the fusion of disk operators
(see Fig.  \ref{OApatch}b).  The commutator between the string and the disk
operators depends on their relative positions.  If the string straddle the
boundary of the disk as in Fig.  \ref{OApatch}c, commutator has a non-trivial
phase as in \eqn{OAZ2c}.  Otherwise (see Fig. \ref{OApatch}d), the string and the
disk operators commute as in \eqn{OAZ2d}.

Since the  string operators have empty bulk, they correspond to patch charge
operators, and the ends of the string operators correspond to charged
particles.  The disk operators have non-trivial bulk, and correspond to patch
$\Z_2$-symmetry operators, which generate the $\Z_2$ symmetry transformations and
select local symmetric operators.

As before, the  patch symmetry transformations can detect the
symmetry charge hidden in the local symmetric operators: when the string
operator $Z_\mathrm{str}$ straddle the boundary of the disk operator $W_\mathrm{patch}$, the two operators have a non-trivial
commutation relation:
\begin{align}
\label{ZWWZ}
Z_\mathrm{str} X_\mathrm{disk} = - X_\mathrm{disk} Z_\mathrm{str}.
\end{align}
This non-trivial commutation relation measures the charge carried by one end of
the string operator.  If we view the order of the operator product as the order
in time, and view the string as world line of a particle in spacetime (see Fig.
\ref{measurecharge}), then the commutation relation \eqn{ZWWZ} can be viewed as
a braiding of the particle around the boundary of the disk operator.  The
charge is measured by such a braiding process.

\subsection{Algebra of patch charge operators and braided fusion higher
category of charge objects}

\begin{figure}[t]
\begin{center}
\includegraphics[scale=0.13]{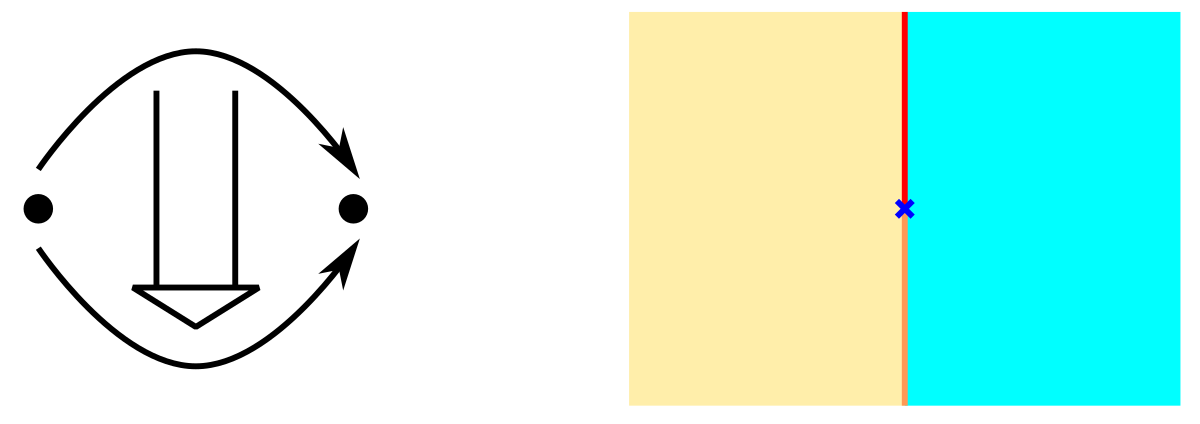}\includegraphics[scale=0.13]{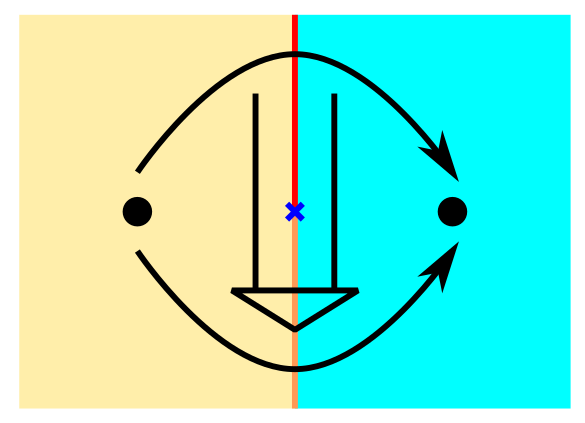}
\\ (a) 
\ \ \ \ \ \ \ \ \ \ \  
\ \ \ \ \ \ \ \ \ \ \ \ \ \   
(b)
\ \ \ \ \ \ \ \ \ \  
\ \ \ \ \ \ \ \ \ \  
(c) \ \ \ \
\end{center}
\caption{ (a) A graphic representation of objects (the points), 1-morphism (the
lines connecting points), and 2-morphism (the disk connecting lines), in a
higher category.  (b) In 2d spacetime (the vertical direction is the time
direction), two world sheets of string-like excitations are separated by two
world lines of point-like excitations.  The  two two world lines of point-like
excitations are separated by an instanton (a local operator).  (c) A higher
category describes the structure of extended excitations: in 2d, object
$\leftrightarrow$ co-dimension-1 excitation (string); 1-morphism
$\leftrightarrow$ co-dimension-2 excitation (particle); 2-morphism
$\leftrightarrow$ co-dimension-3 instanton (local operator).  } \label{MorExc} 
\end{figure}

The properties of the charges of an anomaly-free  symmetry in $n$-dimensional
space can be systematically and fully described by a braided fusion
$n$-category or a local $n$-fusion category.\cite{KZ200514178}  Let us first
give a brief physical introduction of fusion \ml{n}-category (see Fig.
\ref{MorExc}).  A fusion \ml{n}-category can be used to describe extended
physical objects in \ml{n}d space.  For example, in 3d space, 2-dimensional
membranes (co-dimension-1) correspond to the objects in the fusion
\ml{3}-category.  1-dimensional strings (co-dimension-2) correspond
1-morphisms, and 0-dimensional particles (co-dimension-3) correspond
2-morphisms.  The above are physical excitations.  Instantons or local
operators (0-dimensional in spacetime) correspond 3-morphisms, which are top
morphisms.  The physical excitations and local operators form the fusion
\ml{n}-category.

To connect the $\Z_2$ symmetry in 2-dimensional space to a braided fusion
2-category, we view the local symmetric operators \ml{O_i^\mathrm{symm}} as the
2-morphisms, and the end of string operator $Z_{\mathrm{str}}$ (\ie
\ml{\Z_2}-charge) as a 1-morphism \ml{e} in a fusion $2$-category.  Operator
product of string operator can be viewed as fusion of string ends, which gives
rise to the fusion rule of the 1-morphisms \ml{e_i}:
\begin{align}
e\otimes e =  \one, \ \ \
\one \otimes e =  e \otimes\one  =   e.
\end{align}

$e$'s are the point-like \ml{\Z_2}-charges for \ml{\Z_2} symmetry.
Those \ml{\Z_2}-charges can form a 1d quantum liquid state, which
correspond to a string excitation \cite{KW1458}.  Let $s_{\Z_2}$ be a string
excitation that corresponds to the 1d spontaneous symmetry breaking state
formed by the \ml{\Z_2}-charges (which is a state with a non-zero energy gap).
(Note that the \ml{\Z_2}-charges have a $\Z_2$ conservation as implies by the
$\Z_2$ fusion $e\otimes e=\one$.  So they can form a non-trivial 1d gapped
quantum liquid state -- a spontaneous symmetry breaking state.) We have another
string excitation $\one_\text{str}$ which is formed by $\Z_2$ charges along the
string in a gapped symmetric state.  Note that a string with no $\Z_2$ charge is
also a symmetric gapped state. So $\one_\text{str}$ may mean null string, a
string that does not have any thing.  The  string formed by $\Z_2$ charges in
gapped symmetric state and the string formed by nothing are equivalent (\ie
they can deform into each other without closing the energy gap), and both are
denoted as $\one_\text{str}$.

In addition to the point-like excitation $e$, we have another point-like
excitation, denoted as $b_{s}$, which is the domain wall that connects the
string-$s_{\Z_2}$ and string-$\one_\text{str}$.  Since,  string-$\one_\text{str}$ is
trivial (\ie can be nothing), $b_{s}$ can also be viewed as a boundary of
string $s_{\Z_2}$.  The fusion of $e$ and $b_{s}$ gives us the third point-like
excitation $e\otimes b_s$.

The above excitations, plus the $\Z_2$ symmetric local operators form a
symmetric fusion $2$-category denoted as \ml{2\eRep_{\Z_2}}:\cite{KZ200514178}
\begin{enumerate}

\item The string-like excitations $\one_\text{str}$ and $s_{\Z_2}$ are objects
in \ml{2\eRep_{\Z_2}}.

\item The point-like excitations $\one$, $e$, and $b_s$ are 1-morphisms:
\begin{align}
\label{obj1mor}
\one_\text{str} &\stackrel{b_s}{\to} s_{\Z_2}, & 
 s_{\Z_2}&\stackrel{b_s}{\to} \one_\text{str}, & 
\one_\text{str} &\stackrel{e\otimes b_s}{\to} s_{\Z_2}, & 
 s_{\Z_2}&\stackrel{e\otimes b_s}{\to} \one_\text{str},
\nonumber\\
\one_\text{str} &\stackrel{\one}{\to} \one_\text{str}, & 
\one_\text{str} &\stackrel{e}{\to} \one_\text{str}, & 
 s_{\Z_2}&\stackrel{\one}{\to} s_{\Z_2}
 .
\end{align}
The 1-morphisms describe how objects are connected --- in our case, how strings
are connected by point-like domain walls.  The point-like domain walls
connecting trivial string $\one_\text{str}$ to trivial string $\one_\text{str}$
are what we usually call point-like excitations.

\item The symmetric operators $O^\mathrm{symm}$ are 2-morphisms:
\begin{align}
\label{1mor2mor}
\one &\stackrel{O^\mathrm{symm}}{\to} \one, &
e &\stackrel{O^\mathrm{symm}}{\to} e, &
e\otimes b_s &\stackrel{O^\mathrm{symm}}{\to} e\otimes b_s, &
\nonumber\\
b_s &\stackrel{O^\mathrm{symm}}{\to} b_s, &
b_s &\stackrel{O^\mathrm{symm}}{\to} e\otimes b_s, &
e\otimes b_s &\stackrel{O^\mathrm{symm}}{\to} b_s.
\end{align}
The symmetric operators $O^\mathrm{symm}$ describe the possible ways a
point-like excitation can change (\ie possible ``domain walls'' on world lines
of point-like excitations in spacetime) .  We note that, $e\otimes b_s$ and
$b_s$ are connected by 2-morphisms.  Physically, it means that the $\Z_2$
charge $e$ can disappear or appear by itself near $b_s$, by processes induced
by symmetric operators.  This is expected $b_s$ is connected to a spontaneous
symmetry breaking state.  We also note that $e$ is the $\Z_2$ charge which is
not connected to the trivial excitation $\one$ by any 2-morphisms.

\end{enumerate}

Here we would like to introduce a notion elementary-type:\cite{KW1458, KZ200514178}
\begin{DefinitionPH}
Two morphisms (or objects which can be viewed as 0-morphisms) connected by
higher morphisms are said to have the same elementary-type.
\end{DefinitionPH}
We see that \ml{2\eRep_{\Z_2}} has only one elementary-types of objects, which
is the trivial elementary-type, \ie both string-$\one_\text{str}$ and
string-$s_{\Z_2}$ belong to trivial  elementary-type.  \ml{2\eRep_{\Z_2}} has
only three elementary-types of 1-morphisms (particles), $\one$, $e$, and
$b_s\cong e\otimes b_s$.  $e$ is an excitation in the usual physics sense since
it connect string-$\one_\text{str}$ to string-$\one_\text{str}$.  $e$ is not
connected to trivial excitation $\one$ by 2-morphisms, and thus is a
non-trivial elementary excitation.

In the above, we describe the symmetric fusion $2$-category \ml{2\eRep_{\Z_2}}
from the point of view of excitations.  We can also describe the symmetric
fusion $2$-category \ml{2\eRep_{\Z_2}} from the point of view of patch charge
operators, generated by $Z_{\text{str}_{ij}}$.  Note that patch charge
operators from a sub-algreba of the algebra of all t-patch operators.  

To switch from the excitation point of view to operator point of view, we
replace the excitations by the patch charge operators, that create the
corresponding excitations from $\Z_2$ symmetric product state.  Here, the
$\Z_2$ symmetric product state is given by
\begin{align}
|\Psi_\text{symm}\> = \bigotimes_i |0\>_i,\ \ \ \ 
|0\> = \frac{|+1\>+|-1\>}{\sqrt2},
\end{align}
where the $\Z_2$-symmetry action $W$ is given by $|+1\> \leftrightarrow |-1\>$.
This gives rise to a description of symmetric fusion $2$-category
\ml{2\eRep_{\Z_2}} in terms of patch charge operators (\ie t-patch operator
with empty bulk)
\begin{enumerate}

\item The object $\one_\text{str}$ in \ml{2\eRep_{\Z_2}} corresponds to a
disk-operator (a patch-operator with 2-dimensional patch) with empty bulk
\begin{align}
\hat \one_\text{str}(\text{loop}) = \prod_{i'\in \text{loop} = \prt\text{disk}}  \id_{i'}.
\end{align}
where $\id_i$ is the identity operator.  Here $\text{loop}$ is a closed
string, corresponding to the boundary of the disk.  The algebra of the operator
$\hat \one_\text{str}$ 
\begin{align}
\hat \one_\text{str}(\text{loop}) 
\hat \one_\text{str}(\text{loop}) 
=  \hat \one_\text{str}(\text{loop}) .
\end{align}
is consistent with the fusion of the object 
\begin{align}
\one_\text{str} \otimes \one_\text{str} = \one_\text{str}.
\end{align}

The object $s_{\Z_2}$ corresponds to a different disk-operator
with empty bulk
\begin{align}
\hat s_{\Z_2}(\text{loop}) &= 
\prod_{i'\in \text{loop} = \prt\text{disk}} 
P_{+,i'}
+
\prod_{i'\in \text{loop} = \prt\text{disk}} 
P_{-,i'} ,
\nonumber\\
P_\pm &= 
\frac{\id\pm Z}{2}.
\end{align}
(Here $P_\pm$ can be any local operators
that satisfy $P_+\neq P_-$ and $WP_+=P_-W$.)
Again, $s_{\Z_2}$ is a closed string, corresponding to the boundary of the disk.  We
note that string $s_{\Z_2}$ correspond to a spontaneous symmetry breaking state that
has a 2-fold degenerate ground states, $\otimes_i |+1\>_i$ and $\otimes_i
|-1\>_i$.  The operator $\prod_{i'\in \text{loop} = \prt\text{disk}} P_{+,i'} $
creates the state $\otimes_i |+1\>_i$ from $|\Psi_\text{symm}\>$, while the
operator $\prod_{i'\in \text{loop} = \prt\text{disk}} P_{-,i'} $ creates the
state $\otimes_i |-1\>_i$.  A particular superposition of the two states $
\prod_{i'\in \text{loop} = \prt\text{disk}} P_{+,i'} + \prod_{i'\in \text{loop} =
\prt\text{disk}} P_{-,i'} $ is invariant under the $\Z_2$ symmetry
transformation $W$.  The operator $\hat s_{\Z_2}(\text{loop})$ creates such
$\Z_2$ symmetric state, and satisfies
\begin{align}
\hat s_{\Z_2}(\text{loop}) X_\text{disk} = X_\text{disk} \hat s_{\Z_2}(\text{loop}) 
\end{align}
as long as the string is far away from the boundary of patch symmetry operator
$X_\text{disk}$.

The operator algebra
\begin{align}\label{sxs}
&\ \ \ \ \hat s_{\Z_2}(\text{loop}) \hat s_{\Z_2}(\text{loop}')   
\\
& =
\Big(\prod_{i'\in \text{loop} } P_{+,i'}
\prod_{i'\in \text{loop}' } P_{+,i'} 
+
\prod_{i'\in \text{loop} } P_{-,i'}
\prod_{i'\in \text{loop}' } P_{-,i'} 
\Big)
\nonumber\\
& + 
\Big(
\prod_{i'\in \text{loop} } P_{+,i'}
\prod_{i'\in \text{loop}' } P_{-,i'} 
+
\prod_{i'\in \text{loop} } P_{-,i'}
\prod_{i'\in \text{loop}' } P_{+,i'} 
\Big)
\nonumber\\
& \equiv   \hat s_{\Z_2}(\text{loop}''_1) + \hat s_{\Z_2}(\text{loop}''_2).
\nonumber 
\end{align}
implies the following fusion rule for the loop-like object $s_{\Z_2}$: 
\begin{align}
s_{\Z_2} \otimes s_{\Z_2} = s_{\Z_2}\oplus s_{\Z_2} = 2 s_{\Z_2},
\end{align}
which is non-trivial.  Here, we have assumed that the two strings, loop and
loop$'$, are not on top of each other, but are just nearby. Also
\begin{align}
\hat s_{\Z_2}(\text{loop}''_1) &\equiv
\prod_{i'\in \text{loop} }\hskip -1mm P_{+,i'} \hskip -1mm 
\prod_{i'\in \text{loop}' } \hskip -1mm P_{+,i'} 
+
\prod_{i'\in \text{loop} } \hskip -1mm P_{-,i'} \hskip -1mm
\prod_{i'\in \text{loop}' } \hskip -1mm P_{-,i'} 
,
\nonumber\\
\hat s_{\Z_2}(\text{loop}''_2) &\equiv
\prod_{i'\in \text{loop} } \hskip -1mm P_{+,i'} \hskip -1mm
\prod_{i'\in \text{loop}' }\hskip -1mm P_{-,i'} 
+
\prod_{i'\in \text{loop} } \hskip -1mm P_{-,i'} \hskip -1mm
\prod_{i'\in \text{loop}' }  \hskip -1mm P_{+,i'} 
,
\end{align}
and they both create spontaneous symmetry breaking states.

\item The 1-morphisms $\one$, $e$, and $b_s$ (or more precisely, pairs of
1-morphisms) correspond to boundary of open-string operators:
\begin{align}
\hat \one_i \hat \one_j &= \prod_{i' \in \prt \text{str}_{ij}} \id_{i'}=
\id_i\id_j,
\nonumber\\
\hat e_i \hat e_j &= \prod_{i' \in \prt \text{str}_{ij}} Z_{i'}
=Z_iZ_j,
\nonumber\\
\hat b_{s,_i} \boxtimes_s \hat b_{s,j} &= \prod_{i' \in \text{str}_{ij}} 
P_{+,i'} + \prod_{i' \in \text{str}_{ij}} P_{-,i'}
\end{align}
They are consistent with \eqn{obj1mor}, which
describes how objects are connected by the 1-morphisms.

We would like to remark that $\hat \one_i \hat \one_j$ and $\hat e_i \hat e_j$
are t-patch operators with an 1-dimensional patch, while $\hat b_{s,_i}
\boxtimes_s \hat b_{s,j}$ is a t-patch operators with a 2-dimensional patch
(\ie a disk).  The string $s_{\Z_2}$ form a part of the boundary of the disk, and the
string $\one_\mathrm{str}$ form the other part of the boundary.  The two types
of boundaries are connected by the 1-morphism $b_s$.

\item The symmetric operators $O^\mathrm{symm}$ are 2-morphisms.
From the operator algebra
\begin{align}
&\ \ \ \ \hat e_i \hat e_j \big(
\hat b_{s,_i} \boxtimes_s \hat b_{s,j}
\big) 
\nonumber\\
& =  Z_iZ_j\prod_{i' \in \text{str}_{ij}} P_{+,i'} 
+ 
Z_iZ_j\prod_{i' \in \text{str}_{ij}} P_{-,i'}
\nonumber\\
& =  \prod_{i' \in \text{str}_{ij}} P_{+,i'} 
+ 
\prod_{i' \in \text{str}_{ij}} P_{-,i'}
\nonumber\\
&=
\hat b_{s,_i} \boxtimes_s \hat b_{s,j}
\end{align}
we see that we cannot distinguish $b_{s,i}$ from $e_i\otimes b_{s,i}$, \ie they
are connected by identity operator.  This implies the relations $ b_{s,i}
\stackrel{O^\mathrm{symm}}{\to} e_i\otimes b_{s,i}$ and $ e_i\otimes b_{s,i}
\stackrel{O^\mathrm{symm}}{\to} b_{s,i}$, proposed in \eqn{1mor2mor}.  We also
note that operator $\hat \one_i \hat \one_j = \id_i\id_j $ cannot be connected
to operator $\hat e_i \hat e_j =Z_iZ_j $ via local symmetric operators near
vertex-$i$ and vertex-$j$.  This implies that there is no 2-morphisms
connecting $\one$ and $e$.

\end{enumerate}

In our above description of symmetric fusion 2-category $2\eRep_{\Z_2}$, we
include a descendant
excitation\cite{KW1458, LW170404221, GJ190509566, KZ200514178} $s_{\Z_2}$ formed
by elementary excitations $e$.  Such a descendant string excitation $s_{\Z_2}$
is a spontaneous $\Z_2$ symmetry breaking state formed by 1d $e$ gas.  

\begin{figure}[t]
\begin{center}
 \includegraphics[scale=0.8]{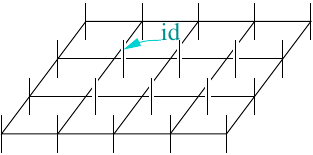} 
\end{center}
\caption{ The structure of a disk-like operator with empty bulk in term of
tensor network.  The short detached vertical lines represent identity operators
on different sites, which given rise to the empty bulk of the disk-like
operator.  The non-trivial string operator on the boundary of the disk may have
a Wess-Zumino form, \ie may be given by a tensor network on the disk bounded by
the string.  } \label{TNop2d} 
\end{figure}

In the above description of operator algebra, we construct the string operators
(or the disk operator with empty bulk) via operators $P^\pm$ on the string.  In
general,  the disk operator with empty bulk is given by a tensor network
operator, whose structure is given in Fig. \ref{TNop2d}.

Since descendant excitations are formed by lower dimensional excitations,
their existence and properties can be derived.  Thus, we may drop all the
descendant excitations and use only elementary
excitations,\cite{KW1458, KZ150201690}\footnote{The elementary excitations are
not formed by lower dimensional excitations. They are defined as the
excitations that do not have any domain wall with the trivial excitations.} to
obtain a simpler description of the symmetric fusion 2-category:
\begin{enumerate}

\item The string-like excitations $\one_\text{str}$ is the only elementary 
object
in \ml{2\eRep_{\Z_2}}.

\item The point-like excitations $\one$ and $e$ are the only elementary
1-morphisms:
\begin{align}
\label{obj1morph}
\nonumber\\
\one_\text{str} &\stackrel{\one}{\to} \one_\text{str}, & 
\one_\text{str} &\stackrel{e}{\to} \one_\text{str}, & 
\end{align}

\item The symmetric operators $O^\mathrm{symm}$ are all the 2-morphisms:
\begin{align}
\label{1mor2morph}
\one &\stackrel{O^\mathrm{symm}}{\to} \one, &
e &\stackrel{O^\mathrm{symm}}{\to} e. 
\end{align}
Note that the elementary morphisms (or objects) $\one$ and $e$ are not
connected to any other elementary morphisms (except themselves) by higher
morphisms.  This defines the \textbf{elementary} morphisms or objects\cite{KW1458, KZ150201690}.

\end{enumerate}

Through the above example, we see that the algebra of the patch charge
operators generated by $Z_{\text{str}_{ij}}$ from a symmetric fusion
2-category $2\eRep_{\Z_2}$.  Such a symmetric fusion 2-category $2\eRep_{\Z_2}$
fully characterize $\Z_2$ symmetry in 2-dimensional space, which is called the
representation category of the symmetry.

Similarly, we can use a fusion 2-category to describe the symmetry
transformations of the $\Z_2$ 0-symmetry, \ie to describe the operator algebra
generated by the patch symmetry operators $X_\text{disk}$.  The boundary of the
disk operators $X_\text{disk}$ are labeled by the group elements in \ml{G=\Z_2},
and correspond to the objects in the fusion 2-category.  Adding the trivial
1-morphisms and the top 2-morphisms formed by the local operators $X_i$ (\ie
the small disk operators), we get a fusion 2-category \ml{2\cVec_{\Z_2}}.  The
fusion 2-category \ml{2\cVec_{\Z_2}} fully describes the \ml{\Z_2} 0-symmetry in
2d space, which is the transformation category of the symmetry.

\subsection{$\Z_2^{(1)}$ 1-symmetry in 2\texorpdfstring{\MakeLowercase{d}}{d} space} 
\label{Z21symm2d}

\begin{figure}[t]
\begin{center}
 \includegraphics[scale=0.6]{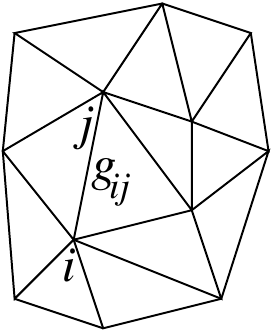}
\end{center}
\caption{A 2d lattice bosonic model, whose degrees of freedom live on the links and are labeled by the elements in a group: $g_{ij}\in G$} \label{TriLatt} 
\end{figure}

\begin{figure}[t]
\begin{center}
 \includegraphics[scale=0.6]{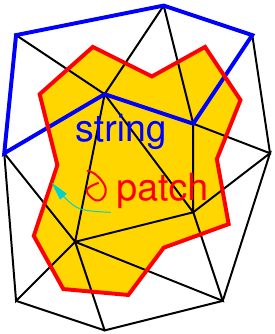}
\end{center}
\caption{A loop formed by links and a patch formed by vertices.
The boundary of the patch is formed by the links of the dual lattice.
} \label{TriLattStrPat} 
\end{figure}

In this section, we are going to discuss a lattice model with the simplest
higher symmetry, and the algebra of its local symmetric operators, as well as
its categorical description.  Let us consider a 2d lattice bosonic quantum
system with two state on every link of the lattice (see Fig. \ref{TriLatt}).
The \ml{\Z_2} {1-symmetry} is defined by the transformations on all the loops
\ml{S^1} (formed by the links, see Fig. \ref{TriLattStrPat}):
\begin{align}
 W(S^1) = \bigotimes_{\<ij\> \in S^1} \t X_{ij} ,
\end{align}
where $ \t X_{ij}$ are the Pauli X-operators acting in the link $\<ij\>$.
Local symmetric operators satisfy
\begin{align}
 W(S^1) O_i^\mathrm{symm} =
O_i^\mathrm{symm}W(S^1), \ \ \forall \text{ loops } S^1.
\end{align}
Such kind of symmetry was called \ml{d}-dimensional gauge-like symmetry
\cite{NOc0605316} or higher form symmetry  \cite{GW14125148}.

\begin{figure}[t]
\begin{center}
 \includegraphics[scale=0.4]{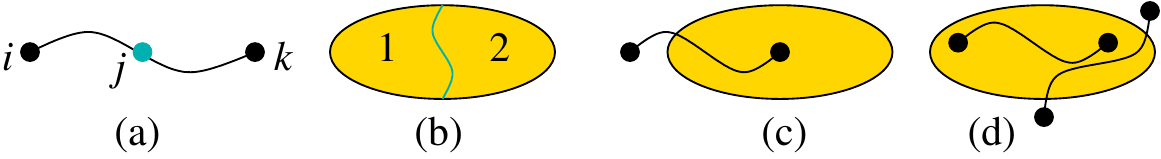} 
\end{center}
\caption{
(a) ``Fusion'' of two string operators.
(b) ``Fusion'' of two disk operators.
(c) Non-trivial ``braiding'' between string operator and disk operator.
(d) Trivial ``braiding'' between string operator and disk operator.
 } \label{OApatch} 
\end{figure}

The algebra of local symmetric operators is generated by
the following open string operators and disk operators:
\begin{align}
\t X_{\mathrm{str}_{ij}} = \bigotimes_{\<i'j'\> \in \mathrm{str}_{ij}}
\t X_{i'j'}
,\ \ \ 
\t Z_{\mathrm{disk}} = \bigotimes_{\<i'j'\> \in \prt \mathrm{disk} }
\t Z_{i'j'}
\end{align}
The key relations of the patch operator algebra are given by (see Fig. \ref{OApatch})
\begin{align}
&\t X_{\mathrm{str}_{ij}} \t X_{\mathrm{str}_{jk}} = \t X_{\mathrm{str}_{ik}},
\ \ \ \
\t Z_{\mathrm{disk}_1} \t Z_{\mathrm{disk}_2} = \t Z_{\mathrm{disk}_{1+2}}
\nonumber \\
& \t X_\mathrm{str} \t Z_\mathrm{disk} = \pm \t Z_\mathrm{disk} \t X_\mathrm{str}, 
\end{align}
where the $\pm$ signs depend on the relation between the string and the disk
(see Fig. \ref{OApatch}(c,d)).  Here and later in this paper, we will ignore
the operators associated with the descendant excitations.  All those descendant
operators are generated by the elementary operators (associated with the
elementary excitations) listed above.

\begin{figure}[t]
\begin{center}
\includegraphics[scale=0.4]{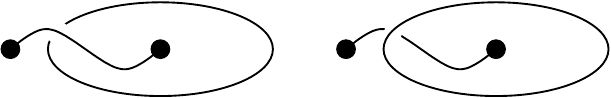}
\end{center}
\caption{ Non-trivial ``braiding'' between string operator and disk operator
measures the 0-symmetry charge carried by boundary of string, if the disk
operator generates a 0-symmetry. It measures the 1-symmetry charge carried by
boundary of disk, if the string operator generates a 1-symmetry.  }
\label{measurecharge} 
\end{figure}

We can also use the patch operators $\t X_\mathrm{str}$ on open strings to
define the 1-symmetry (see Fig. \ref{measurecharge}):
\begin{align}
\t X_\mathrm{str} O_i^\mathrm{symm} =
O_i^\mathrm{symm}
\t X_\mathrm{str},
\end{align}
where $i$ is far away from string ends.  Using such patch symmetry operators,
we can measure the $\Z_2^{(1)}$ 1-charge on the boundary of the disk operator
$\t Z_\mathrm{disk}$:
\begin{align}
\t Z_\mathrm{disk} \t X_\mathrm{str} = - \t X_\mathrm{str} \t
Z_\mathrm{disk}
\end{align}
when the string straddle across the boundary of the disk.  We see that a
$\Z_2^{(1)}$ 1-charge in 2-dimensional space is a 1-dimensional extended object.
In general, an $n$-dimensional charge object correspond to \ml{n}-symmetry, in
any space dimensions.

We can use a fusion 2-category to describe the charges of the $\Z_2^{(1)}$
1-symmetry, \ie to described the operator algebra of the patch charge operators
$\t Z_\text{disk}$.  The 1-dimensional (co-dimension-1) extended charge objects
(the boundary of the disk operators $\t Z_\text{disk}$) are labeled by the
group elements in \ml{G=\Z_2}, and correspond to the objects in the fusion
2-category.  Adding the trivial 1-morphisms and the top 2-morphisms formed by
the local operators $\prod_{i\in \text{small loop}}\t Z_i$ (\ie the small disk
operators), we get a fusion 2-category \ml{2\cVec_{\Z_2}}.  The fusion
2-category \ml{2\cVec_{\Z_2}} fully describes the \ml{\Z_2} 1-symmetry in 2d
space. Such a fusion 2-category \ml{2\cVec_{\Z_2}} is the representation
category of the symmetry.

We can also use a fusion 2-category to describe the symmetry transformations of
the $\Z_2^{(1)}$ 1-symmetry, \ie to describe the operator algebra of the patch
symmetry operators $\t X_\text{str}$.  The boundary of the string operators $\t
X_\text{str}$) are labeled by the representations in \ml{G=\Z_2}, and
correspond to the 1-morphisms in the fusion 2-category.  Adding the trivial
objects and the top 2-morphisms formed by the local operators $\t X_i$ (\ie the
small string operators), we get a fusion 2-category
\ml{2\cRep_{\Z_2}}.\footnote{In this paper, when we refer to $
2\mathcal{R}\text{ep}_G $, we mostly only consider its associated elementary
excitations and the related structures (which correspond to a pre-fusion
2-category).  We do not fully discuss the associated descendent excitations in
$ 2\mathcal{R}\text{ep}_G $. }   The fusion 2-category \ml{2\cRep_{\Z_2}} fully
describes the \ml{\Z_2} 1-symmetry in 2d space. Such a fusion 2-category
\ml{2\cRep_{\Z_2}} is the transformation category of the $\Z_2^{(1)}$
1-symmetry in 2-dimensional space.

\subsection{The equivalence between $\Z_2$ 0-symmetry and $\Z_2^{(1)}$ 1-symmetry
in 2d space } 

We have seen that a $\Z_2$ 0-symmetry can be fully described by a
representations category $2\cRep_{\Z_2}$ or by a transformation category
$2\cVec_{\Z_2}$.  We also see that a $\Z_2^{(1)}$ 1-symmetry can be fully
described by a representations category $2\cVec_{\Z_2}$ or by a transformation
category $2\cRep_{\Z_2}$.  Now it is clear that the two very different looking
symmetries, $\Z_2$ and $\Z_2^{(1)}$, are closely related, \ie they become
identical if we exchange what we call patch charge operators and what we call
patch symmetry operators.

In fact, the two symmetries, $\Z_2$ and $\Z_2^{(1)}$, are indeed equivalent, if
we consider the operator algebras of all local symmetric operators, \ie the
operator algebras generated by both patch charge operators and patch symmetry
operators.  The full operator algebra of  $\Z_2$ symmetry is defined via the
following operator relations
\begin{align}
&Z_{\mathrm{str}_{ij}} Z_{\mathrm{str}_{jk}} = Z_{\mathrm{str}_{ik}},
\ \ \
X_{\mathrm{disk}_1} X_{\mathrm{disk}_2} = X_{\mathrm{disk}_{1+2}}
\nonumber\\
& Z_\mathrm{str} X_\mathrm{disk} = \pm X_\mathrm{disk} Z_\mathrm{str}, 
\end{align}
The full operator algebra of $\Z_2^{(1)}$ symmetry is defined via the following operators relations
\begin{align}
& \t X_{\mathrm{str}_{ij}} \t X_{\mathrm{str}_{jk}} = \t X_{\mathrm{str}_{ik}},
\ 
\t Z_{\mathrm{disk}_1} \t Z_{\mathrm{disk}_2} = \t Z_{\mathrm{disk}_{1+2}}
\nonumber\\
& \t X_\mathrm{str}\t Z_\mathrm{disk}=\pm\t Z_\mathrm{disk} \t X_\mathrm{str}, 
\end{align}
We see that the two operator algebras are isomorphic. 
Thus the $\Z_2$ and $\Z_2^{(1)}$
symmetries have the same \catsymm, which implies that they are
equivalent.  

In fact, the \catsymm\ from the full algebra of extended t-patch operators corresponds to
a non-degenerate braided fusion 2-category $2\eGau_{\Z_2}$ (which describes the
excitations in a 2d $\Z_2$-gauge theory).  The boundary of the disk operators
are labeled by the group elements of $\Z_2$, and correspond to the object in
the braided fusion 2-category $2\eGau_{\Z_2}$.  The ends of the string
operators are labeled by the group representations, and correspond to the
1-morphisms in $2\eGau_{\Z_2}$.  The local symmetric operators (\ie the small
string and small disk operators) correspond to the 2-morphisms in
$2\eGau_{\Z_2}$.  The string-like elementary excitations (the objects) and the
point-like elementary excitations (the 1-morphisms) can fully detect each
others, due to their non-trivial mutual statistics, as implied by the operator
relation
\begin{align}
Z_\mathrm{str} X_\mathrm{disk} = \pm X_\mathrm{disk} Z_\mathrm{str}. 
\end{align}
Thus the  braided fusion 2-category for the full algebra of extended t-patch operators is
non-degenerate. \footnote{The adjective ``full'' here refers to the ``non-degeneracy'' of the associated
braided fusion category.}

A mathematically rigorous proof of this equivalence was presented in
\Rf{D2239}, in terms of the category theoretical notion of Morita equivalence.

\section{A review of holographic theory of (algebraic higher) symmetry}

In the previous sections, we studied many simple examples, trying to
demonstrate a holographic theory of (algebraic higher) symmetry via algebras of
local symmetric operator.  In this section, we are going to present the
holographic theory for generic cases.  Such a holographic theory was developed
in \Rf{KZ200514178} via excitations above the symmetric ground state.  Here we
will present a simplified version, ignoring some subtleties. 

\subsection{Representation category}

We know that symmetries are classified by groups and higher symmetries are
classified by higher groups.  As demonstrated in the last section, it turns out
that algebraic higher symmetries (\ie non-invertible symmetries) are described
by fusion higher categories,\cite{KZ200514178} which is the representation
category\cite{W181202517} generated by patch charge operators that we
introduced in Section \ref{repcat}.  

However, not all fusion higher categories can be representation categories that
describe algebraic higher symmetries.  To identify which fusion higher
category can describe a symmetry, we note that a symmetry is breakable.  The
symmetry breaking will change the fusion higher category into a trivial fusion
higher category $n\cVec$.  This motivate \Rf{KZ200514178} to conjecture that
local fusion higher categories $\cR$ (\ie representation categories generated
by patch charge operators) describe and classify algebraic higher symmetries:
\begin{DefinitionPH}
\label{def:LFC}
A fusion $n$-category $\cR$ equipped with a \textbf{top-faithful} surjective
monoidal functor $\bt$ from $\cR$ to the trivial fusion $n$-category, $\cR
\map{\bt} n\cVec$, is called a \textbf{local fusion $n$-category}.  Here,
\textbf{top-faithful} means that the functor $\bt$ is injective when acting on
the top morphisms (\ie the $n$-morphism in this case). The pair $(\cR,\bt)$
classify anomaly-free algebraic higher symmetries in $n$-dimensional space
(which include anomaly-free symmetries, higher symmetries, and non-invertible
symmetries).  
\end{DefinitionPH}
\noindent To be brief, we usually drop $\bt$ in the pair.
This generalizes the discussion in Section \ref{repcat}.  Physically, the
functor $\bt$ means ``ignore the symmetry'' or ``explicitly break the symmetry
by small perturbations''.  Thus at the top-morphism level, $\bt$ maps local
symmetric operators to local operators, which is a injective map.  At
lower-morphism/object level, the charged excitations in \ml{\cR} are mapped to
the excitations in \ml{n\eVec}. This implies that all the objects and morphisms
in a local fusion higher category \ml{\cR} have integral quantum dimensions.

For example, if we have an \ml{SU(2)} symmetry, then there is a ``charged''
excitation, spin-\ml{1/2} excitation (carrying the 2-dim representation of
\ml{SU(2)}).  If we ignore the \ml{SU(2)} symmetry, such a spin-\ml{1/2}
excitation can be viewed as an accidental degeneracy of two trivial
excitations: 
\begin{align}
\underbrace{\text{spin-}1/2}_{\in \cR} 
\stackrel{\bt}{\to} \underbrace{\one\oplus\one}_{\in n\eVec} .
\end{align}

\begin{figure}[t]
\begin{center}
 \includegraphics[scale=0.6]{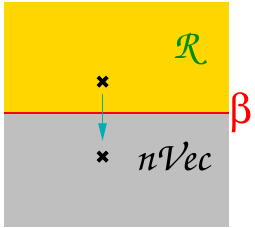} 
\end{center}
\caption{A spacetime picture of the symmetry breaking process $\bt$ (the
vertical direction is the time).  $\bt$ can be viewed as a domain wall between
a product state with the symmetry and a product state with no symmetry.  The
(extended) excitations on the  product state with the symmetry are described by
a fusion higher category $\cR$.  The (extended) excitations on the  product
state with no symmetry are described by the fusion higher category $n\cVec$.
All the top morphisms (the symmetric spacetime instantons or symmetric local
operators) in $\cR$ can go through the domain wall $\bt$ and become the top
morphisms (the spacetime instantons or local operators) in $n\cVec$ without
modification.  } \label{RVec} 
\end{figure}

As we have mentioned above, $\bt$ is a symmetry breaking process. We can also
view $\cR$ as the fusion higher category describing the (extended) excitations
in a symmetric product state with the symmetry.  From this angle, we can view
$\bt$ as a domain wall between $\cR$ and $n\cVec$. The domain wall is
transparent to all the top morphisms in $\cR$ (see Fig. \ref{RVec}).

\subsection{A holographic point view of symmetry} 

Consider two $n$d (algebraic higher) symmetries described by two local fusion
$n$-categories, \ml{\cR} and \ml{\cR'}.  We know that the two symmetries are
equivalent if their algebras of local symmetric operators are isomorphic.  We
have demonstrated that an isomorphic class of local symmetric operator algebras is described by
a braided fusion $n$-category, and called such an isomorphic class as a
\catsymm.  So what is the \catsymm\ for a symmetry
described by local fusion higher categories, \ml{\cR}?

\begin{figure}[t]
\begin{center}
 \includegraphics[scale=0.6]{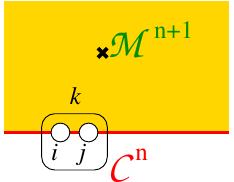} 
\end{center}
\caption{The holographic principle of topological order: \emph{boundary $\cC^n$
uniquely determines bulk $\cM^{n+1}$}. \label{cCcM} }
\end{figure}

To answer this question, let us first review the holographic principle of
topological order: \emph{boundary uniquely determines bulk}.  In physics,
topological orders (\ie gapped quantum liquids) in \ml{n+1}d space are
characterized by their co-dimension-1, co-dimension-2, ...  excitations.  In
other words, such a topological orders are characterized by fusion
\ml{n+1}-category \ml{\cM^{n+1}}

On a \ml{n}-dimensional gapped boundary of the \ml{n+1}-dimensional topological
order, the excitations are described by a fusion \ml{n}-category \ml{\cC^n}.
The holographic principle of topological order state that the boundary
\ml{\cC^n} uniquely determines the bulk \ml{\cM^{n+1}}.  Such a boundary-bulk
relation is given by the center map $\cZ$ in mathematics (see Fig.
\ref{cCcM}):\cite{JS9151, M0319, KW1458, KZ150201690, KZ170200673}s
\begin{align}
\cZ(\cC^n) = \cM^{n+1}.
\end{align}
We see that the physical meaning of ``center'' is ``bulk''. The center map (or
the bulk map) $\cZ$ has a property that the center of a center (or the bulk of
a bulk) is trivial
\begin{align}
 \cZ(\cZ(\cC^n)) = (n+2)\cVec.
\end{align}
This is dual to the well known fact: the boundary of a boundary is trivial.

\begin{figure}[t]
\begin{center}
 \includegraphics[scale=0.6]{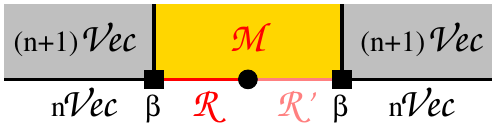} 
\end{center}
\caption{Two symmetries described by
fusion $n$-categories $\cR$ and $\cR'$
are equivalent (\ie have the same \catsymm)
iff they have the same bulk topological order in one higher dimension:
$\eZ(\cR)\cong \eZ(\cR')$.} \label{dEqSymm} 
\end{figure}

It was conjectured that,\cite{KZ200514178} in $n$-dimensional space, the
relation between the representation category $\cR$ generated by all the patch
charge operators and the braided fusion $n$-category $\eM$ (\ie the
\catsymm) generated by all the patch charge operators and the patch
symmetry operators is given by another center map, denoted as
$\eZ$,\cite{KW1458, KZ150201690, KZ170200673} that maps a fusion $n$-category
$\cR$ into a braided fusion $n$-category $\eM$.  The new center map $\eZ$ is
closely related to the previous center map $\cZ$ that maps a fusion
$n$-category $\cR$ into a fusion $(n+1)$-category $\cM$.  This is because both
fusion $(n+1)$-category $\cM$ and  braided fusion $n$-category $\eM$ can be use
to fully describe an anomaly-free topological order in $n+1$-dimensional space.

We note that in an anomaly-free topological order in
$(n+1)$-dimensional space, all the co-dimension-1 excitations are descendant
(\ie formed by lower dim excitations).  Dropping the co-dimension-1 excitations
(called looping \ml{\Om}) maps a fusion $(n+1)$-category $\cM$ into a braided
fusion \ml{n}-category $\eM$: \ml{\eM=\Om \cM}.  Adding back the  descendant
co-dimension-1 excitations is called de-looping followed by Karoubi completion: \ml{\Si \eM=
\cM}.\cite{KW1458, GJ190509566} Thus the anomaly-free topological order can be
described either by the braided fusion \ml{n}-category $\eM$, or by fusion
\ml{(n+1)}-category $\cM$.  The anomaly-free condition of topological order
corresponds to the non-degeneracy condition for the braided fusion \ml{n}-category
$\eM$, which becomes the trivial center condition for the fusion
\ml{(n+1)}-category $\cM$: $\cZ(\cM) = (n+2) \cVec$.
The two kinds of center maps are related by
\begin{align}
 \Si\eZ = \cZ,\ \ \ \ \Om \cZ = \eZ.
\end{align}

This mathematical result provides a macroscopic way to compute the holographic
equivalence classes of symmetries (\ie the topological order in one higher
dimension).  In particular, the two symmetries, described by two
representations categories $\cR$ and $\cR'$, are equivalent, iff they have
equivalent centers (\ie have the same bulk topological order, or have the 
same \catsymm, see Fig.  \ref{dEqSymm})\cite{KZ200514178} 
\begin{align} 
\eZ(\cR) \cong \eZ(\cR').
\end{align}

Not every braided fusion higher category describes a \catsymm. The
operator algebra is formed by \emph{all} the local symmetric operators.  The
condition of \emph{all}, is translated into a condition on the braided fusion
higher category \ml{\eM}: \ml{\eM} must be \emph{non-degenerate}, \ie
satisfying \ml{\eZ(\Si\eM) = (n+1)\eVec}.  Therefore, \textsf{categorical symmetries}
(\ie the isomorphic classes of algebras of symmetric local operator) in \ml{n}d
space are classified by non-degenerate braided fusion \ml{n}-categories
\ml{\eM}.

\subsection{Transformation category -- dual of the representation category}

\begin{figure}[t]
\begin{center}
\includegraphics[scale=0.6]{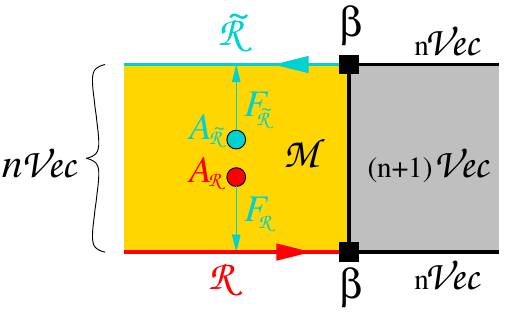} 
\end{center}
\caption{$\cR\boxtimes_{\cM} \t\cR^\mathrm{rev}$ is a fusion $n$-category that
describes the excitations in a slab of topological order in $(n+1)$-dimensional
space.  One boundary of the slab has excitations described by fusion
$n$-category $\cR$.  The other boundary of the slab has excitations described
by fusion $n$-category $\t \cR^\mathrm{rev}$.  The condition
$\cR\boxtimes_{\cM} \t\cR^\mathrm{rev}=n\cVec$ ensure that all the excitations
on the boundary $\cR$ and $\t \cR$ comes from symmetry described by the bulk
$\eM$. In other words, all the excitations on the boundary are symmetry
charges. There is no topological excitations.  $F_\cR$ is the forgetful functor
that maps bulk excitations described by $\eM$ to boundary excitations described
by $\cR$.  $A_\cR$ is a Lagrangian condensable algebra formed by bulk
excitations, which are mapped to trivial excitations on the boundary $\cR$.
$A_\cR$, with a trivial action on the symmetric boundary, correspond to the
patch symmetry operators on the boundary.  The non-trivial excitations $\cR$ on
the $\cR$-boundary are created by the patch charge operator.  $\cR$ and $\t
\cR$ are dual to each other is all the bulk excitations either condense on the
$\cR$-boundary or  $\t\cR$-boundary.  } \label{RdR} 
\end{figure}

Instead of representation category generated by patch charge operators, we can
also use transformation category generated by patch symmetry operators to fully
describe an (algebraic higher) symmetry. We believe both characterizations are
complete characterizations.  This belief is supported by the following result
\cite{KZ200514178}:
\begin{Proposition}
Consider two fusion $n$-category $\cR$ and $\t\cR$, such that
$\eM=\eZ(\cR)=\eZ(\t\cR)$. If $ n\cVec = \cR\boxtimes_{\eM} \t\cR^\mathrm{rev}
$ (see Fig. \ref{RdR}), then both $\cR$ and $\t\cR$ are local fusion
$n$-categories.  Furthermore, for each $\cR$, $\t\cR$ is unique.
We say that $\t\cR$ is the \textbf{dual} of $\cR$.
\end{Proposition}

For example, an $n$d bosonic lattice model with a finite symmetry $G$  has a
representations category $n\cRep_G$ and a transformation category is
$n\cVec_G$.  $n\cVec_G$ happens to be the dual of $n\cRep_G$.  Such a bosonic
model has a dual lattice model with a dual symmetry $G_\text{rep}^{(n-1)}$ (see
\Rf{KZ200514178} for an explicit construction).  The representations category
of the dual symmetry $G_\text{rep}^{(n-1)}$ is $n\cVec_G$, and the
transformation category of the dual symmetry is $n\cRep_G$.  This example
illustrates the dual relation between the representations category and the
transformation category.

Putting the representation category  and the transformation category together -- \ie
combining the algebras of patch charge operators and patch symmetry
operators -- gives us the full algebra of local symmetric
operators. This algebra contains the full information of the \catsymm, which represents the essence
of symmetry. From this point of view, symmetry and dual symmetry have the same \catsymm\
and are equivalent.  They only differ by swapping the names for patch charge
operators and patch symmetry operators.

\subsection{A simple example}

In this subsection, we are going to discuss a simple example, to illustrate the
above abstract discussions.

\subsubsection{Holographic view of 2d $\Z_2$ 0-symmetry}

As we have discussed in Section \ref{Z2symm2d}, the representation category of
2d \ml{\Z_2} 0-symmetry is a fusion 2-category \ml{\cR = 2\cRep_{\Z_2}}.  The
transformation category of 2d \ml{\Z_2} 0-symmetry is a fusion 2-category
\ml{\t\cR = 2\cVec_{\Z_2}}.  It has the \catsymm\
\ml{2\eGau_{\Z_2} =\eZ(2\cRep_{\Z_2}) = \eZ(2\cVec_{\Z_2})}, which is the 3d
topological order described by \ml{\Z_2} gauge theory.  In the following, we
will use the holographic picture to understand the above
results.

\begin{figure}[t]
\begin{center}
 \includegraphics[scale=0.6]{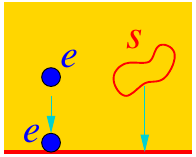}
\end{center}
\caption{A boundary of 3d $\Z_2$ topological order $\eM = 2\eGau_{\Z_2}$ induced
by $s$-string condensation.  The boundary excitations is described by fusion
2-category $\cR=2\cRep_{\Z_2}$.} \label{Z2bndry} 
\end{figure}

The \emph{elementary} excitations in  3d \ml{\Z_2}-gauge theory include
point-like excitations \ml{e} (the bosonic \ml{\Z_2} charge) and string-like
excitations \ml{s} (the bosonic \ml{\Z_2}-flux string), as well as
the trivial excitations $\one$ and $\one_\text{str}$.
They satisfy the fusion rule:
\begin{align}
e\otimes e =\one
 \ \ \ s\otimes s =\one_\text{str} 
\end{align}
The string-like excitation \ml{s} corresponds to the flux line in the 3d
\ml{\Z_2}-gauge theory, which is an elementary excitation.  The 3d
\ml{\Z_2}-gauge theory also has a non-elementary excitation (\ie descendant)
string-like excitation, $s_{\Z_2}$, which is a $\Z_2$
spontaneous-symmetry-break state formed by the $e$-particles.  Here we ignore
all the descendant excitations.

\ml{\cR=2\cRep_{\Z_2}} is a boundary of \ml{2\eGau_{\Z_2}}, induced by the
\ml{\Z_2}-flux loop condensation, so on the boundary \ml{s \sim
\one_\text{str}}. The boundary excitations then are described by
\ml{\{\one,e\}=2\cRep_{\Z_2}}.  Fig. \ref{Z2bndry} represents the picture that
a symmetry characterized by representation category \ml{\cR=2\cRep_{\Z_2}}
has the \catsymm\ $2\eGau_{\Z_2}$.  The Lagrangian condensible algebra is
generated by $s$, which corresponds to the transformation category $\t\cR =
2\cVec_{\Z_2}$. Thus Fig. \ref{Z2bndry} also represents the picture that a
symmetry characterized by transformation category \ml{\t\cR=2\cVec_{\Z_2}} has
the \catsymm\ $2\eGau_{\Z_2}$.

\subsubsection{Holographic view of 2d $\Z_2$ 1-symmetry}

As we have discussed in Section \ref{Z21symm2d}, the representation category of
2d \ml{\Z_2} 0-symmetry is a fusion 2-category \ml{\t\cR = 2\cVec_{\Z_2}}.  The
transformation category of 2d \ml{\Z_2} 0-symmetry is a fusion 2-category
\ml{\cR = 2\cRep_{\Z_2}}.  It belongs to \catsymm\ \ml{2\eGau_{\Z_2}
=\eZ(2\cRep_{\Z_2}) = \eZ(2\cVec_{\Z_2})}, which is the 3d topological order
described by \ml{\Z_2} gauge theory.  

\begin{figure}[t]
\begin{center}
 \includegraphics[scale=0.6]{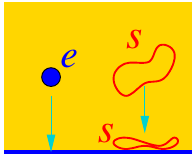}
\end{center}
\caption{A boundary of 3d $\Z_2$ topological order $\eM = 2\eGau_{\Z_2}$ induced
by $e$-particle condensation.  The boundary excitations is described by fusion
2-category $\t\cR=2\cVec_{\Z_2}$.} \label{Z2bndry1} 
\end{figure}

\ml{\t\cR=2\cVec_{\Z_2}} is a boundary of \ml{2\eGau_{\Z_2}}, induced by the
\ml{\Z_2}-charge condensation, so on the boundary \ml{e \sim \one}. The boundary
excitations then are described by \ml{\{\one_\text{str},s\}=2\cVec_{\Z_2}}.
Fig. \ref{Z2bndry} represents the picture that a symmetry characterized by
representation category \ml{\t\cR=2\cVec_{\Z_2}} has the \catsymm\ $2\eGau_{\Z_2}$.  The Lagrangian condensible algebra is generated by $e$,
which corresponds to the transformation category $\cR = 2\cRep_{\Z_2}$. Thus
Fig. \ref{Z2bndry1} also represents the picture that a symmetry characterized by
transformation category \ml{\cR=2\cRep_{\Z_2}} has the \catsymm\
$2\eGau_{\Z_2}$.

\subsubsection{Symmetry $\cR = 2\cRep_{\Z_2}$ and dual-symmetry
$\t\cR = 2\cVec_{\Z_2}$}

\begin{figure}[t]
\begin{center}
 \includegraphics[scale=0.6]{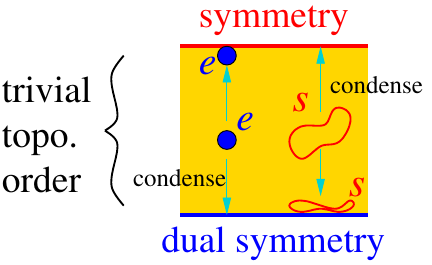}
\end{center}
\caption{All the non-trivial excitations in the bulk $2\eGau_{\Z_2}$, either
condense on the \ml{2\cRep_{\Z_2}}-boundary ($s$ condense) or condense on the
\ml{2\cVec_{\Z_2}}-boundary ($e$ condense).  Thus the slab has no topological
excitations and correspond to a trivial topological order.} \label{RdRcond} 
\end{figure}
 
In 2d space, \ml{\Z_2} 0-symmetry and \ml{\Z_2^{(1)}} 1-symmetry are
equivalent, and are dual to each other.  This means that \ml{\Z_2} 0-symmetry
and \ml{\Z_2^{(1)}} 1-symmetry have the same 3d bulk topological order
$2\eGau_{\Z_2}$ (\ie have the same \catsymm).  When we consider a slab 3d bulk
topological order $2\eGau_{\Z_2}$ with one boundary being $2\cRep_{\Z_2}$ and
the other boundary being $2\cVec_{\Z_2}$, then all the non-trivial excitations
in the bulk, either condense on the \ml{2\cRep_{\Z_2}}-boundary or condense on
the \ml{2\cVec_{\Z_2}}-boundary (see Fig. \ref{RdRcond}).  So the slab is
actually a trivial 2d topological order.  This implies that $2\cRep_{\Z_2}$ and
$2\cVec_{\Z_2}$ are dual to each other.

\section{A derivation of topological holographic principle}

In this paper, we have derived a holographic point of view of symmetry. For a
lattice system with a symmetry, we concentrate on the algebra of local
symmetric operators, and its irreducible representation -- the symmetric
sub-Hilbert space.  The symmetric sub-Hilbert space does not have a tensor
product decomposition, which indicates a (non-invertible) gravitational
anomaly.\footnote{Here, we view a gravitational anomaly is an obstruction to
have a lattice realization \emph{without symmetry}.}  Since the
(non-invertible) gravitational anomaly corresponds to a topological order in
one higher dimension (for finite symmetries) \cite{W1313, KW1458}, the symmetric
sub-Hilbert space, plus the algebra of local symmetric operators in it, gives
rise to a topological order in one higher dimension.

The above is just some vague ideas.  In this paper, we out line a way to
compute this topological order in one higher dimension, using the algebra of
local symmetric operators.  This approach is very general.  Even if we do not
know the symmetry transformation and do not know the symmetric sub-Hilbert
space, but if we know the set of local operators and its algebra, then we can
compute the bulk topological order, by compute the braided fusion (higher)
category from the operator algebra.

Under such a general setting, our approach can be viewed a derivation of
topological holographic principle, which can be simply stated as:
\emph{boundary determines the bulk}.  The usual holographic principle in
AdS/CFT refers to boundary conformal field theory (CFT) with a global symmetry
determines a bulk quantum gravity with a gauge theory in an anti-de Sitter
(AdS) space in one higher dimension.  The topological holographic principle
here refers to boundary quantum field theory determines a bulk topological
order in one higher dimension.  In this paper, we make the above statement more
precise by treating quantum field theory as an algebra of local operators.  As
was shown in this paper, from the algebra of local operators, we can determine
a non-degenerate braided fusion (higher) category, which in turn determine the
bulk topological order (provided that the braided fusion (higher) category is
finite). This corresponds to a derivation of the topological holographic
principle.

We may also consider one of the many boundaries of a topological order.  The
boundary is more precisely described by an algebra of boundary local operators,
which create all the low energy boundary excitations.\footnote{Here, we may
assume the bulk topological order to have an infinite energy gap.  Then any
finite energy excitations can be viewed as boundaries excitations.} Then, from
the boundary operator algebra, we can determine a braided fusion (higher)
category which determine the bulk topological order, up to an invertible
topological order.  The invertible topological order correspond to the usual
invertible gravitational anomaly of the boundary theory, which is also
determined by the boundary.  This way, we showed that \frmbox{\emph{boundary
theory uniquely determines the topological bulk}, } which is the topological
holographic principle.

In this paper, we try to use topological order to describe generalized symmetry
(which can go beyond group and higher group) in one lower dimension.  We like
to remark that, it is the non-invertible topological order that is close to
symmetry.  The invertible topological order (and the associated usual more
familiar invertible gravitational anomaly) is furthest from symmetry.  At
moment, it is not clear should we generalize the symmetry even more to include
the ones associated with invertible topological order in one higher dimension,
or should we use equivalent classes of bulk topological order up to invertible
topological order to describe generalized symmetry.

\section{Equivalent symmetries}
\label{eqsymm}

One application of the holographic theory of symmetry is to identify
equivalence between symmetries, higher symmetries, anomalous (higher)
symmetries, algebraic (higher) symmetries, and gravitational anomalies.  All
those (anomalous and/or higher) symmetries and gravitational anomalies impose
constraint on the low energy dynamics of the system.  They are equivalent if
they impose the identical constraint.  Such an equivalence was called
holo-equivalence in \Rf{KZ200514178}, to stress its connection holographic
picture.

As we have discussed in this paper, two symmetries (described by representation
categories $\cR$ and $\cR'$) are equivalent if they have the same
\catsymm, \ie have the same bulk topological order:
\begin{align}
\eZ(\cR) \cong \eZ(\cR').
\end{align}
In practice, if we know a (higher) symmetry $\cR$ is realized as a boundary of
a SPT state or a symmetric product state, then the \catsymm\ is
simply the bulk topological order obtained by gauging the (higher) symmetry in
the bulk SPT state or the symmetric product state.  We can identify many
equivalent symmetries this way.

\subsection{Some known examples}

First, let us list some known examples.  In \ml{n}d space, \ml{Z_N^{(m)}}
\ml{m}-symmetry can be realized by a boundary of $(n+1)$d product state with
\ml{Z_N^{(m)}} \ml{m}-symmetry.  Thus the \catsymm\ of $n$d
\ml{Z_N^{(m)}} \ml{m}-symmetry is the $(n+1)$d \ml{Z_N} \ml{(m+1)}-gauge
theory.  In $(n+1)$-dimensional space, \ml{Z_N} \ml{(m+1)}-gauge theory and
\ml{Z_N} \ml{(n-m)}-gauge theory correspond to the same topological order.
Therefore, in \ml{n}d space, \ml{Z_N^{(m)}} \ml{m}-symmetry is equivalent to
\ml{Z_N^{(n-m-1)}} \ml{(n-m-1)}-symmetry: 
\begin{align}
\label{ZNZN}
Z_N^{(m)} \sim Z_N^{(n-m-1)}. 
\end{align}
Furthermore, the two symmetries are dual to each other.

Using the similar argument, we can obtain the following results 
\begin{itemize}
\item
 In 2d,  \ml{\Z_3\times \Z_2 \sim \Z_3^{(1)} \times \Z_2}.
This is actually a direct application of \eqn{ZNZN}.

\item
 In 2d,  \ml{S_3 = \Z_3\rtimes \Z_2 \sim \Z_3^{(1)} \rtimes
Z_2}.\cite{JW191213492}  This is the twisted version of the above.
\ml{\Z_3^{(1)} \rtimes \Z_2} is a non-trivial mix of \ml{\Z_3^{(1)}} 1-symmetry
and \ml{\Z_2} 0-symmetry.  The charge objects of \ml{\Z_3^{(1)}} are strings
labeled by \ml{s,\bar s}.  The \ml{\Z_2} 0-symmetry exchange \ml{s} and \ml{\bar
s}.

\item
In 1d, an anomalous \ml{\Z_2\times \Z_2\times \Z_2} symmetry is equivalent to
\ml{D_4} symmetry, for a very different reason than the above
two examples.\cite{GNm0603191, WW1454}

\end{itemize}

\subsection{Equivalence between anomalous and anomaly-free $\Z_n$ and
$\Z_{n_1}\times \Z_{n_2}$ symmetries in 1-dimensional space}

In Section \ref{Z2Z2aZ4}, we find an equivalence between 1d $\Z_4$ symmetry and
$\Z_2\times \Z_2$ symmetry with the mixed anomaly.  In this section, we would like to
generalize that result.  An 1d anomalous $\Z_n$ symmetry is realized by a
boundary of 2d $\Z_n$ SPT state. After gauging the $\Z_n$ symmetry in the 2d SPT
state, we obtain a 2d Abelian bosonic topological order, which is classified by
even $K$-matrices.\cite{WZ9290} In the present case, the corresponding
topological order is given by\cite{W1447}
\begin{align}
 K = \begin{pmatrix}
 -2m & n\\
   n & 0\\
\end{pmatrix}
\end{align}
where $m \in H^3(\Z_n; \RZ)=\Z_n$ charactering the $\Z_n$ anomaly ($m=0$ for
anomaly-free).  We will label the anomalous $\Z_n$ symmetry by $(n;m)$.

Similarly, the anomalous 1d $\Z_{n_1}\times \Z_{n_2}$ symmetry is realized by a
boundary of 2d $\Z_{n_1}\times \Z_{n_2}$ SPT state.  After gauging the
$\Z_{n_1}\times \Z_{n_2}$ symmetry, we obtain a 2d Abelian topological order
characterized by\cite{W1447}
\begin{align}
 K=
\begin{pmatrix}
 -2m_2 &  n_1 &    -m_{12} &   0\\ 
   n_1  &  0  &     0  &  0\\ 
  -m_{12}  &  0  & -2m_1  & n_2\\ 
    0  &  0  &    n_2  &  0\\
\end{pmatrix},
\end{align}
where $m_1\in \Z_{n_1}$ describing the anomaly of the $\Z_{n_1}$ symmetry,
$m_2\in \Z_{n_2}$ describing the anomaly of the $\Z_{n_2}$ symmetry, and
$m_{12}\in \Z_{\text{gcd}(n_1,n_2)}$ describing the mixed anomaly of the
$\Z_{n_1} \times \Z_{n_2}$ symmetry.  We will label the anomalous $\Z_{n_1} \times
\Z_{n_2}$ symmetry by $(n_1,n_2;m_1,m_{12},m_2)$.

By computing the $S, T$ matrices of the 2d topological
orders\cite{W9039, W150605768} described by $K$-matrices, we can identify a set
of $K$-matrices that give rise to the same 2d topological order, and hence
correspond to equivalent symmetries. This allows us to find the following sets
of equivalent symmetries:
\begin{itemize}

\item (2,2;0,0,1), (2,2;1,0,0), (2,2;1,0,1) 
\item (4;0), (2,2;0,1,0), (2,2;0,1,1), (2,2;1,1,0) 
\item (5;2), (5;3) 
\item (5;1), (5;4) 
\item (6;1), (2,3;1,0,1) 
\item (6;5), (2,3;1,0,2) 
\item (6;3), (2,3;1,0,0) 
\item (6;4), (2,3;0,0,1) 
\item (6;2), (2,3;0,0,2) 
\item (6;0), (2,3;0,0,0) 
\item (7;3), (7;5), (7;6) 
\item (7;1), (7;2), (7;4) 
\item (2,4;0,0,1), (2,4;1,0,1) 
\item (2,4;0,0,3), (2,4;1,0,3) 
\item (2,4;1,1,1), (2,4;1,1,3) 
\item (8;0), (2,4;0,1,0), (2,4;0,1,2), (2,4;1,1,0), (2,4;1,1,2) 
\item (8;4), (2,4;0,1,1), (2,4;0,1,3) 
\item (3,3;1,0,1), (3,3;1,1,2), (3,3;1,2,2), (3,3;2,1,1), (3,3;2,2,1), (3,3;2,0,2) 
\item (3,3;0,0,1), (3,3;1,0,0), (3,3;1,1,1), (3,3;1,2,1) 
\item (9;1), (9;4), (9;7) 
\item (9;2), (9;5), (9;8) 
\item (3,3;0,0,2), (3,3;2,0,0), (3,3;2,1,2), (3,3;2,2,2) 
\item (9;0), (3,3;0,1,0), (3,3;0,2,0), (3,3;0,1,1), (3,3;0,2,1), (3,3;0,1,2), (3,3;0,2,2), (3,3;1,1,0), (3,3;1,2,0), (3,3;1,0,2), (3,3;2,1,0), (3,3;2,2,0), (3,3;2,0,1) 
\item (10;3), (10;7) 
\item (10;1), (10;9) 
\item (10;2), (10;8) 
\item (10;4), (10;6) 
\item (11;2), (11;6), (11;7), (11;8), (11;10) 
\item (11;1), (11;3), (11;4), (11;5), (11;9) 
\item (12;1), (3,4;1,0,1) 
\item (12;7), (3,4;1,0,3) 
\item (12;5), (3,4;2,0,1) 
\item (12;11), (3,4;2,0,3) 
\item (12;9), (3,4;0,0,1) 
\item (12;3), (3,4;0,0,3) 
\item (12;10), (3,4;1,0,2) 
\item (12;4), (3,4;1,0,0) 
\item (12;2), (3,4;2,0,2) 
\item (12;8), (3,4;2,0,0) 
\item (12;6), (3,4;0,0,2) 
\item (12;0), (3,4;0,0,0) 
\item (13;2), (13;5), (13;6), (13;7), (13;8), (13;11) 
\item (13;1), (13;3), (13;4), (13;9), (13;10), (13;12) 
\item (14;3), (14;5), (14;13) 
\item (14;1), (14;9), (14;11) 
\item (14;6), (14;10), (14;12) 
\item (14;2), (14;4), (14;8) 
\item (15;7), (15;13) 
\item (15;1), (15;4) 
\item (15;2), (15;8) 
\item (15;11), (15;14) 
\item (15;3), (15;12) 
\item (15;6), (15;9) 
\item (4,4;1,0,1), (4,4;1,2,2), (4,4;2,2,1) 
\item (4,4;1,0,2), (4,4;1,2,3), (4,4;2,0,1), (4,4;2,0,3), (4,4;3,2,1), (4,4;3,0,2) 
\item (4,4;2,2,3), (4,4;3,2,2), (4,4;3,0,3) 
\item (4,4;0,0,1), (4,4;1,0,0), (4,4;1,2,1) 
\item (4,4;0,2,1), (4,4;0,2,3), (4,4;1,2,0), (4,4;1,0,3), (4,4;3,2,0), (4,4;3,0,1) 
\item (4,4;0,0,3), (4,4;3,0,0), (4,4;3,2,3) 
\item (4,4;0,0,2), (4,4;2,0,0), (4,4;2,0,2) 
\item (4,4;0,2,0), (4,4;0,2,2), (4,4;2,2,0) 
\item (4,4;1,1,1), (4,4;1,3,1), (4,4;1,1,3), (4,4;1,3,3), (4,4;3,1,1), (4,4;3,3,1), (4,4;3,1,3), (4,4;3,3,3) 
\item (16;1), (16;9) 
\item (16;5), (16;13) 
\item (16;7), (16;15) 
\item (16;3), (16;11) 
\item (16;0), (4,4;0,1,0), (4,4;0,3,0), (4,4;0,1,1), (4,4;0,3,1), (4,4;0,1,2), (4,4;0,3,2), (4,4;0,1,3), (4,4;0,3,3), (4,4;1,1,0), (4,4;1,3,0), (4,4;1,1,2), (4,4;1,3,2), (4,4;2,1,0), (4,4;2,3,0), (4,4;2,1,1), (4,4;2,3,1), (4,4;2,1,2), (4,4;2,3,2), (4,4;2,1,3), (4,4;2,3,3), (4,4;3,1,0), (4,4;3,3,0), (4,4;3,1,2), (4,4;3,3,2)

\end{itemize}
We see that the two symmetries $(4;0)$ and $(2,2;0,1,0)$ are equivalent.  This
is the equivalence between \ml{\Z_2\times \Z_2} symmetry with the mixed anomaly
and \ml{\Z_4} symmetry in 1d discussed in Section \ref{Z2Z2aZ4}, where we also
find a duality transformation, that maps a lattice model with anomalous
\ml{\Z_2\times \Z_2} symmetry to another lattice model with $\Z_4$ symmetry.
We believe that, in general, for any pair of equivalent symmetries, there is a
lattice duality transformation, that maps a lattice model with one symmetry to
another lattice model with the other equivalent symmetry.  Each pair of the
equivalent symmetries in the above list implies a lattice duality map.

We also see
that $(4;0)$ and $(2,2;0,1,1)$ are equivalent.  Thus the  \ml{\Z_4} symmetry is
also equivalent to \ml{\Z_2\times \Z_2} symmetry with the mixed anomaly and an
anomaly in one of the $\Z_2$ symmetry.  More generally, it appears that
\ml{\Z_n\times \Z_n} symmetry with a particular mixed anomaly is equivalent to
\ml{\Z_{n^2}} symmetry.
It is also interesting to note that, for $\Z_p$ group ($2 < p$ = prime),
its $p-1$ anomalous symmetries form just two equivalent classes, and
its anomaly-free symmetry form its own equivalent class.

\section{Summary -- The essence of a symmetry}

\begin{figure}[t]
\begin{center}
\includegraphics[scale=0.15]{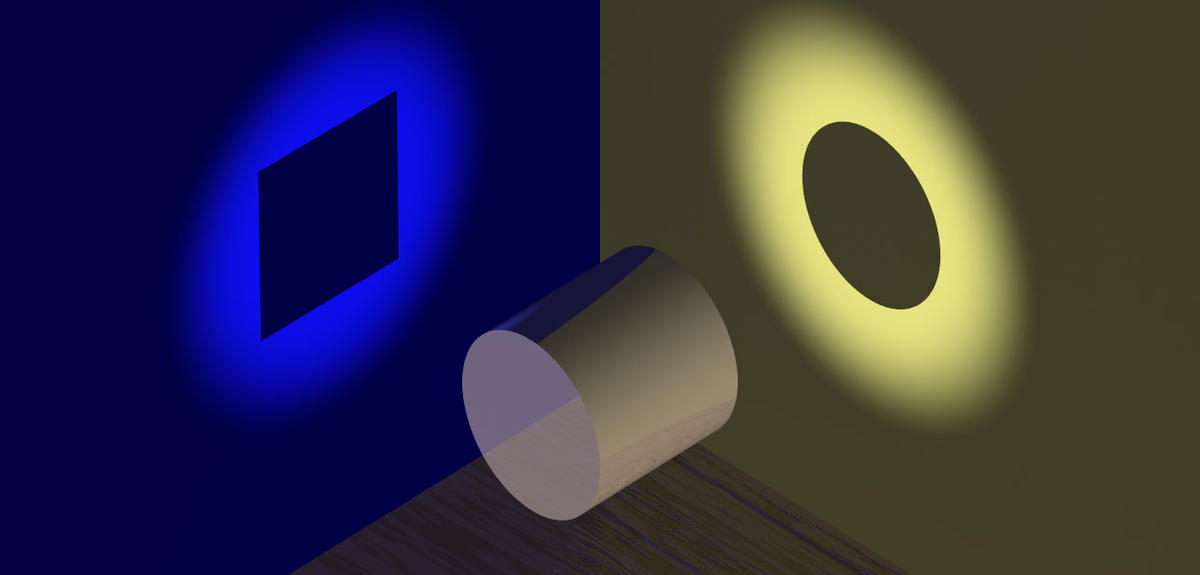}
\end{center}
\caption{The same topological order (in one higher dimension) can have
different shadows, which correspond to equivalent symmetries, and gives rise to
the notion of \catsymm.  }
\label{dualview} 
\end{figure}

With so many equivalences between symmetries labeled by (higher) groups and
anomalies, it is clear that group, higher group, anomalies, local fusion higher
categories, \etc\ are not the best notions to describe a symmetry. The algebra of local symmetric
operators provides a more fundamental description of symmetries and (invertible and
non-invertible) anomalies of quantum many body systems. In this paper, we show that this algebra contains a special subset of extended operators, dubbed t-patch operators, whose algebraic relations encode the data of a non-degenerate braided fusion $n$-category. This category
happens to capture the universal data of a topological order in one higher dimension. So, this point of view
leads to a holographic theory of symmetries and anomalies.

With this, we re-iterate our slogan: "Finite symmetry (with or without anomaly) is the shadow of topological order in one
higher dimension" (see Fig. \ref{dualview}). The topological order in one
higher dimension -- the \catsymm\  --
captures the essence of the symmetry.
We end the paper by listing different aspects of
\catsymm:\\
A \catsymm\ is
\begin{itemize}

\item
a symmetry plus its dual symmetry \cite{JW191213492, KZ200514178}.

\item
a non-invertible gravitational anomaly \cite{KW1458, FV14095723, M14107442, KZ150201690, KZ170200673, JW190513279}.

\item
a class of isomorphic algebras of local symmetric operators.

\item a non-degenerate braided fusion higher category.

\item a topological order in one higher dimension
\cite{FT180600008, TW191202817, JW191213492, KZ200514178}.

\end{itemize}

~

We would like to thank Yu-An Chen, Michael DeMarco, Wenjie Ji, Kyle Kawagoe, Sal Pace, and Carolyn Zhang for useful discussions.  This
work is partially supported by NSF DMR-2022428 and by the Simons Collaboration
on Ultra-Quantum Matter, which is a grant from the Simons Foundation (651446,
XGW).

\appendix
\allowdisplaybreaks

\section{Local symmetric operator algebra and non-degenerate braided fusion 3-category --
a 3-dimensional example without symmetry}
\label{example3d}

Let us discuss an example to illustrate the definitions in Section
\ref{bosonsys}, for the case without any symmetry.  We assume the space to be
3-dimensional. On each vertex-$i$, we have two degrees of freedom labeled by
elements in $\Z_2 \equiv \{+1,-1\}$, \ie the local Hilbert space $\cV_i$ on a
vertex 2-dimensional.  The algebra  of local operators is then generated by
$X_i, Z_i$ acting on $\cV_i$: 
\begin{align} 
\cA=\{X_i, Z_i, X_iZ_i, X_iX_j, Z_iZ_j,\cdots\}
\end{align} 
where $i,j$ are near each other, and the Pauli-$X, Z$ operators are
defined by 
\begin{align} 
X |\pm1\>=|\mp1\>,\ \ \  Z |\pm1\>=\pm|\pm1\>.
\end{align}

Our local symmetric operator algebra (after the closure by the extended operators) is
generated by the following \emph{t-patch operators}:
\begin{enumerate}
\item
0-dimensional t-patch operators, $X_i, Z_i$.

\item
1-dimensional t-patch operators -- string operators, 
\begin{align}
 X_{\mathrm{str}_{ij}} = X_iX_j,\ \ \
 Z_{\mathrm{str}_{ij}} = Z_iZ_j,
\end{align}
where the string$_{ij}$ connects the vertex-$i$ and vertex-$j$.  The string
operators must have an empty bulk to commute with the 0-dimensional t-patch
operators, when they are far away from the ends of the strings.

\item
2-dimensional t-patch operators -- disk operators, 
\begin{align}
 X_{\mathrm{disk}} &= \prod_{i\in \prt \mathrm{disk} }X_i, &
 Z_{\mathrm{disk}} &= \prod_{i\in \prt \mathrm{disk} }Z_i,
\nonumber\\
 O_{\mathrm{disk}} &= \prod_{i\in \prt \mathrm{disk} }O_i,
\end{align}
where $O_i$ can be any local operators.

\item
3-dimensional t-patch operators -- ball operators, 
\begin{align}
 X_{\mathrm{ball}} &= \prod_{i\in \prt \mathrm{ball} }X_i, \ \ \ \ \ \
 Z_{\mathrm{ball}} = \prod_{i\in \prt \mathrm{ball} }Z_i,
\nonumber\\
 O_{\mathrm{ball}} &= 
\sum_{\{m_i\} } \Psi(\{m_i\})
\prod_{i\in \prt \mathrm{ball} }O_i(m_i).
\end{align}
where $O_i(m_i)$ can be any local operators.  For example, $O_i(0) = \id$ and
$O_i(1) = X_i$.  (More precisely, $O_{\mathrm{ball}}$ is a tensor network
operator on the boundary of the ball, $\prt$ball.)
\end{enumerate}
We see that the t-patch operators all have empty bulk, \ie are patch charge
operators.  There is no patch symmetry operators.  This implies that
our bosonic system has no symmetry.  

If some t-patch operators have non-trivial bulk, then our system will have
non-trivial symmetry, as we see in the examples in Section IV and beyond of the
main text.  In fact, the non-trivial bulk of the t-patch operators will
generate the corresponding symmetries, higher symmetries, and/or non-invertible
higher symmetries.  

We believe that the above algebra of extended t-patch operators is closely related of a
braided fusion 3-category $3\eVec$.  At moment, we can only give a very rough
description of this connection.  A 3-category is formed by 0-morphisms (also
called objects), 1-morphisms, 2-morphisms, and 3-morphisms (also called top
morphisms).  All those morphisms have relations between them.  In fact, the
collection of all relations between $n$-morphisms is the collection of all
$(n+1)$-morphisms.  The ball operators correspond to the objects, the disk
operators the 1-morphisms, the string operators the 2-morphisms, and the local
operators the top 3-morphisms.  The difference of two ball operators are given
by the disk operators, the difference of two disk operators are given by the
string operators, etc.

For example, if two string operators $O_{\text{str}_{ij}}$ and
$O'_{\text{str}_{ij}}$ are related by local operators $O_i$ and $O_j$:
\begin{align}
 O'_{\text{str}_{ij}} = O_i O_j O_{\text{str}_{ij}},
\end{align}
we say the 2-morphism $O_{\text{str}_{ij}}$ connects to
the 2-morphism $O'_{\text{str}_{ij}}$
 via the 3-morphism $ O_i O_j$ on the left:
\begin{align}
 O_{\text{str}_{ij}} \xrightarrow{L: O_i O_j}  O'_{\text{str}_{ij}}.
\end{align}
Similarly, if
 $O_{\text{str}_{ij}}$ and $O'_{\text{str}_{ij}}$ are
related by local operators $O_i$ and $O_j$ on the right:
\begin{align}
 O'_{\text{str}_{ij}} = O_{\text{str}_{ij}} O_i O_j ,
\end{align}
we also say the 2-morphism $O_{\text{str}_{ij}}$ connects to
the 2-morphism $O'_{\text{str}_{ij}}$
 via the 3-morphism $ O_i O_j$:
\begin{align}
 O_{\text{str}_{ij}} \xrightarrow{R: O_i O_j}  O'_{\text{str}_{ij}}.
\end{align}

The 3-morphisms connecting 2-morphisms allow us to defined the notion of
\textbf{simple} 2-morphisms.  A  2-morphism $O_{\text{str}_{ij}}$ is simple if
an existence of 3-morphism $O_{\text{str}_{ij}} \xrightarrow{f}
O'_{\text{str}_{ij}}$ always implies an existence of 3-morphism
$O'_{\text{str}_{ij}} \xrightarrow{g} O_{\text{str}_{ij}}$ in the opposite
direction.  It turns out that $X_{\text{str}_{ij}}$ and $Z_{\text{str}_{ij}}$
introduced above are not simple.  The following string operators are simple
\begin{align}
P_{\text{str}_{ij}}^\pm = P_i^\pm P_j^\pm, \ \ \ 
P_i^\pm=\frac{1\pm Z_i}{2},\ \ \
P_j^\pm=\frac{1\pm Z_j}{2}.
\end{align}
Certainly, the notion of simpleness applies to all morphisms.

If two 2-morphisms, $O_{\text{str}_{ij}}$ and $ O'_{\text{str}_{ij}}$, satisfy
\begin{align}
 O_{\text{str}_{ij}} &\xrightarrow{f}  O'_{\text{str}_{ij}},
&
 O'_{\text{str}_{ij}} &\xrightarrow{g}  O_{\text{str}_{ij}},
\nonumber\\
 O_{\text{str}_{ij}} &\xrightarrow{f\circ g=\id}  O_{\text{str}_{ij}},
&
 O'_{\text{str}_{ij}} &\xrightarrow{g\circ f=\id}  O'_{\text{str}_{ij}},
\end{align}
then we say the two 2-morphisms are isomorphic.  In the above example, $
O_{\text{str}_{ij}} \xrightarrow{L: O_i O_j}  O'_{\text{str}_{ij}} $, if $O_i$
and $O_j$ are invertible, then the 2-morphism $O'_{\text{str}_{ij}}$ connects
to the 2-morphism $O_{\text{str}_{ij}}$ via the 3-morphism $ O_i^{-1}
O_j^{-1}$:
\begin{align}
 O_{\text{str}_{ij}} &= O_i^{-1} O_j^{-1} O'_{\text{str}_{ij}},
\nonumber\\
\text{or }\ \ \ \
O'_{\text{str}_{ij}} &\xrightarrow{L: O_i^{-1} O_j^{-1}}  O_{\text{str}_{ij}}
\end{align}
In this case, the two 2-morphisms $O_{\text{str}_{ij}}$ and
$O'_{\text{str}_{ij}}$ are isomorphic.

The isomorphic relations between 2-morphisms is an equivalent relation.  For
example $P_{\text{str}_{ij}}^- \cong P_{\text{str}_{ij}}^+$.  Although there
are infinite many simple 2-morphisms in our example, there is only one
equivalence class of simple 2-morphisms. A representative in this equivalence
class is given by $ P_{\text{str}_{ij}}^- = P_i^- P_j^- $.  

In this paper, when we refer to objects and morphisms, we usually refer to the
equivalence classes of objects and morphisms, under the isomorphisms discussed
above.  Combining the definition of simpleness and isomorphism, we see that two
simple morphisms cannot be connected by a higher morphism if they are not
isomorphic.  In other words, different types of morphisms (\ie  different
equivalence classes of morphisms) cannot be connected by a higher morphism.

We would like to stress that although the t-patch operator considered above all have
an empty bulk, the tensor network operator on the boundary can have a
Wess-Zumino form.  For example, $O_{\mathrm{ball}}$ is a tensor network
operator on the boundary of the ball, but it can be defined \ie defined by a
tensor network on an extension of $\prt$ball in one higher dimension. Such a
tensor network can be viewed as a spacetime path integral on the ball, which
can give rise to a topologically ordered state on $\prt$ball described by wave
function $\Psi(\{m_i\})$.  We see that we can have infinitely many types of ball
operators, each type corresponds to a topological order in 2-dimensional space.
Since there is no non-trivial topological order in 0- and 1-dimensional space,
thus we have only one type of string-like t-patch operators and one type of
membrane-like t-patch operators.  Such a structure matches the structure of
braided fusion 3-category $3\eVec$.\cite{GJ190509566, KZ200514178}

\section{Detailed calculations}

\subsection{Calculation of $ F(s,s,s) $}\label{Fsymbdeets}
To compute the F-symbol $ F(s,s,s) $, described in \eqn{Fsss}, we refer to Fig. \ref{Fsym-KL} and substitute $ a=b=c=s $. Using the definitions in equations \ref{anomZ2patch} and \ref{Ts}, this picture translates to the following calculation: 
\begin{equation*}
\scalebox{0.85}{\begin{tikzpicture}
	\begin{pgfonlayer}{nodelayer}
		\node [style=none] (1) at (13.25, 4.75) {};
		\node [style=none] (2) at (-7, 4.75) {};
		\node [style=none] (3) at (-6, 5.25) {};
		\node [style=none] (4) at (-6, 4.25) {};
		\node [style=none] (7) at (-4, 5.25) {};
		\node [style=none] (8) at (-2, 5.25) {};
		\node [style=none] (9) at (0, 5.25) {};
		\node [style=none] (10) at (0, 4.25) {};
		\node [style=none] (11) at (-2, 4.25) {};
		\node [style=none] (12) at (-4, 4.25) {};
		\node [style=none] (13) at (2, 5.25) {};
		\node [style=none] (14) at (4, 5.25) {};
		\node [style=none] (15) at (6, 5.25) {};
		\node [style=none] (17) at (2, 4.25) {};
		\node [style=none] (18) at (4, 4.25) {};
		\node [style=none] (19) at (6, 4.25) {};
		\node [style=none] (30) at (-2, 1.75) {$X$};
		\node [style=none] (31) at (0, 1.75) {$ZX$};
		\node [style=none] (37) at (-1.75, 0.75) {};
		\node [style=none] (38) at (-0.25, 0.75) {};
		\node [style=none] (39) at (0.25, 0.75) {};
		\node [style=none] (40) at (1.75, 0.75) {};
		\node [style=none] (43) at (-2.25, 0.75) {};
		\node [style=none] (44) at (-3.75, 0.75) {};
		\node [style=cyan rect] (45) at (-3, 0.75) {$s$};
		\node [style=cyan rect] (47) at (-1, 0.75) {$s$};
		\node [style=cyan rect] (48) at (1, 0.75) {$s$};
		\node [style=none] (50) at (-4, 1.75) {$X$};
		\node [style=none] (51) at (-6, 1.75) {$O^\dagger$};
		\node [style=none] (52) at (0, -0.25) {$O$};
		\node [style=none] (54) at (4, 2.75) {$X$};
		\node [style=none] (55) at (6, 2.75) {$ZX$};
		\node [style=none] (58) at (4.25, 1.75) {};
		\node [style=none] (59) at (5.75, 1.75) {};
		\node [style=none] (60) at (6.25, 1.75) {};
		\node [style=none] (61) at (7.75, 1.75) {};
		\node [style=none] (62) at (3.75, 1.75) {};
		\node [style=none] (63) at (2.25, 1.75) {};
		\node [style=cyan rect] (64) at (3, 1.75) {$s$};
		\node [style=cyan rect] (66) at (5, 1.75) {$s$};
		\node [style=cyan rect] (67) at (7, 1.75) {$s$};
		\node [style=none] (68) at (2, 2.75) {$X$};
		\node [style=none] (69) at (0, 2.75) {$O^\dagger$};
		\node [style=none] (70) at (6, 0.75) {$O$};
		\node [style=none] (71) at (8, 5.25) {};
		\node [style=none] (72) at (10, 5.25) {};
		\node [style=none] (73) at (12, 5.25) {};
		\node [style=none] (74) at (8, 4.25) {};
		\node [style=none] (75) at (10, 4.25) {};
		\node [style=none] (76) at (12, 4.25) {};
		\node [style=none] (77) at (4, -2.25) {$X$};
		\node [style=none] (78) at (6, -2.25) {$XZ$};
		\node [style=none] (79) at (4.25, -1.25) {};
		\node [style=none] (80) at (5.75, -1.25) {};
		\node [style=none] (81) at (6.25, -1.25) {};
		\node [style=none] (82) at (7.75, -1.25) {};
		\node [style=none] (83) at (3.75, -1.25) {};
		\node [style=none] (84) at (2.25, -1.25) {};
		\node [style=cyan rect] (85) at (3, -1.25) {$s$};
		\node [style=cyan rect] (86) at (5, -1.25) {$s$};
		\node [style=cyan rect] (87) at (7, -1.25) {$s$};
		\node [style=none] (88) at (2, -2.25) {$X$};
		\node [style=none] (89) at (6, -0.25) {$O^\dagger$};
		\node [style=none] (90) at (0, -2.25) {$O$};
		\node [style=none] (91) at (4, -4.25) {$X$};
		\node [style=none] (92) at (10, -4.25) {$XZ$};
		\node [style=none] (93) at (8.25, -3.25) {};
		\node [style=none] (94) at (9.75, -3.25) {};
		\node [style=none] (95) at (10.25, -3.25) {};
		\node [style=none] (96) at (11.75, -3.25) {};
		\node [style=none] (97) at (3.75, -3.25) {};
		\node [style=none] (98) at (2.25, -3.25) {};
		\node [style=cyan rect] (99) at (3, -3.25) {$s$};
		\node [style=cyan rect] (100) at (9, -3.25) {$s$};
		\node [style=cyan rect] (101) at (11, -3.25) {$s$};
		\node [style=none] (102) at (2, -4.25) {$X$};
		\node [style=none] (103) at (10, -2.25) {$O^\dagger$};
		\node [style=none] (104) at (0, -4.25) {$O$};
		\node [style=none] (105) at (7.75, -3.25) {};
		\node [style=none] (106) at (6.25, -3.25) {};
		\node [style=cyan rect] (107) at (7, -3.25) {$s$};
		\node [style=none] (108) at (8, -4.25) {$X$};
		\node [style=none] (109) at (5.75, -3.25) {};
		\node [style=none] (110) at (4.25, -3.25) {};
		\node [style=cyan rect] (111) at (5, -3.25) {$s$};
		\node [style=none] (112) at (6, -4.25) {$X$};
		\node [style=none] (113) at (3, -5.5) {=};
		\node [style=none] (114) at (-2, -10.25) {$X$};
		\node [style=none] (115) at (0, -10.25) {$ZX$};
		\node [style=none] (116) at (-1.75, -11.25) {};
		\node [style=none] (117) at (-0.25, -11.25) {};
		\node [style=none] (118) at (0.25, -11.25) {};
		\node [style=none] (119) at (1.75, -11.25) {};
		\node [style=none] (120) at (-2.25, -11.25) {};
		\node [style=none] (121) at (-3.75, -11.25) {};
		\node [style=cyan rect] (122) at (-3, -11.25) {$s$};
		\node [style=cyan rect] (123) at (-1, -11.25) {$s$};
		\node [style=cyan rect] (124) at (1, -11.25) {$s$};
		\node [style=none] (125) at (-4, -10.25) {$X$};
		\node [style=none] (126) at (-6, -10.25) {$O^\dagger$};
		\node [style=none] (127) at (0, -12.25) {$O$};
		\node [style=none] (156) at (4, -8.25) {$X$};
		\node [style=none] (157) at (10, -8.25) {$XZ$};
		\node [style=none] (158) at (8.25, -7.25) {};
		\node [style=none] (159) at (9.75, -7.25) {};
		\node [style=none] (160) at (10.25, -7.25) {};
		\node [style=none] (161) at (11.75, -7.25) {};
		\node [style=none] (162) at (3.75, -7.25) {};
		\node [style=none] (163) at (2.25, -7.25) {};
		\node [style=cyan rect] (164) at (3, -7.25) {$s$};
		\node [style=cyan rect] (165) at (9, -7.25) {$s$};
		\node [style=cyan rect] (166) at (11, -7.25) {$s$};
		\node [style=none] (167) at (2, -8.25) {$X$};
		\node [style=none] (168) at (0, -8.25) {$O$};
		\node [style=none] (169) at (7.75, -7.25) {};
		\node [style=none] (170) at (6.25, -7.25) {};
		\node [style=cyan rect] (171) at (7, -7.25) {$s$};
		\node [style=none] (172) at (8, -8.25) {$X$};
		\node [style=none] (173) at (5.75, -7.25) {};
		\node [style=none] (174) at (4.25, -7.25) {};
		\node [style=cyan rect] (175) at (5, -7.25) {$s$};
		\node [style=none] (176) at (6, -8.25) {$X$};
		\node [style=none] (177) at (10, -6.25) {$O^\dagger$};
		\node [style=none] (178) at (-6, -7.25) {$(-1)\cdot$};
		\node [style=none] (179) at (-6, 6.25) {0};
		\node [style=none] (180) at (0, 6.25) {1};
		\node [style=none] (181) at (6, 6.25) {2};
		\node [style=none] (182) at (10, 6.25) {3};
	\end{pgfonlayer}
	\begin{pgfonlayer}{edgelayer}
		\draw (2.center) to (1.center);
		\draw (3.center) to (4.center);
		\draw (7.center) to (12.center);
		\draw (8.center) to (11.center);
		\draw (9.center) to (10.center);
		\draw (13.center) to (17.center);
		\draw (14.center) to (18.center);
		\draw (15.center) to (19.center);
		\draw [style=thicc blue, in=180, out=0] (44.center) to (43.center);
		\draw [style=thicc blue] (37.center) to (38.center);
		\draw [style=thicc blue] (39.center) to (40.center);
		\draw [style=thicc blue, in=180, out=0] (63.center) to (62.center);
		\draw [style=thicc blue] (58.center) to (59.center);
		\draw [style=thicc blue] (60.center) to (61.center);
		\draw (71.center) to (74.center);
		\draw (72.center) to (75.center);
		\draw (73.center) to (76.center);
		\draw [style=thicc blue, in=180, out=0] (84.center) to (83.center);
		\draw [style=thicc blue] (79.center) to (80.center);
		\draw [style=thicc blue] (81.center) to (82.center);
		\draw [style=thicc blue, in=180, out=0] (98.center) to (97.center);
		\draw [style=thicc blue] (93.center) to (94.center);
		\draw [style=thicc blue] (95.center) to (96.center);
		\draw [style=thicc blue, in=180, out=0] (106.center) to (105.center);
		\draw [style=thicc blue, in=180, out=0] (110.center) to (109.center);
		\draw [style=thicc blue, in=180, out=0] (121.center) to (120.center);
		\draw [style=thicc blue] (116.center) to (117.center);
		\draw [style=thicc blue] (118.center) to (119.center);
		\draw [style=thicc blue, in=180, out=0] (163.center) to (162.center);
		\draw [style=thicc blue] (158.center) to (159.center);
		\draw [style=thicc blue] (160.center) to (161.center);
		\draw [style=thicc blue, in=180, out=0] (170.center) to (169.center);
		\draw [style=thicc blue, in=180, out=0] (174.center) to (173.center);
	\end{pgfonlayer}
\end{tikzpicture}}
\end{equation*}
This tells us that $ F(s,s,s)=-1 $. Note that our operator ordering convention is top-to-bottom and left-to-right (when in the same row).

\subsection{Self-statistics of $ s $ particles}\label{ssstat}
We express Fig \ref{exchange} in equations as
\begin{equation}\label{hopalgs}
\begin{split}
&\ \ \ \ T_s(0\rightarrow 1) T_s(1\rightarrow 2)
T_s(3\to 1) 
\\
&=W_{\mathrm{patch}_{01}} W_{\mathrm{patch}_{12}} W_{\mathrm{patch}_{13}}^\dag \\
&=\ee^{\ii \th_s} W_{\mathrm{patch}_{13}}^\dag  W_{\mathrm{patch}_{12}} W_{\mathrm{patch}_{01}}\\
&=\ee^{\ii \th_s} 
T_s(3\to 1) 
T_s(1\rightarrow 2) T_s(0\rightarrow 1)
\end{split}
\end{equation}
The l.h.s. can be simplified as
\begin{equation*}
\scalebox{0.85}{\input{s-self-statistics-1.tikz}}
\end{equation*}
while the r.h.s. can be simplified as
\begin{equation*}
\scalebox{0.85}{\begin{tikzpicture}
	\begin{pgfonlayer}{nodelayer}
		\node [style=none] (1) at (13, 0) {};
		\node [style=none] (2) at (-7, 0) {};
		\node [style=none] (3) at (-6, 0.5) {};
		\node [style=none] (4) at (-6, -0.5) {};
		\node [style=none] (7) at (-4, 0.5) {};
		\node [style=none] (8) at (-2, 0.5) {};
		\node [style=none] (9) at (0, 0.5) {};
		\node [style=none] (10) at (0, -0.5) {};
		\node [style=none] (11) at (-2, -0.5) {};
		\node [style=none] (12) at (-4, -0.5) {};
		\node [style=none] (13) at (2, 0.5) {};
		\node [style=none] (14) at (4, 0.5) {};
		\node [style=none] (15) at (6, 0.5) {};
		\node [style=none] (17) at (2, -0.5) {};
		\node [style=none] (18) at (4, -0.5) {};
		\node [style=none] (19) at (6, -0.5) {};
		\node [style=none] (60) at (-2, -6.75) {$X$};
		\node [style=none] (61) at (0, -6.75) {$ZX$};
		\node [style=none] (62) at (-1.75, -7.75) {};
		\node [style=none] (63) at (-0.25, -7.75) {};
		\node [style=none] (64) at (0.25, -7.75) {};
		\node [style=none] (65) at (1.75, -7.75) {};
		\node [style=none] (66) at (-2.25, -7.75) {};
		\node [style=none] (67) at (-3.75, -7.75) {};
		\node [style=cyan rect] (68) at (-3, -7.75) {$s$};
		\node [style=cyan rect] (69) at (-1, -7.75) {$s$};
		\node [style=cyan rect] (70) at (1, -7.75) {$s$};
		\node [style=none] (71) at (-4, -6.75) {$X$};
		\node [style=none] (72) at (-6, -6.75) {$O^\dagger$};
		\node [style=none] (73) at (0, -8.75) {$O$};
		\node [style=none] (74) at (8, 0.5) {};
		\node [style=none] (75) at (10, 0.5) {};
		\node [style=none] (76) at (12, 0.5) {};
		\node [style=none] (77) at (8, -0.5) {};
		\node [style=none] (78) at (10, -0.5) {};
		\node [style=none] (79) at (12, -0.5) {};
		\node [style=none] (80) at (4, -4.75) {$X$};
		\node [style=none] (81) at (6, -4.75) {$ZX$};
		\node [style=none] (82) at (4.25, -5.75) {};
		\node [style=none] (83) at (5.75, -5.75) {};
		\node [style=none] (84) at (6.25, -5.75) {};
		\node [style=none] (85) at (7.75, -5.75) {};
		\node [style=none] (86) at (3.75, -5.75) {};
		\node [style=none] (87) at (2.25, -5.75) {};
		\node [style=cyan rect] (88) at (3, -5.75) {$s$};
		\node [style=cyan rect] (89) at (5, -5.75) {$s$};
		\node [style=cyan rect] (90) at (7, -5.75) {$s$};
		\node [style=none] (91) at (2, -4.75) {$X$};
		\node [style=none] (92) at (0, -4.75) {$O^\dagger$};
		\node [style=none] (93) at (6, -6.75) {$O$};
		\node [style=none] (94) at (4, -3.75) {$X$};
		\node [style=none] (95) at (10, -3.75) {$XZ$};
		\node [style=none] (96) at (8.25, -2.75) {};
		\node [style=none] (97) at (9.75, -2.75) {};
		\node [style=none] (98) at (10.25, -2.75) {};
		\node [style=none] (99) at (11.75, -2.75) {};
		\node [style=none] (100) at (3.75, -2.75) {};
		\node [style=none] (101) at (2.25, -2.75) {};
		\node [style=cyan rect] (102) at (3, -2.75) {$s$};
		\node [style=cyan rect] (103) at (9, -2.75) {$s$};
		\node [style=cyan rect] (104) at (11, -2.75) {$s$};
		\node [style=none] (105) at (2, -3.75) {$X$};
		\node [style=none] (106) at (0, -3.75) {$O$};
		\node [style=none] (107) at (7.75, -2.75) {};
		\node [style=none] (108) at (6.25, -2.75) {};
		\node [style=cyan rect] (109) at (7, -2.75) {$s$};
		\node [style=none] (110) at (8, -3.75) {$X$};
		\node [style=none] (111) at (5.75, -2.75) {};
		\node [style=none] (112) at (4.25, -2.75) {};
		\node [style=cyan rect] (113) at (5, -2.75) {$s$};
		\node [style=none] (114) at (6, -3.75) {$X$};
		\node [style=none] (115) at (10, -1.75) {$O^\dagger$};
		\node [style=none] (116) at (-6, 1.5) {0};
		\node [style=none] (117) at (0, 1.5) {1};
		\node [style=none] (118) at (6, 1.5) {2};
		\node [style=none] (119) at (10, 1.5) {3};
		\node [style=none] (120) at (3.25, -10.25) {=};
		\node [style=none] (273) at (4, -13.75) {$X$};
		\node [style=none] (274) at (10, -13.75) {$XZ$};
		\node [style=none] (275) at (8.25, -12.75) {};
		\node [style=none] (276) at (9.75, -12.75) {};
		\node [style=none] (277) at (10.25, -12.75) {};
		\node [style=none] (278) at (11.75, -12.75) {};
		\node [style=none] (279) at (3.75, -12.75) {};
		\node [style=none] (280) at (2.25, -12.75) {};
		\node [style=cyan rect] (281) at (3, -12.75) {$s$};
		\node [style=cyan rect] (282) at (9, -12.75) {$s$};
		\node [style=cyan rect] (283) at (11, -12.75) {$s$};
		\node [style=none] (284) at (2, -13.75) {$X$};
		\node [style=none] (285) at (0, -13.75) {$O$};
		\node [style=none] (286) at (7.75, -12.75) {};
		\node [style=none] (287) at (6.25, -12.75) {};
		\node [style=cyan rect] (288) at (7, -12.75) {$s$};
		\node [style=none] (289) at (8, -13.75) {$X$};
		\node [style=none] (290) at (5.75, -12.75) {};
		\node [style=none] (291) at (4.25, -12.75) {};
		\node [style=cyan rect] (292) at (5, -12.75) {$s$};
		\node [style=none] (293) at (6, -13.75) {$X$};
		\node [style=none] (294) at (10, -11.75) {$O^\dagger$};
		\node [style=none] (295) at (-2, -15) {$X$};
		\node [style=none] (296) at (6, -15) {$ZX$};
		\node [style=none] (297) at (-1.75, -16) {};
		\node [style=none] (298) at (-0.25, -16) {};
		\node [style=none] (299) at (6.25, -16) {};
		\node [style=none] (300) at (7.75, -16) {};
		\node [style=none] (301) at (-2.25, -16) {};
		\node [style=none] (302) at (-3.75, -16) {};
		\node [style=cyan rect] (303) at (-3, -16) {$s$};
		\node [style=cyan rect] (304) at (-1, -16) {$s$};
		\node [style=cyan rect] (305) at (7, -16) {$s$};
		\node [style=none] (306) at (-4, -15) {$X$};
		\node [style=none] (307) at (-6, -15) {$O^\dagger$};
		\node [style=none] (308) at (6, -17) {$O$};
		\node [style=none] (309) at (2, -15) {$X$};
		\node [style=none] (310) at (2.25, -16) {};
		\node [style=none] (311) at (3.75, -16) {};
		\node [style=none] (312) at (1.75, -16) {};
		\node [style=none] (313) at (0.25, -16) {};
		\node [style=cyan rect] (314) at (1, -16) {$s$};
		\node [style=cyan rect] (315) at (3, -16) {$s$};
		\node [style=none] (316) at (0, -15) {$X$};
		\node [style=none] (317) at (4, -15) {$X$};
		\node [style=none] (318) at (4.25, -16) {};
		\node [style=none] (319) at (5.75, -16) {};
		\node [style=cyan rect] (320) at (5, -16) {$s$};
		\node [style=none] (321) at (-6, -12.75) {$\ii\ \ \cdot$};
	\end{pgfonlayer}
	\begin{pgfonlayer}{edgelayer}
		\draw (2.center) to (1.center);
		\draw (3.center) to (4.center);
		\draw (7.center) to (12.center);
		\draw (8.center) to (11.center);
		\draw (9.center) to (10.center);
		\draw (13.center) to (17.center);
		\draw (14.center) to (18.center);
		\draw (15.center) to (19.center);
		\draw [style=thicc blue, in=180, out=0] (67.center) to (66.center);
		\draw [style=thicc blue] (62.center) to (63.center);
		\draw [style=thicc blue] (64.center) to (65.center);
		\draw (74.center) to (77.center);
		\draw (75.center) to (78.center);
		\draw (76.center) to (79.center);
		\draw [style=thicc blue, in=180, out=0] (87.center) to (86.center);
		\draw [style=thicc blue] (82.center) to (83.center);
		\draw [style=thicc blue] (84.center) to (85.center);
		\draw [style=thicc blue, in=180, out=0] (101.center) to (100.center);
		\draw [style=thicc blue] (96.center) to (97.center);
		\draw [style=thicc blue] (98.center) to (99.center);
		\draw [style=thicc blue, in=180, out=0] (108.center) to (107.center);
		\draw [style=thicc blue, in=180, out=0] (112.center) to (111.center);
		\draw [style=thicc blue, in=180, out=0] (280.center) to (279.center);
		\draw [style=thicc blue] (275.center) to (276.center);
		\draw [style=thicc blue] (277.center) to (278.center);
		\draw [style=thicc blue, in=180, out=0] (287.center) to (286.center);
		\draw [style=thicc blue, in=180, out=0] (291.center) to (290.center);
		\draw [style=thicc blue, in=180, out=0] (302.center) to (301.center);
		\draw [style=thicc blue] (297.center) to (298.center);
		\draw [style=thicc blue] (299.center) to (300.center);
		\draw [style=thicc blue, in=180, out=0] (313.center) to (312.center);
		\draw [style=thicc blue] (310.center) to (311.center);
		\draw [style=thicc blue] (318.center) to (319.center);
	\end{pgfonlayer}
\end{tikzpicture}}
\end{equation*}
Comparing the two, we can see that the self-statistics phase $ \ee^{\ii\theta_s} $ equals $ \ii $, \ie $ \th_s=\pi/2 $. Thus, the $ s $ particles have semionic self-statistics.

\subsection{Mutual and self statistics of $ m $, $ \t m $, $ s $ particles in $ \Z_2 \times \Z_2$ with mixed anomaly}\label{ZtZdeets}
First we calculate the mutual statistics of $ m $ and $ \t m $, as discussed in \eqn{mutualmm}. Representing it pictorially, we find
\begin{equation*}
\scalebox{.85}{\begin{tikzpicture}
	\begin{pgfonlayer}{nodelayer}
		\node [style=none] (1) at (7, 0) {};
		\node [style=none] (2) at (-11, 0) {};
		\node [style=none] (3) at (-10, 0.5) {};
		\node [style=none] (4) at (-10, -0.5) {};
		\node [style=none] (7) at (-8, 0.5) {};
		\node [style=none] (8) at (-6, 0.5) {};
		\node [style=none] (9) at (-4, 0.5) {};
		\node [style=none] (10) at (-4, -0.5) {};
		\node [style=none] (11) at (-6, -0.5) {};
		\node [style=none] (12) at (-8, -0.5) {};
		\node [style=none] (13) at (-2, 0.5) {};
		\node [style=none] (14) at (0, 0.5) {};
		\node [style=none] (15) at (2, 0.5) {};
		\node [style=none] (17) at (-2, -0.5) {};
		\node [style=none] (18) at (0, -0.5) {};
		\node [style=none] (19) at (2, -0.5) {};
		\node [style=none] (29) at (-6, -2.75) {$X$};
		\node [style=none] (30) at (-4, -2.75) {$X$};
		\node [style=none] (31) at (-2, -2.75) {$X$};
		\node [style=none] (32) at (0, -2.75) {$\t O^\dagger$};
		\node [style=none] (50) at (-8, -2.75) {$X$};
		\node [style=none] (60) at (4, 0.5) {};
		\node [style=none] (61) at (4, -0.5) {};
		\node [style=none] (62) at (6, 0.5) {};
		\node [style=none] (65) at (6, -0.5) {};
		\node [style=none] (68) at (-8, 1.5) {0};
		\node [style=none] (69) at (-8, -1.75) {$\t O$};
		\node [style=none] (70) at (-4, 1.5) {1};
		\node [style=none] (71) at (0, 1.5) {2};
		\node [style=none] (76) at (-1.75, -5) {};
		\node [style=none] (77) at (-0.25, -5) {};
		\node [style=none] (78) at (0.25, -5) {};
		\node [style=none] (79) at (1.75, -5) {};
		\node [style=none] (80) at (2.25, -5) {};
		\node [style=none] (81) at (3.75, -5) {};
		\node [style=none] (82) at (-2.25, -5) {};
		\node [style=none] (83) at (-3.75, -5) {};
		\node [style=cyan rect] (84) at (-3, -5) {$s$};
		\node [style=cyan rect] (85) at (-1, -5) {$s$};
		\node [style=cyan rect] (86) at (1, -5) {$s$};
		\node [style=cyan rect] (87) at (3, -5) {$s$};
		\node [style=none] (88) at (-2, -4) {$\t X$};
		\node [style=none] (92) at (4, 1.5) {3};
		\node [style=none] (93) at (-2, -6.5) {=};
		\node [style=none] (94) at (-1.75, -21.5) {};
		\node [style=none] (95) at (-0.25, -21.5) {};
		\node [style=none] (96) at (0.25, -21.5) {};
		\node [style=none] (97) at (1.75, -21.5) {};
		\node [style=none] (98) at (2.25, -21.5) {};
		\node [style=none] (99) at (3.75, -21.5) {};
		\node [style=none] (100) at (-2.25, -21.5) {};
		\node [style=none] (101) at (-3.75, -21.5) {};
		\node [style=cyan rect] (102) at (-3, -21.5) {$s$};
		\node [style=cyan rect] (103) at (-1, -21.5) {$s$};
		\node [style=cyan rect] (104) at (1, -21.5) {$s$};
		\node [style=cyan rect] (105) at (3, -21.5) {$s$};
		\node [style=none] (110) at (-6, -22.5) {$X$};
		\node [style=none] (111) at (-4, -22.5) {$X$};
		\node [style=none] (112) at (-2, -22.5) {$X$};
		\node [style=none] (113) at (0, -22.5) {$\t O^\dagger$};
		\node [style=none] (114) at (-8, -22.5) {$X$};
		\node [style=none] (115) at (-8, -21.5) {$\t O$};
		\node [style=none] (116) at (-10, -20.5) {$\ii$};
		\node [style=none] (117) at (-6, -10) {$X$};
		\node [style=none] (118) at (-4, -10) {$X$};
		\node [style=none] (119) at (-2, -10) {$X$};
		\node [style=none] (120) at (0, -10) {$\t O^\dagger$};
		\node [style=none] (121) at (-8, -10) {$X$};
		\node [style=none] (122) at (-8, -9) {$\t O$};
		\node [style=none] (123) at (-1.75, -11.25) {};
		\node [style=none] (124) at (-0.25, -11.25) {};
		\node [style=none] (125) at (0.25, -11.25) {};
		\node [style=none] (126) at (1.75, -11.25) {};
		\node [style=none] (127) at (2.25, -11.25) {};
		\node [style=none] (128) at (3.75, -11.25) {};
		\node [style=none] (129) at (-2.25, -11.25) {};
		\node [style=none] (130) at (-3.75, -11.25) {};
		\node [style=cyan rect] (131) at (-3, -11.25) {$s$};
		\node [style=cyan rect] (132) at (-1, -11.25) {$s$};
		\node [style=cyan rect] (133) at (1, -11.25) {$s$};
		\node [style=cyan rect] (134) at (3, -11.25) {$s$};
		\node [style=none] (139) at (0, -9) {$-\ii \t Z$};
		\node [style=none] (140) at (-2, -12.5) {=};
		\node [style=none] (141) at (-6, -18) {$X$};
		\node [style=none] (142) at (-4, -18) {$X$};
		\node [style=none] (143) at (-2, -18) {$X$};
		\node [style=none] (144) at (0, -18) {$\t O^\dagger$};
		\node [style=none] (145) at (-8, -18) {$X$};
		\node [style=none] (146) at (-8, -17) {$\t O$};
		\node [style=none] (147) at (-1.75, -16) {};
		\node [style=none] (148) at (-0.25, -16) {};
		\node [style=none] (149) at (0.25, -16) {};
		\node [style=none] (150) at (1.75, -16) {};
		\node [style=none] (151) at (2.25, -16) {};
		\node [style=none] (152) at (3.75, -16) {};
		\node [style=none] (153) at (-2.25, -16) {};
		\node [style=none] (154) at (-3.75, -16) {};
		\node [style=cyan rect] (155) at (-3, -16) {$s$};
		\node [style=cyan rect] (156) at (-1, -16) {$s$};
		\node [style=cyan rect] (157) at (1, -16) {$s$};
		\node [style=cyan rect] (158) at (3, -16) {$s$};
		\node [style=none] (163) at (0, -14.75) {$-\ii \t Z$};
		\node [style=none] (164) at (-2, -19.25) {=};
		\node [style=none] (165) at (0, -17) {$-\t Z$};
		\node [style=none] (166) at (0, -4) {$\t X$};
		\node [style=none] (167) at (2, -4) {$\t X$};
		\node [style=none] (168) at (4, -4) {$\t X$};
		\node [style=none] (169) at (-2, -8) {$\t X$};
		\node [style=none] (170) at (0, -8) {$\t X$};
		\node [style=none] (171) at (2, -8) {$\t X$};
		\node [style=none] (172) at (4, -8) {$\t X$};
		\node [style=none] (173) at (-2, -13.75) {$\t X$};
		\node [style=none] (174) at (0, -13.75) {$\t X$};
		\node [style=none] (175) at (2, -13.75) {$\t X$};
		\node [style=none] (176) at (4, -13.75) {$\t X$};
		\node [style=none] (177) at (-2, -20.5) {$\t X$};
		\node [style=none] (178) at (0, -20.5) {$\t X$};
		\node [style=none] (179) at (2, -20.5) {$\t X$};
		\node [style=none] (180) at (4, -20.5) {$\t X$};
	\end{pgfonlayer}
	\begin{pgfonlayer}{edgelayer}
		\draw (2.center) to (1.center);
		\draw (3.center) to (4.center);
		\draw (7.center) to (12.center);
		\draw (8.center) to (11.center);
		\draw (9.center) to (10.center);
		\draw (13.center) to (17.center);
		\draw (14.center) to (18.center);
		\draw (15.center) to (19.center);
		\draw (60.center) to (61.center);
		\draw (62.center) to (65.center);
		\draw [style=thicc blue, in=180, out=0] (83.center) to (82.center);
		\draw [style=thicc blue] (76.center) to (77.center);
		\draw [style=thicc blue] (78.center) to (79.center);
		\draw [style=thicc blue] (80.center) to (81.center);
		\draw [style=thicc blue, in=180, out=0] (101.center) to (100.center);
		\draw [style=thicc blue] (94.center) to (95.center);
		\draw [style=thicc blue] (96.center) to (97.center);
		\draw [style=thicc blue] (98.center) to (99.center);
		\draw [style=thicc blue, in=180, out=0] (130.center) to (129.center);
		\draw [style=thicc blue] (123.center) to (124.center);
		\draw [style=thicc blue] (125.center) to (126.center);
		\draw [style=thicc blue] (127.center) to (128.center);
		\draw [style=thicc blue, in=180, out=0] (154.center) to (153.center);
		\draw [style=thicc blue] (147.center) to (148.center);
		\draw [style=thicc blue] (149.center) to (150.center);
		\draw [style=thicc blue] (151.center) to (152.center);
	\end{pgfonlayer}
\end{tikzpicture}}
\end{equation*}
This proves \eqn{mutualmm}. Now, recall that $ s $ is a bound state of $ m$ and $ \t m $. In other words, 
\begin{equation}\label{Wsdef}
W^s_{\text{patch}_{ij}} \stackrel{\text{def}}{=} W_{\text{patch}_{ij}}\cdot \t W_{\text{patch}_{ij}} 
\end{equation}
Then the self-statistics calculation shown in Fig \ref{exchange} corresponds to the computation of the phase in the following sequence of operations:
\begin{equation}\label{ZtZ-s-self}
\scalebox{.85}{\begin{tikzpicture}
	\begin{pgfonlayer}{nodelayer}
		\node [style=none] (1) at (7, 6) {};
		\node [style=none] (2) at (-7, 6) {};
		\node [style=none] (3) at (-6, 6.5) {};
		\node [style=none] (4) at (-6, 5.5) {};
		\node [style=none] (7) at (-4, 6.5) {};
		\node [style=none] (8) at (-2, 6.5) {};
		\node [style=none] (9) at (0, 6.5) {};
		\node [style=none] (10) at (0, 5.5) {};
		\node [style=none] (11) at (-2, 5.5) {};
		\node [style=none] (12) at (-4, 5.5) {};
		\node [style=none] (13) at (2, 6.5) {};
		\node [style=none] (14) at (4, 6.5) {};
		\node [style=none] (15) at (6, 6.5) {};
		\node [style=none] (17) at (2, 5.5) {};
		\node [style=none] (18) at (4, 5.5) {};
		\node [style=none] (19) at (6, 5.5) {};
		\node [style=none] (21) at (-6, 4.5) {$m \tilde m$};
		\node [style=none] (22) at (6, 4.5) {$m\tilde m$};
		\node [style=none] (24) at (-6, 7.5) {0};
		\node [style=none] (25) at (-2, 7.5) {1};
		\node [style=none] (26) at (2, 7.5) {2};
		\node [style=none] (27) at (6, 7.5) {3};
		\node [style=none] (30) at (-6, 1.75) {};
		\node [style=none] (31) at (-2.75, 0.5) {};
		\node [style=none] (32) at (-6, 4) {};
		\node [style=none] (33) at (-2, 0.25) {$m\tilde m$};
		\node [style=none] (34) at (5.25, 4.5) {};
		\node [style=none] (35) at (-2, 3.25) {};
		\node [style=none] (36) at (2, 1.75) {};
		\node [style=none] (37) at (2, 0.75) {};
		\node [style=none] (38) at (2, 0.25) {$m\tilde m$};
		\node [style=none] (39) at (1.5, -4.75) {};
		\node [style=none] (40) at (1.5, -0.25) {};
		\node [style=none] (41) at (2.5, -4.75) {};
		\node [style=none] (42) at (2.5, -0.25) {};
		\node [style=none] (43) at (4.5, -3.5) {};
		\node [style=none] (44) at (-1.5, -0.75) {};
		\node [style=none] (45) at (5.75, -3.5) {};
		\node [style=none] (46) at (-0.75, -0.5) {};
		\node [style=none] (47) at (5.5, -4) {$m\tilde m$};
		\node [style=none] (48) at (2, -5.25) {$m\tilde m$};
		\node [style=none] (49) at (1, -5.5) {};
		\node [style=none] (50) at (-5, -6.75) {};
		\node [style=none] (51) at (-6, -7) {$m \tilde m$};
		\node [style=red hazy circle] (52) at (1.5, -1.5) {};
		\node [style=red hazy circle] (53) at (2.5, -2.5) {};
		\node [style=none] (54) at (5.75, -6.5) {};
		\node [style=none] (55) at (5.75, -4.25) {};
		\node [style=none] (56) at (5.75, -7) {$m \tilde m$};
	\end{pgfonlayer}
	\begin{pgfonlayer}{edgelayer}
		\draw (2.center) to (1.center);
		\draw (3.center) to (4.center);
		\draw (7.center) to (12.center);
		\draw (8.center) to (11.center);
		\draw (9.center) to (10.center);
		\draw (13.center) to (17.center);
		\draw (14.center) to (18.center);
		\draw (15.center) to (19.center);
		\draw [style=arrow-to-dest] (30.center) to (31.center);
		\draw [style=arrow-to-dest] (32.center) to (30.center);
		\draw [style=arrow-to-dest] (34.center) to (35.center);
		\draw [style=arrow-to-dest] (35.center) to (36.center);
		\draw [style=arrow-to-dest] (36.center) to (37.center);
		\draw [style=arrow-to-dest] (40.center) to (39.center);
		\draw [style=arrow-to-dest] (42.center) to (41.center);
		\draw [style=arrow-to-dest] (44.center) to (43.center);
		\draw [style=arrow-to-dest] (46.center) to (45.center);
		\draw [style=arrow-to-dest] (49.center) to (50.center);
		\draw [style=arrow-to-dest] (55.center) to (54.center);
	\end{pgfonlayer}
\end{tikzpicture}}
\end{equation}
From the above picture, it is clear that the computation of self-statistics of $ s $ particles is equivalent to the computation of mutual statistics of $ m $ and $ \t m $ particles. 

\section{Global action of 1+1D $ \Z_2\times \Z_2 $ symmetry with mixed anomaly}\label{z2z2global}

Symmetry protected topological (SPT) states in $ d $ space dimensions are
associated with anomalous symmetry actions on their $ (d-1) $-dimensional
boundary. Such non-onsite action of the symmetry encodes a 't Hooft anomaly of
the symmetry, when considered exclusively on the boundary. In \Rf{WW170506728}
(Section 4), the authors wrote down an exactly soluble path integral model
(also known as cocycle model\cite{W161201418}) to realize SPT states in general
$ d $ space dimensions. These were then used to construct the corresponding
anomalous symmetry action for the boundary effective theory. This framework
then provides us with a recipe to write down a representative symmetry action
for any anomalous symmetry in any number of dimensions. In particular, we can
use this recipe to write down the anomalous (non-onsite) symmetry action for
the 1+1D bosonic theory having a $ \Z_2 \times \Z_2  $ symmetry with a mixed
anomaly. For this we must consider an SPT state in 2+1D that is protected by $
\Z_2\times \Z_2 $ symmetry.\footnote{We use the additive presentation of the $ \Z_2 $ group in this appendix.} The path integral is defined on a 3-manifold $ M^3
$ with boundary $ M^2=\partial M^3 $, and involves a 3-cocycle $ \nu_3 $. In
Euclidean signature, the integrand of the path integral reads
\begin{equation}\label{pathint}
\ee^{-\int_{M^3} \cL_{\text{Bulk}}d^3x}=\prod_{M^3} \ee^{2\pi \ii \nu_3(g_i,g_j,g_k,g_l)}
\end{equation}
where the ordered collections $ (i,j,k,l) $ are the tetrahedra belonging to the triangulation of $ M^3 $. For the effective boundary theory, one can simplify the bulk so that it contains a single point. This reduces to an effectively $ 1+1D $ path integral due to properties of the cocycle which we will not go into here -- the interested reader is directed to section 4.2 of \Rf{WW170506728}. This path integral still has the original protecting symmetry of the SPT state, however it is no longer realized in an on-site manner. In Hamiltonian formalism, this symmetry action on the 1+1D boundary is given by
\begin{equation}\label{Ug}
U(g) \ket{\{g_i\}}= \prod_{(i,j)} \ee^{2\pi \ii \nu_3(g_i,g_j,g^*,-g+ g^*)} \ket{\{g + g_i\}} \
\end{equation}
where $ (i,j) $ are nearest neighbors on the 1d spatial boundary, $ -g $ denotes the inverse of the group element $ g \in \Z_2\times \Z_2$, and $ g^* $ is an arbitrary reference group element, which can be taken to be the identity element of the symmetry group without any loss of generality. One choice of $ \nu_3 $ that encodes the mixed anomaly of two $ \Z_2 $ symmetries is
\begin{equation}\label{nu3def}
\nu_3 = a_1 \smile a_2 \smile a_2
\end{equation}
with $ a=dg $ taking values on links, and the subscripts on $ a $ labeling the two $ \Z_2 $ groups. Using equations \ref{Ug} and \ref{nu3def} allows us to write down the global symmetry generators in equations \ref{z2z2W1} and \ref{z2z2W2}.

\section{2\texorpdfstring{\MakeLowercase{d}}{d} non-Abelian symmetry and its dual} 

In the section \ref{2dZ2}, we see that a 2d $\Z_2$ 0-symmetry is equivalent to
its dual, a 2d $\Z_2^{(1)}$ 1-symmetry. The dual symmetry is obtained by
exchanging patch charge operators and patch symmetry operators.  We note that
for a symmetry described by a non-Abelian finite group $G$, it also has patch
charge operators and patch symmetry operators.  Naturally, one may ask what is
the dual of the $G$ symmetry? Are symmetry and dual symmetry equivalent?  In
this section, we will discuss briefly the algebra of local symmetric operators
for a non-Abelian symmetry, and the dual of a non-Abelian symmetry.  Although
our discussion is far from complete, it suggests that the dual of the $G$
0-symmetry (whose charge-objects form a fusion 2-category $2\cRep_G$) is a
symmetry whose charge-objects form a fusion 2-category $2\cVec_G$.  In other
words, the symmetries $2\cRep_G$ and  $2\cVec_G$ are dual to each other.

\subsection{The $G$ 0-symmetry in 2d space} 

Let us consider a bosonic quantum system, whose degrees of freedoms live on the
vertices and are labeled by a non-Abelian group $G$.  In other words, the total
Hilbert space is given by \ml{\cV = \bigotimes_{i} \cV_{i} } (\ml{\cV_{i} =
\mathrm{span}\{ |g_{i}\>\ \big|\ g_{i} \in G\}} ).

The $G$ 0-symmetry is defined by the transformations on the whole 2d space
\begin{align}
T_h = \prod_{i} T_i(h) ,\ \ \ h\in G,
\end{align}
where $T_i(h)$ acts on  $\cV_{i}$:
\begin{align}
 T_i(h) |g_i\> =|hg_i\>.
\end{align}
The associated t-patch symmetry operator is given by
\begin{align}
 \hat \chi_\text{disk} = \sum_{h\in \chi}  \hat h_\text{disk},
\end{align}
where $ \hat h_\text{disk} = \prod_{i\in \text{disk}} T_i(h)$ and
$\chi$ is a conjugacy class of $G$.  Note that here we need to sum over
conjugacy class as required by the transparency condition (\ie
$T_\text{disk}(\chi)$ must carry vanishing total charge):
\begin{align}
 T_\text{disk}(\chi)T_\text{disk}(\chi')  =
  T_\text{disk'}(\chi') T_\text{disk}(\chi),
\end{align}
where the boundaries, $\prt$disk and $\prt$disk', are far away (\ie do not
intersect).  
Local symmetric operators satisfy
\begin{align}
\hat \chi_\text{disk} O_i^\mathrm{symm} 
= O_i^\mathrm{symm}\hat \chi_\text{disk} , \ \ \
\forall\ \chi,
\end{align}
where $i$ is far away from $\prt \text{disk}$.

The patch charge operators, with empty bulk,  are given by
\begin{align}
 \hat R_{\text{str}_{ij}} = \Tr[R(\hat g_i) R(\hat g_j^{-1})],\ \ \ \ 
\hat g_i |g_i\> = g_i |g_i\>,
\end{align}
where $R$ is an irreducible matrix representation of $G$.
The transparency condition requires us to take the trace: 
\begin{align}
 \hat \chi_\text{disk} \hat R_{\text{str}_{ij}}=
  \hat R_{\text{str}_{ij}} \hat \chi_\text{disk},
\end{align}
where $\text{str}_{ij}$ is far away from $\prt \text{disk}$.  But the one end
of string operator carries a non-zero charge, which can be seen by trying to
calculate the commutation between $\hat \chi_\text{disk}$ and $\hat
R_{\text{str}_{ij}}$ with one end of string, $i$ inside the disk and the other
end of string, $j$ outside the disk: 
\begin{align}
\label{chiR}
&\ \ \ \
\hat \chi_\mathrm{disk} \hat R_{\mathrm{str}_{ij}}
\\
&= 
\Big(\sum_{h\in \chi} \prod_{i \in \mathrm{disk} } T_i(h) \Big)
\Tr \big( R(\hat g_{i}) R(\hat g_{j}^{-1}) \big) 
\nonumber\\
&=
\Big( 
\sum_{h\in \chi} 
\Tr \big( R(h) R(\hat g_{i})  R(\hat g_{j}^{-1}) \big) 
\prod_{i \in \mathrm{disk} } T_i(h) \Big)
\nonumber 
\end{align}
We see that commutator is complicated. In fact, they do not even form a proper
commutator.  The non-trivial relation indicates that the ends of string carries
non-trivial charge.  But for a non-Abelian group $G$, the charge is not
described by a simple phase factor.  


We know that the algebra generated by the patch charge operators $ \hat
R_{\text{str}_{ij}}$ and patch symmetry operators $\hat\chi_\text{disk}$ should
correspond to a 3d topological order.  For the present case, such a 3d
topological order should be the one described by the $G$-gauge theory.  The 1d
boundary of the disk operator $\hat\chi_\text{disk}$ corresponds to the flux
loop in the $G$-gauge theory.  When $G$ is non-Abelian, a single flux loop in
$G$-gauge theory is not labeled by a group element in $G$, but rather by a
conjugacy class $\chi$.  
For $n$ flux loops with gauge flux described by $h_1,\cdots,h_n$, the distinct
physical states that labeled the conjugacy class $[h_1,\cdots,h_n] =\{
hh_1h^{-1},\cdots,hh_nh^{-1} | h \in G\}$ For large $n$, the number of distinct
physical states is of order $|G|^n$.  In this case, we may say the  gauge flux
is labeled by the group elements of $G$.

Similarly, if we consider a more general patch symmetry operators formed by $n$
disks, it is given by
\begin{align}
 \hat \chi_{n\text{-disk}} = 
\hskip -4mm
\sum_{h_1,\cdots,h_n\in [h_1,\cdots,h_n] }  
\hskip -4mm
(\hat h_1)_{\text{disk}_1}
\cdots
(\hat h_n)_{\text{disk}_n}
.
\end{align}
We see that the number of generalized patch symmetry operators is of order
$|G|^n$.  We may say $n$ disk-like patch symmetry operators are labeled by the
elements in $G^n$, and each disk-like patch symmetry operators are labeled by
the elements in $G$.  This agrees with the picture from the gauge flux.

The  patch charge operator $ \hat R_{\text{str}_{ij}}$ corresponds to the
charge excitations in the 3d $G$-gauge theory on $S^0$, \ie on two points with
one carries charge $R$ and the other charge $\bar R$. Here $R$ is a
representation of $G$ and $\bar R$ is its charge conjugate.  The fusion of the
charges is given by the fusion of $G$-representations
\begin{align}
\label{Rfuse}
R_1\otimes R_2 &= \bigoplus_{R_3} N_{R_1, R_2}^{R_3} R_3.
\end{align}

To measure the charge in the $G$-gauge theory, let us braid a charge $R$ around
a single flux $\chi$.  When $G$ is non-Abelian, both the charge $R$ and the
flux $\chi$ can be degenerate.  The degeneracy of the charge $R$ is dim$(R)$.
The degeneracy of the flux $\chi$ is is the number of group elements in
conjugacy class $\chi$, $|\chi|$.  With those degeneracies, the braiding of a
charge around a flux loop is not simply a phase factor.  This is why the
commutation \eqn{chiR} is complicated.  

The above correspondence suggests that the \catsymm\ of a 2d
$G$ 0-symmetry is a 3d topological order described by a $G$-gauge theory, which
will be denoted as $2\eGau_{G}$.  $2\eGau_{G}$ can also be viewed as a
non-degenerate braided fusion 2-category describing the point-like excitations
(the $G$-gauge charge) and string-like excitations (the $G$-gauge flux) in the
3d $G$-gauge theory.  Thus the \catsymm\ of a 2d $G$ 0-symmetry is a
non-degenerate braided fusion 2-category $2\eGau_{G}$.

We would like to mention that the patch charge operators $\hat R_\text{str}$
should generate a symmetric fusion 2-category $2\eRep_{G}$.  The patch symmetry
operators $\hat \chi_\text{disk}$ should generate a braided fusion 2-category
$2\eVec_{G}$.  

We would like to remark that the simple objects in $2\eVec_{G}$ are labeled by
the elements $g$ of the group $G$.  The boundary of a single  patch symmetry
operator $\hat \chi_\text{disk}$ correspond to a composite object $\chi =
\bigoplus_{g\in \chi} g$, where $\chi$ is a conjugacy class of $G$.  On the
other hand, the boundary of $n$  patch symmetry operators $\hat
\chi_{n\text{-disk}}$, in the large $n$ limit, correspond to the simple objects
in  $2\eVec_{G}$.  Since both 0-symmetry $G$ and the algebraic 1-symmetry
$G^{(1)}_\text{rep}$ have the same \catsymm, they are equivalent symmetries.
The class of quantum systems with 0-symmetry $G$ and the class of quantum
systems with algebraic 1-symmetry $G^{(1)}_\text{rep}$ will have a 1-to-1
correspondence, so that the corresponding quantum systems have identical local
low energy properties.

\subsection{The $G^{(1)}_\mathrm{rep}$ 1-symmetry in 2d space} 

The $\Z_2^{(1)}$ 1-symmetry discussed before is described by a higher group.
In this section, we are going to study a symmetry that is beyond higher group
since the symmetry transformation is not invertible.  Such a symmetry is called
algebraic higher symmetry in \Rf{KZ200514178}.

Let us consider a bosonic quantum system, whose degrees of freedoms live on the
links and are labeled by by an non-Abelian group $G$.  In other words, the
total Hilbert space is given by \ml{\cV = \bigotimes_{\<ij\>} \cV_{ij} }
(\ml{\cV_{ij} = \mathrm{span}\{ |g_{ij}\>\ \big|\ g_{ij} \in G\}} ) 

The symmetry is defined by the transformations on all the
loops $S^1$:
\begin{align}
W_R(S^1) = \Tr \prod_{\<ij\> \in S^1} R(\hat g_{ij}) ,\ \ \
\hat g_{ij} |g_{ij}\>=g_{ij} |g_{ij}\>,
\end{align}
for all matrix representation $R$ of \ml{G}.  Local symmetric operators
satisfy
\begin{align}
W_R(S^1) O_i^\mathrm{symm} = O_i^\mathrm{symm}W_R(S^1) , \ \ \
\forall\ \ S^1, R .
\end{align}
We will call such a symmetry as \ml{G^{(1)}_\mathrm{rep}} 1-symmetry.

The algebra of local symmetric operators is generated by the following two
kinds of operators:
\begin{align}
\hat R_{\mathrm{str}_{ij}} &= \Tr \big( R(\hat g_{ik}) R(\hat g_{kl}) \cdots R(\hat g_{mj})
\big) ,
\nonumber\\
 \hat \chi_\text{disk} &= \sum_{h\in \chi}  \hat h_\text{disk},
\end{align}
where $\chi$ is a conjugacy class of $G$,  $ \hat h_\text{disk} = \prod_{i\in
\text{disk}} T_i(h)$, and the $T_i(h)$ operator (for $h\in G$) is defined as
\begin{align} 
T_i(h) |\cdots,g_{ki}, g_{ij} \cdots\> = |\cdots,g_{ki}h^{-1},
hg_{ij} \cdots\>. 
\end{align}
One can check that the above patch operators are t-patch operators, satisfying
the transparency condition \eqn{pcom}.  The trace in the definition of $\hat
R_{\mathrm{str}_{ij}}$ and the sum over  conjugacy class in the definition of
$\hat \chi_{\mathrm{disk}}$ are important to ensure the transparency property.

The symmetry transformations  $W_R(S^1) = \Tr \prod_{\<ij\> \in S^1} R(g_{ij})$
are not invertible. They form a more general algebra
\begin{align}
W_{R_1}(S^1) W_{R_2}(S^1)
&= \sum_{R_3} N_{R_1, R_2}^{R_3} W_{R_3}(S^1),
\end{align}
where $N_{R_1, R_2}^{R_3}$ is the fusion coefficients of the irreducible
representations, $R_1, R_2, R_3$, of $G$ (see \eqn{Rfuse}).  Thus the symmetry generated by
$W_{R}(S^1)$'s is a new kind of symmetry.

We would like to remark that non-invertible symmetry also exist in 1-dimensional
space, which can be constructed in a very similar way.  In 1d, the
non-invertible symmetry is still described by the transformation \ml{W_R(S^1) =
\Tr \prod_{\<ij\> \in S^1} R(g_{ij}) }, which correspond to an non-invertible
0-symmetry denoted as $G_\mathrm{rep}$. Those 1d beyond-group symmetries have
been studied under the name (1) topological
defect-lines/twisted-boundary-conditions in 1+1D {(spacetime dimension)} CFT
\cite{PZh0011021, CSh0107001, FSh0204148, CY180204445}; (2) fusion category
symmetry,\cite{TW191202817, I210315588}; (3) quantum group
symmetry,\cite{Q200509072}; \etc.

Now let us go back to 2-dimensional space.  We can use the t-patch operators
\ml{\hat R_\mathrm{str}} on open strings to define the 1-symmetry, \ie to
select the local symmetric operators:
\begin{align}
\hat R_\mathrm{str} O_i^\mathrm{symm} =
O_i^\mathrm{symm}
\hat R_\mathrm{str} ,\ \ 
i \text{ far away from string ends} 
\end{align}

The  patch charge operator $\hat \chi_\mathrm{disk}$ carry vanishing 1-charge
since
\begin{align}
\hat R_\mathrm{str}  
\hat \chi_\mathrm{disk}
=
\hat \chi_\mathrm{disk}
\hat R_\mathrm{str} 
\end{align}
if the disk of $\hat \chi_\mathrm{disk}$ is far away from the string  ends of
$\hat R_\mathrm{str}$.  However, a segment of the boundary of the disk operator
$\hat \chi_\mathrm{disk}$ can carry a non zero 1-charge. 
To measure such a
1-charge, we try to compute the commutator
\begin{align}
\label{chiR1}
&\ \ \ \
\hat \chi_\mathrm{disk} \hat R_{\mathrm{str}_{ij}}
\\
&= 
\Big(\sum_{h\in \chi} \prod_{i \in \mathrm{disk} } T_i(h) \Big)
\Tr \big( R(g_{ik}) R(g_{kl}) \cdots R(g_{mj}) \big) 
\nonumber\\
&=
\Big( 
\sum_{h\in \chi} 
\Tr \big( R(h) R(g_{ik}) R(g_{kl}) \cdots R(g_{mj}) \big) 
\prod_{i \in \mathrm{disk} } T_i(h) \Big)
\nonumber 
\end{align}
assuming one end of string, $i$, is inside the disk and the other end of
string, $j$, is outside the disk.  We see that commutator is complicated.  The
non-trivial relation at least indicates that the boundary of the disk carries
non-trivial 1-charge.  But for a non-Abelian group $G$, the 1-charge is not
described by a simple phase factor.  This, in fact, is an expected result.

The above discussion suggests the algebra of local symmetric operator for 2d
algebraic 1-symmetry $G^{(1)}_\text{rep}$ is isomorphic to the algebra of local
symmetric operator from 2d 0-symmetry $G$.  To see this more clearly, we remove
the trace and the sum over conjugacy class
in equations \eqref{chiR} and \eqref{chiR1}, and rewrite them as
\begin{align}
\label{chiRa}
&\ \ \ \
\hat h_\mathrm{disk} \hat R_{\mathrm{str}_{ij}}^{\al\bt}
\nonumber\\
&= 
\Big( \prod_{i \in \mathrm{disk} } T_i(h) \Big)
\big( R(g_{i}) R(g_{j}^{-1}) \big)^{\al\bt} 
\nonumber\\
&=
\Big( 
\sum_{\ga} 
R(h)^{\al\ga} \big(  R(g_{i})  R(g_{j}^{-1}) \big)^{\ga\bt} 
\prod_{i \in \mathrm{disk} } T_i(h) \Big)
\nonumber \\
&= \sum_{ \ga} 
R(h)^{\al\ga} 
\hat R_{\mathrm{str}_{ij}}^{\ga\bt}
\hat h_\mathrm{disk} 
\end{align}
and
\begin{align}
\label{chiR1a}
&\ \ \ \
\hat h_\mathrm{disk} \hat R_{\mathrm{str}_{ij}}^{\al\bt}
\nonumber \\
&= 
\Big( \prod_{i \in \mathrm{disk} } T_i(h) \Big)
\big( R(g_{ik}) R(g_{kl}) \cdots R(g_{mj}) \big)^{\al\bt} 
\nonumber\\
&=
\Big( 
\sum_{ \ga} 
R(h)^{\al\ga} \big(  R(g_{ik}) R(g_{kl}) \cdots R(g_{mj}) \big)^{\ga\bt} 
\prod_{i \in \mathrm{disk} } T_i(h) \Big)
\nonumber \\
&= \sum_{ \ga} 
R(h)^{\al\ga} 
\hat R_{\mathrm{str}_{ij}}^{\ga\bt}
\hat h_\mathrm{disk} 
\end{align}
The above two equations have the same form, suggesting that the two operator
algebras are isomorphic.  In this case, the 2d algebraic 1-symmetry
$G^{(1)}_\text{rep}$ also has the \catsymm\ $2\eGau_G$, the
3d $G$-gauge theory.  The only difference is that, for 2d algebraic 1-symmetry
$G^{(1)}_\text{rep}$, the patch symmetry operators generate a symmetric fusion
2-category $2\eRep_G$, while the patch charge operators generate a braided
fusion 2-category $2\eVec_G$.  So compare to 2d 0-symmetry $G$, the patch
symmetry operators and the  patch charge operators are switched.

\bibliography{./LoaBfc.bib,../../bib/all,../../bib/allnew,../../bib/publst,../../bib/publstnew,../../bib/ACrefs}

\begin{thebibliography}{100}%
\makeatletter
\providecommand \@ifxundefined [1]{%
 \@ifx{#1\undefined}
}%
\providecommand \@ifnum [1]{%
 \ifnum #1\expandafter \@firstoftwo
 \else \expandafter \@secondoftwo
 \fi
}%
\providecommand \@ifx [1]{%
 \ifx #1\expandafter \@firstoftwo
 \else \expandafter \@secondoftwo
 \fi
}%
\providecommand \natexlab [1]{#1}%
\providecommand \enquote  [1]{``#1''}%
\providecommand \bibnamefont  [1]{#1}%
\providecommand \bibfnamefont [1]{#1}%
\providecommand \citenamefont [1]{#1}%
\providecommand \href@noop [0]{\@secondoftwo}%
\providecommand \href [0]{\begingroup \@sanitize@url \@href}%
\providecommand \@href[1]{\@@startlink{#1}\@@href}%
\providecommand \@@href[1]{\endgroup#1\@@endlink}%
\providecommand \@sanitize@url [0]{\catcode `\\12\catcode `\$12\catcode
  `\&12\catcode `\#12\catcode `\^12\catcode `\_12\catcode `\%12\relax}%
\providecommand \@@startlink[1]{}%
\providecommand \@@endlink[0]{}%
\providecommand \url  [0]{\begingroup\@sanitize@url \@url }%
\providecommand \@url [1]{\endgroup\@href {#1}{\urlprefix }}%
\providecommand \urlprefix  [0]{URL }%
\providecommand \Eprint [0]{\href }%
\providecommand \doibase [0]{https://doi.org/}%
\providecommand \selectlanguage [0]{\@gobble}%
\providecommand \bibinfo  [0]{\@secondoftwo}%
\providecommand \bibfield  [0]{\@secondoftwo}%
\providecommand \translation [1]{[#1]}%
\providecommand \BibitemOpen [0]{}%
\providecommand \bibitemStop [0]{}%
\providecommand \bibitemNoStop [0]{.\EOS\space}%
\providecommand \EOS [0]{\spacefactor3000\relax}%
\providecommand \BibitemShut  [1]{\csname bibitem#1\endcsname}%
\let\auto@bib@innerbib\@empty
\bibitem [{\citenamefont {Nussinov}\ and\ \citenamefont
  {Ortiz}(2009{\natexlab{a}})}]{NOc0605316}%
  \BibitemOpen
  \bibfield  {author} {\bibinfo {author} {\bibfnamefont {Z.}~\bibnamefont
  {Nussinov}}\ and\ \bibinfo {author} {\bibfnamefont {G.}~\bibnamefont
  {Ortiz}},\ }\bibfield  {title} {\bibinfo {title} {Sufficient symmetry
  conditions for topological quantum order},\ }\href
  {https://doi.org/10.1073/pnas.0803726105} {\bibfield  {journal} {\bibinfo
  {journal} {Proceedings of the National Academy of Sciences}\ }\textbf
  {\bibinfo {volume} {106}},\ \bibinfo {pages} {16944} (\bibinfo {year}
  {2009}{\natexlab{a}})},\ \Eprint {https://arxiv.org/abs/cond-mat/0605316}
  {arXiv:cond-mat/0605316} \BibitemShut {NoStop}%
\bibitem [{\citenamefont {Nussinov}\ and\ \citenamefont
  {Ortiz}(2009{\natexlab{b}})}]{NOc0702377}%
  \BibitemOpen
  \bibfield  {author} {\bibinfo {author} {\bibfnamefont {Z.}~\bibnamefont
  {Nussinov}}\ and\ \bibinfo {author} {\bibfnamefont {G.}~\bibnamefont
  {Ortiz}},\ }\bibfield  {title} {\bibinfo {title} {A symmetry principle for
  topological quantum order},\ }\href
  {https://doi.org/10.1016/j.aop.2008.11.002} {\bibfield  {journal} {\bibinfo
  {journal} {Ann. Phys.}\ }\textbf {\bibinfo {volume} {324}},\ \bibinfo {pages}
  {977} (\bibinfo {year} {2009}{\natexlab{b}})},\ \Eprint
  {https://arxiv.org/abs/cond-mat/0702377} {arXiv:cond-mat/0702377}
  \BibitemShut {NoStop}%
\bibitem [{\citenamefont {{Kapustin}}\ and\ \citenamefont
  {{Thorngren}}(2017)}]{KT13094721}%
  \BibitemOpen
  \bibfield  {author} {\bibinfo {author} {\bibfnamefont {A.}~\bibnamefont
  {{Kapustin}}}\ and\ \bibinfo {author} {\bibfnamefont {R.}~\bibnamefont
  {{Thorngren}}},\ }\bibfield  {title} {\bibinfo {title} {Higher symmetry and
  gapped phases of gauge theories},\ }in\ \href
  {https://doi.org/10.1007/978-3-319-59939-7_5} {\emph {\bibinfo {booktitle}
  {Algebra, Geometry, and Physics in the 21st Century. Progress in Mathematics.
  Progress in Mathematics}}},\ Vol.\ \bibinfo {volume} {324},\ \bibinfo
  {editor} {edited by\ \bibinfo {editor} {\bibfnamefont {D.}~\bibnamefont
  {Auroux}}, \bibinfo {editor} {\bibfnamefont {L.}~\bibnamefont {Katzarkov}},
  \bibinfo {editor} {\bibfnamefont {T.}~\bibnamefont {Pantev}}, \bibinfo
  {editor} {\bibfnamefont {Y.}~\bibnamefont {Soibelman}},\ and\ \bibinfo
  {editor} {\bibfnamefont {Y.}~\bibnamefont {Tschinkel}}}\ (\bibinfo
  {publisher} {Birkh\"auser, Cham.},\ \bibinfo {year} {2017})\ pp.\ \bibinfo
  {pages} {177--202},\ \Eprint {https://arxiv.org/abs/1309.4721}
  {arXiv:1309.4721} \BibitemShut {NoStop}%
\bibitem [{\citenamefont {Gaiotto}\ \emph {et~al.}(2015)\citenamefont
  {Gaiotto}, \citenamefont {Kapustin}, \citenamefont {Seiberg},\ and\
  \citenamefont {Willett}}]{GW14125148}%
  \BibitemOpen
  \bibfield  {author} {\bibinfo {author} {\bibfnamefont {D.}~\bibnamefont
  {Gaiotto}}, \bibinfo {author} {\bibfnamefont {A.}~\bibnamefont {Kapustin}},
  \bibinfo {author} {\bibfnamefont {N.}~\bibnamefont {Seiberg}},\ and\ \bibinfo
  {author} {\bibfnamefont {B.}~\bibnamefont {Willett}},\ }\bibfield  {title}
  {\bibinfo {title} {Generalized global symmetries},\ }\href
  {https://doi.org/10.1007/jhep02(2015)172} {\bibfield  {journal} {\bibinfo
  {journal} {J. High Energ. Phys.}\ }\textbf {\bibinfo {volume}
  {2015}}\bibfield  {number} {\bibinfo  {number} { (2)},\ \bibinfo {pages}
  {172}},\ }\Eprint {https://arxiv.org/abs/1412.5148} {arXiv:1412.5148}
  \BibitemShut {NoStop}%
\bibitem [{\citenamefont {{Alvarez-Gaum{\'e}}}\ and\ \citenamefont
  {{Witten}}(1984)}]{AW8469}%
  \BibitemOpen
  \bibfield  {author} {\bibinfo {author} {\bibfnamefont {L.}~\bibnamefont
  {{Alvarez-Gaum{\'e}}}}\ and\ \bibinfo {author} {\bibfnamefont
  {E.}~\bibnamefont {{Witten}}},\ }\bibfield  {title} {\bibinfo {title}
  {{Gravitational anomalies}},\ }\href
  {https://doi.org/10.1016/0550-3213(84)90066-X} {\bibfield  {journal}
  {\bibinfo  {journal} {Nuclear Physics B}\ }\textbf {\bibinfo {volume}
  {234}},\ \bibinfo {pages} {269} (\bibinfo {year} {1984})}\BibitemShut
  {NoStop}%
\bibitem [{\citenamefont {Witten}(1985)}]{W8597}%
  \BibitemOpen
  \bibfield  {author} {\bibinfo {author} {\bibfnamefont {E.}~\bibnamefont
  {Witten}},\ }\bibfield  {title} {\bibinfo {title} {Global gravitational
  anomalies},\ }\href {https://doi.org/10.1007/bf01212448} {\bibfield
  {journal} {\bibinfo  {journal} {Commun.Math. Phys.}\ }\textbf {\bibinfo
  {volume} {100}},\ \bibinfo {pages} {197} (\bibinfo {year}
  {1985})}\BibitemShut {NoStop}%
\bibitem [{\citenamefont {'t~Hooft}(1980)}]{H8035}%
  \BibitemOpen
  \bibfield  {author} {\bibinfo {author} {\bibfnamefont {G.}~\bibnamefont
  {'t~Hooft}},\ }\bibfield  {title} {\bibinfo {title} {Naturalness, chiral
  symmetry, and spontaneous chiral symmetry breaking},\ }in\ \href
  {https://doi.org/10.1007/978-1-4684-7571-5_9} {\emph {\bibinfo {booktitle}
  {Recent Developments in Gauge Theories. NATO Advanced Study Institutes Series
  (Series B. Physics)}}},\ Vol.~\bibinfo {volume} {59},\ \bibinfo {editor}
  {edited by\ \bibinfo {editor} {\bibfnamefont {G.}~\bibnamefont {'t~Hooft~et
  al.}}}\ (\bibinfo  {publisher} {Springer, Boston, MA.},\ \bibinfo {year}
  {1980})\ pp.\ \bibinfo {pages} {135--157}\BibitemShut {NoStop}%
\bibitem [{\citenamefont {Hastings}\ and\ \citenamefont {Wen}(2005)}]{HW0541}%
  \BibitemOpen
  \bibfield  {author} {\bibinfo {author} {\bibfnamefont {M.~B.}\ \bibnamefont
  {Hastings}}\ and\ \bibinfo {author} {\bibfnamefont {X.-G.}\ \bibnamefont
  {Wen}},\ }\bibfield  {title} {\bibinfo {title} {Quasiadiabatic continuation
  of quantum states: {The} stability of topological ground-state degeneracy and
  emergent gauge invariance},\ }\href
  {https://doi.org/10.1103/physrevb.72.045141} {\bibfield  {journal} {\bibinfo
  {journal} {Phys. Rev. B}\ }\textbf {\bibinfo {volume} {72}},\ \bibinfo
  {pages} {045141} (\bibinfo {year} {2005})},\ \Eprint
  {https://arxiv.org/abs/cond-mat/0503554} {arXiv:cond-mat/0503554}
  \BibitemShut {NoStop}%
\bibitem [{\citenamefont {{Baez}}\ and\ \citenamefont
  {{Schreiber}}(2007)}]{BSm0511710}%
  \BibitemOpen
  \bibfield  {author} {\bibinfo {author} {\bibfnamefont {J.~C.}\ \bibnamefont
  {{Baez}}}\ and\ \bibinfo {author} {\bibfnamefont {U.}~\bibnamefont
  {{Schreiber}}},\ }\bibfield  {title} {\bibinfo {title} {{Higher Gauge
  Theory}},\ }\href@noop {} {\bibfield  {journal} {\bibinfo  {journal}
  {Contemp. Math. 431, AMS, Providence, Rhode Island}\ ,\ \bibinfo {pages} {7}}
  (\bibinfo {year} {2007})},\ \Eprint {https://arxiv.org/abs/math/0511710}
  {math/0511710} \BibitemShut {NoStop}%
\bibitem [{\citenamefont {Girelli}\ \emph {et~al.}(2008)\citenamefont
  {Girelli}, \citenamefont {Pfeiffer},\ and\ \citenamefont
  {Popescu}}]{GP07083051}%
  \BibitemOpen
  \bibfield  {author} {\bibinfo {author} {\bibfnamefont {F.}~\bibnamefont
  {Girelli}}, \bibinfo {author} {\bibfnamefont {H.}~\bibnamefont {Pfeiffer}},\
  and\ \bibinfo {author} {\bibfnamefont {E.~M.}\ \bibnamefont {Popescu}},\
  }\bibfield  {title} {\bibinfo {title} {Topological higher gauge theory:
  {From} {BF} to {BFCG} theory},\ }\href {https://doi.org/10.1063/1.2888764}
  {\bibfield  {journal} {\bibinfo  {journal} {J. Math. Phys.}\ }\textbf
  {\bibinfo {volume} {49}},\ \bibinfo {pages} {032503} (\bibinfo {year}
  {2008})},\ \Eprint {https://arxiv.org/abs/0708.3051} {arXiv:0708.3051}
  \BibitemShut {NoStop}%
\bibitem [{\citenamefont {Baez}\ and\ \citenamefont
  {Huerta}(2010)}]{BH10034485}%
  \BibitemOpen
  \bibfield  {author} {\bibinfo {author} {\bibfnamefont {J.~C.}\ \bibnamefont
  {Baez}}\ and\ \bibinfo {author} {\bibfnamefont {J.}~\bibnamefont {Huerta}},\
  }\bibfield  {title} {\bibinfo {title} {An invitation to higher gauge
  theory},\ }\href {https://doi.org/10.1007/s10714-010-1070-9} {\bibfield
  {journal} {\bibinfo  {journal} {Gen Relativ Gravit}\ }\textbf {\bibinfo
  {volume} {43}},\ \bibinfo {pages} {2335} (\bibinfo {year} {2010})},\ \Eprint
  {https://arxiv.org/abs/1003.4485} {arXiv:1003.4485} \BibitemShut {NoStop}%
\bibitem [{\citenamefont {Sharpe}(2015)}]{S150804770}%
  \BibitemOpen
  \bibfield  {author} {\bibinfo {author} {\bibfnamefont {E.}~\bibnamefont
  {Sharpe}},\ }\bibfield  {title} {\bibinfo {title} {Notes on generalized
  global symmetries in {QFT}},\ }\href {https://doi.org/10.1002/prop.201500048}
  {\bibfield  {journal} {\bibinfo  {journal} {Fortschr. Phys.}\ }\textbf
  {\bibinfo {volume} {63}},\ \bibinfo {pages} {659} (\bibinfo {year} {2015})},\
  \Eprint {https://arxiv.org/abs/1508.04770} {arXiv:1508.04770} \BibitemShut
  {NoStop}%
\bibitem [{\citenamefont {{Thorngren}}\ and\ \citenamefont {{von
  Keyserlingk}}(2015)}]{TK151102929}%
  \BibitemOpen
  \bibfield  {author} {\bibinfo {author} {\bibfnamefont {R.}~\bibnamefont
  {{Thorngren}}}\ and\ \bibinfo {author} {\bibfnamefont {C.}~\bibnamefont {{von
  Keyserlingk}}},\ }\bibfield  {title} {\bibinfo {title} {{Higher {SPT's} and a
  generalization of anomaly in-flow}},\ }\href@noop {} {\  (\bibinfo {year}
  {2015})},\ \Eprint {https://arxiv.org/abs/1511.02929} {arXiv:1511.02929}
  \BibitemShut {NoStop}%
\bibitem [{\citenamefont {Bullivant}\ \emph {et~al.}(2017)\citenamefont
  {Bullivant}, \citenamefont {Calcada}, \citenamefont {K\'ad\'ar},
  \citenamefont {Martin},\ and\ \citenamefont {Martins}}]{BM160606639}%
  \BibitemOpen
  \bibfield  {author} {\bibinfo {author} {\bibfnamefont {A.}~\bibnamefont
  {Bullivant}}, \bibinfo {author} {\bibfnamefont {M.}~\bibnamefont {Calcada}},
  \bibinfo {author} {\bibfnamefont {Z.}~\bibnamefont {K\'ad\'ar}}, \bibinfo
  {author} {\bibfnamefont {P.}~\bibnamefont {Martin}},\ and\ \bibinfo {author}
  {\bibfnamefont {J.~a.~F.}\ \bibnamefont {Martins}},\ }\bibfield  {title}
  {\bibinfo {title} {Topological phases from higher gauge symmetry in
  3+1-dimensions},\ }\href {https://doi.org/10.1103/physrevb.95.155118}
  {\bibfield  {journal} {\bibinfo  {journal} {Phys. Rev. B}\ }\textbf {\bibinfo
  {volume} {95}},\ \bibinfo {pages} {155118} (\bibinfo {year} {2017})},\
  \Eprint {https://arxiv.org/abs/1606.06639} {arXiv:1606.06639} \BibitemShut
  {NoStop}%
\bibitem [{\citenamefont {Ibieta-Jimenez}\ \emph {et~al.}(2020)\citenamefont
  {Ibieta-Jimenez}, \citenamefont {Petrucci}, \citenamefont {Xavier},\ and\
  \citenamefont {Teotonio-Sobrinho}}]{CT171104186}%
  \BibitemOpen
  \bibfield  {author} {\bibinfo {author} {\bibfnamefont {J.}~\bibnamefont
  {Ibieta-Jimenez}}, \bibinfo {author} {\bibfnamefont {M.}~\bibnamefont
  {Petrucci}}, \bibinfo {author} {\bibfnamefont {L.~Q.}\ \bibnamefont
  {Xavier}},\ and\ \bibinfo {author} {\bibfnamefont {P.}~\bibnamefont
  {Teotonio-Sobrinho}},\ }\bibfield  {title} {\bibinfo {title} {Topological
  entanglement entropy in d-dimensions for abelian higher gauge theories},\
  }\href {https://doi.org/10.1007/jhep03(2020)167} {\bibfield  {journal}
  {\bibinfo  {journal} {J. High Energ. Phys.}\ }\textbf {\bibinfo {volume}
  {2020}}\bibfield  {number} {\bibinfo  {number} { (3)}},\ }\Eprint
  {https://arxiv.org/abs/1711.04186} {arXiv:1711.04186} \BibitemShut {NoStop}%
\bibitem [{\citenamefont {Tachikawa}(2020)}]{T171209542}%
  \BibitemOpen
  \bibfield  {author} {\bibinfo {author} {\bibfnamefont {Y.}~\bibnamefont
  {Tachikawa}},\ }\bibfield  {title} {\bibinfo {title} {On gauging finite
  subgroups},\ }\bibfield  {journal} {\bibinfo  {journal} {SciPost Phys.}\
  }\textbf {\bibinfo {volume} {8}},\ \href
  {https://doi.org/10.21468/scipostphys.8.1.015} {10.21468/scipostphys.8.1.015}
  (\bibinfo {year} {2020}),\ \Eprint {https://arxiv.org/abs/1712.09542}
  {arXiv:1712.09542} \BibitemShut {NoStop}%
\bibitem [{\citenamefont {{Bhardwaj}}\ and\ \citenamefont
  {{Tachikawa}}(2018)}]{BT170402330}%
  \BibitemOpen
  \bibfield  {author} {\bibinfo {author} {\bibfnamefont {L.}~\bibnamefont
  {{Bhardwaj}}}\ and\ \bibinfo {author} {\bibfnamefont {Y.}~\bibnamefont
  {{Tachikawa}}},\ }\bibfield  {title} {\bibinfo {title} {{On finite symmetries
  and their gauging in two dimensions}},\ }\href
  {https://doi.org/10.1007/JHEP03(2018)189} {\bibfield  {journal} {\bibinfo
  {journal} {Journal of High Energy Physics}\ }\textbf {\bibinfo {volume}
  {2018}},\ \bibinfo {pages} {189} (\bibinfo {year} {2018})},\ \Eprint
  {https://arxiv.org/abs/1704.02330} {arXiv:1704.02330} \BibitemShut {NoStop}%
\bibitem [{\citenamefont {Parzygnat}(2019)}]{P180201139}%
  \BibitemOpen
  \bibfield  {author} {\bibinfo {author} {\bibfnamefont {A.~J.}\ \bibnamefont
  {Parzygnat}},\ }\bibfield  {title} {\bibinfo {title} {Two-dimensional algebra
  in lattice gauge theory},\ }\href {https://doi.org/10.1063/1.5078532}
  {\bibfield  {journal} {\bibinfo  {journal} {J. Math. Phys.}\ }\textbf
  {\bibinfo {volume} {60}},\ \bibinfo {pages} {043506} (\bibinfo {year}
  {2019})},\ \Eprint {https://arxiv.org/abs/1802.01139} {arXiv:1802.01139}
  \BibitemShut {NoStop}%
\bibitem [{\citenamefont {Delcamp}\ and\ \citenamefont
  {Tiwari}(2018)}]{DT180210104}%
  \BibitemOpen
  \bibfield  {author} {\bibinfo {author} {\bibfnamefont {C.}~\bibnamefont
  {Delcamp}}\ and\ \bibinfo {author} {\bibfnamefont {A.}~\bibnamefont
  {Tiwari}},\ }\bibfield  {title} {\bibinfo {title} {From gauge to higher gauge
  models of topological phases},\ }\href
  {https://doi.org/10.1007/jhep10(2018)049} {\bibfield  {journal} {\bibinfo
  {journal} {J. High Energ. Phys.}\ }\textbf {\bibinfo {volume}
  {2018}}\bibfield  {number} {\bibinfo  {number} { (10)},\ \bibinfo {pages}
  {49}},\ }\Eprint {https://arxiv.org/abs/1802.10104} {arXiv:1802.10104}
  \BibitemShut {NoStop}%
\bibitem [{\citenamefont {{Lake}}(2018)}]{L180207747}%
  \BibitemOpen
  \bibfield  {author} {\bibinfo {author} {\bibfnamefont {E.}~\bibnamefont
  {{Lake}}},\ }\bibfield  {title} {\bibinfo {title} {{Higher-form symmetries
  and spontaneous symmetry breaking}},\ }\href@noop {} {\  (\bibinfo {year}
  {2018})},\ \Eprint {https://arxiv.org/abs/1802.07747} {arXiv:1802.07747}
  \BibitemShut {NoStop}%
\bibitem [{\citenamefont {C\'ordova}\ \emph {et~al.}(2019)\citenamefont
  {C\'ordova}, \citenamefont {Dumitrescu},\ and\ \citenamefont
  {Intriligator}}]{CI180204790}%
  \BibitemOpen
  \bibfield  {author} {\bibinfo {author} {\bibfnamefont {C.}~\bibnamefont
  {C\'ordova}}, \bibinfo {author} {\bibfnamefont {T.~T.}\ \bibnamefont
  {Dumitrescu}},\ and\ \bibinfo {author} {\bibfnamefont {K.}~\bibnamefont
  {Intriligator}},\ }\bibfield  {title} {\bibinfo {title} {Exploring 2-group
  global symmetries},\ }\href {https://doi.org/10.1007/jhep02(2019)184}
  {\bibfield  {journal} {\bibinfo  {journal} {J. High Energ. Phys.}\ }\textbf
  {\bibinfo {volume} {2019}}\bibfield  {number} {\bibinfo  {number} { (2)},\
  \bibinfo {pages} {184}},\ }\Eprint {https://arxiv.org/abs/1802.04790}
  {arXiv:1802.04790} \BibitemShut {NoStop}%
\bibitem [{\citenamefont {Hofman}\ and\ \citenamefont
  {Iqbal}(2019)}]{HI180209512}%
  \BibitemOpen
  \bibfield  {author} {\bibinfo {author} {\bibfnamefont {D.}~\bibnamefont
  {Hofman}}\ and\ \bibinfo {author} {\bibfnamefont {N.}~\bibnamefont {Iqbal}},\
  }\bibfield  {title} {\bibinfo {title} {Goldstone modes and photonization for
  higher form symmetries},\ }\href
  {https://doi.org/10.21468/scipostphys.6.1.006} {\bibfield  {journal}
  {\bibinfo  {journal} {SciPost Phys.}\ }\textbf {\bibinfo {volume} {6}},\
  \bibinfo {pages} {006} (\bibinfo {year} {2019})},\ \Eprint
  {https://arxiv.org/abs/1802.09512} {arXiv:1802.09512} \BibitemShut {NoStop}%
\bibitem [{\citenamefont {Bouzid}\ and\ \citenamefont
  {Tahiri}(2018)}]{BT180300529}%
  \BibitemOpen
  \bibfield  {author} {\bibinfo {author} {\bibfnamefont {B.}~\bibnamefont
  {Bouzid}}\ and\ \bibinfo {author} {\bibfnamefont {M.}~\bibnamefont
  {Tahiri}},\ }\bibfield  {title} {\bibinfo {title} {2-connections, a lattice
  point of view},\ }\href {https://doi.org/10.5506/aphyspolb.49.1885}
  {\bibfield  {journal} {\bibinfo  {journal} {Acta Phys. Pol. B}\ }\textbf
  {\bibinfo {volume} {49}},\ \bibinfo {pages} {1885} (\bibinfo {year}
  {2018})},\ \Eprint {https://arxiv.org/abs/1803.00529} {arXiv:1803.00529}
  \BibitemShut {NoStop}%
\bibitem [{\citenamefont {Benini}\ \emph {et~al.}(2019)\citenamefont {Benini},
  \citenamefont {C\'ordova},\ and\ \citenamefont {Hsin}}]{BH180309336}%
  \BibitemOpen
  \bibfield  {author} {\bibinfo {author} {\bibfnamefont {F.}~\bibnamefont
  {Benini}}, \bibinfo {author} {\bibfnamefont {C.}~\bibnamefont {C\'ordova}},\
  and\ \bibinfo {author} {\bibfnamefont {P.-S.}\ \bibnamefont {Hsin}},\
  }\bibfield  {title} {\bibinfo {title} {On 2-group global symmetries and their
  anomalies},\ }\href {https://doi.org/10.1007/jhep03(2019)118} {\bibfield
  {journal} {\bibinfo  {journal} {J. High Energ. Phys.}\ }\textbf {\bibinfo
  {volume} {2019}}\bibfield  {number} {\bibinfo  {number} { (3)},\ \bibinfo
  {pages} {118}},\ }\Eprint {https://arxiv.org/abs/1803.09336}
  {arXiv:1803.09336} \BibitemShut {NoStop}%
\bibitem [{\citenamefont {Nikolaus}\ and\ \citenamefont
  {Waldorf}(2019)}]{NW180400677}%
  \BibitemOpen
  \bibfield  {author} {\bibinfo {author} {\bibfnamefont {T.}~\bibnamefont
  {Nikolaus}}\ and\ \bibinfo {author} {\bibfnamefont {K.}~\bibnamefont
  {Waldorf}},\ }\bibfield  {title} {\bibinfo {title} {Higher geometry for
  non-geometric t-duals},\ }\bibfield  {journal} {\bibinfo  {journal} {Commun.
  Math. Phys.}\ }\href {https://doi.org/10.1007/s00220-019-03496-3}
  {10.1007/s00220-019-03496-3} (\bibinfo {year} {2019}),\ \Eprint
  {https://arxiv.org/abs/1804.00677} {arXiv:1804.00677} \BibitemShut {NoStop}%
\bibitem [{\citenamefont {Zhu}\ \emph {et~al.}(2019)\citenamefont {Zhu},
  \citenamefont {Lan},\ and\ \citenamefont {Wen}}]{ZW180809394}%
  \BibitemOpen
  \bibfield  {author} {\bibinfo {author} {\bibfnamefont {C.}~\bibnamefont
  {Zhu}}, \bibinfo {author} {\bibfnamefont {T.}~\bibnamefont {Lan}},\ and\
  \bibinfo {author} {\bibfnamefont {X.-G.}\ \bibnamefont {Wen}},\ }\bibfield
  {title} {\bibinfo {title} {Topological nonlinear $\sigma$-model, higher gauge
  theory, and a systematic construction of {3+1D} topological orders for boson
  systems},\ }\href {https://doi.org/10.1103/physrevb.100.045105} {\bibfield
  {journal} {\bibinfo  {journal} {Phys. Rev. B}\ }\textbf {\bibinfo {volume}
  {100}},\ \bibinfo {pages} {045105} (\bibinfo {year} {2019})},\ \Eprint
  {https://arxiv.org/abs/1808.09394} {arXiv:1808.09394} \BibitemShut {NoStop}%
\bibitem [{\citenamefont {Harlow}\ and\ \citenamefont
  {Ooguri}(2021)}]{HO181005338}%
  \BibitemOpen
  \bibfield  {author} {\bibinfo {author} {\bibfnamefont {D.}~\bibnamefont
  {Harlow}}\ and\ \bibinfo {author} {\bibfnamefont {H.}~\bibnamefont
  {Ooguri}},\ }\bibfield  {title} {\bibinfo {title} {Symmetries in quantum
  field theory and quantum gravity},\ }\href@noop {} {\bibfield  {journal}
  {\bibinfo  {journal} {Communications in Mathematical Physics}\ }\textbf
  {\bibinfo {volume} {383}},\ \bibinfo {pages} {1669} (\bibinfo {year}
  {2021})}\BibitemShut {NoStop}%
\bibitem [{\citenamefont {Wan}\ \emph {et~al.}(2020)\citenamefont {Wan},
  \citenamefont {Wang},\ and\ \citenamefont {Zheng}}]{WW181211968}%
  \BibitemOpen
  \bibfield  {author} {\bibinfo {author} {\bibfnamefont {Z.}~\bibnamefont
  {Wan}}, \bibinfo {author} {\bibfnamefont {J.}~\bibnamefont {Wang}},\ and\
  \bibinfo {author} {\bibfnamefont {Y.}~\bibnamefont {Zheng}},\ }\bibfield
  {title} {\bibinfo {title} {New higher anomalies, su (n) yang--mills gauge
  theory and cpn- 1 sigma model},\ }\href@noop {} {\bibfield  {journal}
  {\bibinfo  {journal} {Annals of Physics}\ }\textbf {\bibinfo {volume}
  {414}},\ \bibinfo {pages} {168074} (\bibinfo {year} {2020})}\BibitemShut
  {NoStop}%
\bibitem [{\citenamefont {{Wan}}\ and\ \citenamefont
  {{Wang}}(2019)}]{WW181211967}%
  \BibitemOpen
  \bibfield  {author} {\bibinfo {author} {\bibfnamefont {Z.}~\bibnamefont
  {{Wan}}}\ and\ \bibinfo {author} {\bibfnamefont {J.}~\bibnamefont {{Wang}}},\
  }\bibfield  {title} {\bibinfo {title} {{Non-Abelian Gauge Theories, Sigma
  Models, Higher Anomalies, Symmetries, and Cobordisms}},\ }\href@noop {}
  {\bibfield  {journal} {\bibinfo  {journal} {Annals of Mathematical Sciences
  and Applications}\ }\textbf {\bibinfo {volume} {4}},\ \bibinfo {pages} {107}
  (\bibinfo {year} {2019})},\ \Eprint {https://arxiv.org/abs/1812.11967}
  {arXiv:1812.11967} \BibitemShut {NoStop}%
\bibitem [{\citenamefont {{Guo}}\ \emph {et~al.}(2020)\citenamefont {{Guo}},
  \citenamefont {{Ohmori}}, \citenamefont {{Putrov}}, \citenamefont {{Wan}},\
  and\ \citenamefont {{Wang}}}]{GW181211959}%
  \BibitemOpen
  \bibfield  {author} {\bibinfo {author} {\bibfnamefont {M.}~\bibnamefont
  {{Guo}}}, \bibinfo {author} {\bibfnamefont {K.}~\bibnamefont {{Ohmori}}},
  \bibinfo {author} {\bibfnamefont {P.}~\bibnamefont {{Putrov}}}, \bibinfo
  {author} {\bibfnamefont {Z.}~\bibnamefont {{Wan}}},\ and\ \bibinfo {author}
  {\bibfnamefont {J.}~\bibnamefont {{Wang}}},\ }\bibfield  {title} {\bibinfo
  {title} {{Fermionic Finite-Group Gauge Theories and Interacting
  Symmetric/Crystalline Orders via Cobordisms}},\ }\href@noop {} {\bibfield
  {journal} {\bibinfo  {journal} {Communications in Mathematical Physics}\
  }\textbf {\bibinfo {volume} {376}},\ \bibinfo {pages} {1073} (\bibinfo {year}
  {2020})},\ \Eprint {https://arxiv.org/abs/1812.11959} {arXiv:1812.11959}
  \BibitemShut {NoStop}%
\bibitem [{\citenamefont {Wan}\ and\ \citenamefont {Wang}(2019)}]{WW181211955}%
  \BibitemOpen
  \bibfield  {author} {\bibinfo {author} {\bibfnamefont {Z.}~\bibnamefont
  {Wan}}\ and\ \bibinfo {author} {\bibfnamefont {J.}~\bibnamefont {Wang}},\
  }\bibfield  {title} {\bibinfo {title} {Adjoint {QCD4} , deconfined critical
  phenomena, symmetry-enriched topological quantum field theory, and higher
  symmetry extension},\ }\href {https://doi.org/10.1103/physrevd.99.065013}
  {\bibfield  {journal} {\bibinfo  {journal} {Phys. Rev. D}\ }\textbf {\bibinfo
  {volume} {99}},\ \bibinfo {pages} {065013} (\bibinfo {year} {2019})},\
  \Eprint {https://arxiv.org/abs/1812.11955} {arXiv:1812.11955} \BibitemShut
  {NoStop}%
\bibitem [{\citenamefont {Wen}(2019)}]{W181202517}%
  \BibitemOpen
  \bibfield  {author} {\bibinfo {author} {\bibfnamefont {X.-G.}\ \bibnamefont
  {Wen}},\ }\bibfield  {title} {\bibinfo {title} {Emergent (anomalous) higher
  symmetries from topological order and from dynamical electromagnetic field in
  condensed matter systems},\ }\href
  {https://doi.org/10.1103/physrevb.99.205139} {\bibfield  {journal} {\bibinfo
  {journal} {Phys. Rev. B}\ }\textbf {\bibinfo {volume} {99}},\ \bibinfo
  {pages} {205139} (\bibinfo {year} {2019})},\ \Eprint
  {https://arxiv.org/abs/1812.02517} {arXiv:1812.02517} \BibitemShut {NoStop}%
\bibitem [{\citenamefont {Callan}\ and\ \citenamefont {Harvey}(1985)}]{CH8527}%
  \BibitemOpen
  \bibfield  {author} {\bibinfo {author} {\bibfnamefont {C.}~\bibnamefont
  {Callan}}\ and\ \bibinfo {author} {\bibfnamefont {J.}~\bibnamefont
  {Harvey}},\ }\bibfield  {title} {\bibinfo {title} {Anomalies and fermion zero
  modes on strings and domain walls},\ }\href
  {https://doi.org/10.1016/0550-3213(85)90489-4} {\bibfield  {journal}
  {\bibinfo  {journal} {Nucl. Phys. B}\ }\textbf {\bibinfo {volume} {250}},\
  \bibinfo {pages} {427} (\bibinfo {year} {1985})}\BibitemShut {NoStop}%
\bibitem [{\citenamefont {Witten}(1989)}]{W8951}%
  \BibitemOpen
  \bibfield  {author} {\bibinfo {author} {\bibfnamefont {E.}~\bibnamefont
  {Witten}},\ }\bibfield  {title} {\bibinfo {title} {Quantum field theory and
  the {Jones} polynomial},\ }\href {https://doi.org/10.1007/bf01217730}
  {\bibfield  {journal} {\bibinfo  {journal} {Commun.Math. Phys.}\ }\textbf
  {\bibinfo {volume} {121}},\ \bibinfo {pages} {351} (\bibinfo {year}
  {1989})}\BibitemShut {NoStop}%
\bibitem [{\citenamefont {Wen}(1991)}]{W9125}%
  \BibitemOpen
  \bibfield  {author} {\bibinfo {author} {\bibfnamefont {X.-G.}\ \bibnamefont
  {Wen}},\ }\bibfield  {title} {\bibinfo {title} {Gapless boundary excitations
  in the quantum hall states and in the chiral spin states},\ }\href
  {https://doi.org/10.1103/physrevb.43.11025} {\bibfield  {journal} {\bibinfo
  {journal} {Phys. Rev. B}\ }\textbf {\bibinfo {volume} {43}},\ \bibinfo
  {pages} {11025} (\bibinfo {year} {1991})}\BibitemShut {NoStop}%
\bibitem [{\citenamefont {Kapustin}\ and\ \citenamefont
  {Thorngren}(2014)}]{KT14043230}%
  \BibitemOpen
  \bibfield  {author} {\bibinfo {author} {\bibfnamefont {A.}~\bibnamefont
  {Kapustin}}\ and\ \bibinfo {author} {\bibfnamefont {R.}~\bibnamefont
  {Thorngren}},\ }\bibfield  {title} {\bibinfo {title} {Anomalous discrete
  symmetries in three dimensions and group cohomology},\ }\href
  {https://doi.org/10.1103/physrevlett.112.231602} {\bibfield  {journal}
  {\bibinfo  {journal} {Phys. Rev. Lett.}\ }\textbf {\bibinfo {volume} {112}},\
  \bibinfo {pages} {231602} (\bibinfo {year} {2014})},\ \Eprint
  {https://arxiv.org/abs/1404.3230} {arXiv:1404.3230} \BibitemShut {NoStop}%
\bibitem [{\citenamefont {Witten}\ and\ \citenamefont
  {Yonekura}(2022)}]{WY190908775}%
  \BibitemOpen
  \bibfield  {author} {\bibinfo {author} {\bibfnamefont {E.}~\bibnamefont
  {Witten}}\ and\ \bibinfo {author} {\bibfnamefont {K.}~\bibnamefont
  {Yonekura}},\ }\bibfield  {title} {\bibinfo {title} {Anomaly inflow and the
  $\eta$-invariant},\ }in\ \href@noop {} {\emph {\bibinfo {booktitle} {Memorial
  Volume for Shoucheng Zhang}}}\ (\bibinfo  {publisher} {World Scientific},\
  \bibinfo {year} {2022})\ pp.\ \bibinfo {pages} {283--352}\BibitemShut
  {NoStop}%
\bibitem [{\citenamefont {Ji}\ and\ \citenamefont {Wen}(2019)}]{JW190513279}%
  \BibitemOpen
  \bibfield  {author} {\bibinfo {author} {\bibfnamefont {W.}~\bibnamefont
  {Ji}}\ and\ \bibinfo {author} {\bibfnamefont {X.-G.}\ \bibnamefont {Wen}},\
  }\bibfield  {title} {\bibinfo {title} {Non-invertible anomalies and
  mapping-class-group transformation of anomalous partition functions},\ }\href
  {https://doi.org/10.1103/PhysRevResearch.1.033054} {\bibfield  {journal}
  {\bibinfo  {journal} {Phys. Rev. Research}\ }\textbf {\bibinfo {volume}
  {1}},\ \bibinfo {pages} {033054} (\bibinfo {year} {2019})},\ \Eprint
  {https://arxiv.org/abs/1905.13279} {arXiv:1905.13279} \BibitemShut {NoStop}%
\bibitem [{\citenamefont {Wen}(2013)}]{W1313}%
  \BibitemOpen
  \bibfield  {author} {\bibinfo {author} {\bibfnamefont {X.-G.}\ \bibnamefont
  {Wen}},\ }\bibfield  {title} {\bibinfo {title} {Classifying gauge anomalies
  through symmetry-protected trivial orders and classifying gravitational
  anomalies through topological orders},\ }\href
  {https://doi.org/10.1103/physrevd.88.045013} {\bibfield  {journal} {\bibinfo
  {journal} {Phys. Rev. D}\ }\textbf {\bibinfo {volume} {88}},\ \bibinfo
  {pages} {045013} (\bibinfo {year} {2013})},\ \Eprint
  {https://arxiv.org/abs/1303.1803} {arXiv:1303.1803} \BibitemShut {NoStop}%
\bibitem [{\citenamefont {Kong}\ and\ \citenamefont {Wen}(2014)}]{KW1458}%
  \BibitemOpen
  \bibfield  {author} {\bibinfo {author} {\bibfnamefont {L.}~\bibnamefont
  {Kong}}\ and\ \bibinfo {author} {\bibfnamefont {X.-G.}\ \bibnamefont {Wen}},\
  }\bibfield  {title} {\bibinfo {title} {Braided fusion categories,
  gravitational anomalies, and the mathematical framework for topological
  orders in any dimensions},\ }\href@noop {} {\  (\bibinfo {year} {2014})},\
  \Eprint {https://arxiv.org/abs/1405.5858} {arXiv:1405.5858} \BibitemShut
  {NoStop}%
\bibitem [{\citenamefont {Kong}\ \emph {et~al.}(2015)\citenamefont {Kong},
  \citenamefont {Wen},\ and\ \citenamefont {Zheng}}]{KZ150201690}%
  \BibitemOpen
  \bibfield  {author} {\bibinfo {author} {\bibfnamefont {L.}~\bibnamefont
  {Kong}}, \bibinfo {author} {\bibfnamefont {X.-G.}\ \bibnamefont {Wen}},\ and\
  \bibinfo {author} {\bibfnamefont {H.}~\bibnamefont {Zheng}},\ }\bibfield
  {title} {\bibinfo {title} {Boundary-bulk relation for topological orders as
  the functor mapping higher categories to their centers},\ }\href@noop {} {\
  (\bibinfo {year} {2015})},\ \Eprint {https://arxiv.org/abs/1502.01690}
  {arXiv:1502.01690} \BibitemShut {NoStop}%
\bibitem [{\citenamefont {Fiorenza}\ and\ \citenamefont
  {Valentino}(2015)}]{FV14095723}%
  \BibitemOpen
  \bibfield  {author} {\bibinfo {author} {\bibfnamefont {D.}~\bibnamefont
  {Fiorenza}}\ and\ \bibinfo {author} {\bibfnamefont {A.}~\bibnamefont
  {Valentino}},\ }\bibfield  {title} {\bibinfo {title} {Boundary conditions for
  topological quantum field theories, anomalies and projective modular
  functors},\ }\href {https://doi.org/10.1007/s00220-015-2371-3} {\bibfield
  {journal} {\bibinfo  {journal} {Commun. Math. Phys.}\ }\textbf {\bibinfo
  {volume} {338}},\ \bibinfo {pages} {1043} (\bibinfo {year} {2015})},\ \Eprint
  {https://arxiv.org/abs/1409.5723} {arXiv:1409.5723} \BibitemShut {NoStop}%
\bibitem [{\citenamefont {Monnier}(2015)}]{M14107442}%
  \BibitemOpen
  \bibfield  {author} {\bibinfo {author} {\bibfnamefont {S.}~\bibnamefont
  {Monnier}},\ }\bibfield  {title} {\bibinfo {title} {{Hamiltonian} anomalies
  from extended field theories},\ }\href
  {https://doi.org/10.1007/s00220-015-2369-x} {\bibfield  {journal} {\bibinfo
  {journal} {Commun. Math. Phys.}\ }\textbf {\bibinfo {volume} {338}},\
  \bibinfo {pages} {1327} (\bibinfo {year} {2015})},\ \Eprint
  {https://arxiv.org/abs/1410.7442} {arXiv:1410.7442} \BibitemShut {NoStop}%
\bibitem [{\citenamefont {{Kong}}\ \emph
  {et~al.}(2020{\natexlab{a}})\citenamefont {{Kong}}, \citenamefont {{Lan}},
  \citenamefont {{Wen}}, \citenamefont {{Zhang}},\ and\ \citenamefont
  {{Zheng}}}]{KZ200308898}%
  \BibitemOpen
  \bibfield  {author} {\bibinfo {author} {\bibfnamefont {L.}~\bibnamefont
  {{Kong}}}, \bibinfo {author} {\bibfnamefont {T.}~\bibnamefont {{Lan}}},
  \bibinfo {author} {\bibfnamefont {X.-G.}\ \bibnamefont {{Wen}}}, \bibinfo
  {author} {\bibfnamefont {Z.-H.}\ \bibnamefont {{Zhang}}},\ and\ \bibinfo
  {author} {\bibfnamefont {H.}~\bibnamefont {{Zheng}}},\ }\bibfield  {title}
  {\bibinfo {title} {Classification of topological phases with finite internal
  symmetries in all dimensions},\ }\href
  {https://doi.org/doi.org/10.1007/JHEP09(2020)093} {\bibfield  {journal}
  {\bibinfo  {journal} {J. High Energ. Phys.}\ }\textbf {\bibinfo {volume}
  {2020}}\bibfield  {number} {\bibinfo  {number} { (9)},\ \bibinfo {pages}
  {93}},\ }\Eprint {https://arxiv.org/abs/2003.08898} {arXiv:2003.08898}
  \BibitemShut {NoStop}%
\bibitem [{\citenamefont {{Kong}}\ \emph
  {et~al.}(2020{\natexlab{b}})\citenamefont {{Kong}}, \citenamefont {{Lan}},
  \citenamefont {{Wen}}, \citenamefont {{Zhang}},\ and\ \citenamefont
  {{Zheng}}}]{KZ200514178}%
  \BibitemOpen
  \bibfield  {author} {\bibinfo {author} {\bibfnamefont {L.}~\bibnamefont
  {{Kong}}}, \bibinfo {author} {\bibfnamefont {T.}~\bibnamefont {{Lan}}},
  \bibinfo {author} {\bibfnamefont {X.-G.}\ \bibnamefont {{Wen}}}, \bibinfo
  {author} {\bibfnamefont {Z.-H.}\ \bibnamefont {{Zhang}}},\ and\ \bibinfo
  {author} {\bibfnamefont {H.}~\bibnamefont {{Zheng}}},\ }\bibfield  {title}
  {\bibinfo {title} {{Algebraic higher symmetry and categorical symmetry: A
  holographic and entanglement view of symmetry}},\ }\href
  {https://doi.org/10.1103/PhysRevResearch.2.043086} {\bibfield  {journal}
  {\bibinfo  {journal} {Physical Review Research}\ }\textbf {\bibinfo {volume}
  {2}},\ \bibinfo {pages} {043086} (\bibinfo {year} {2020}{\natexlab{b}})},\
  \Eprint {https://arxiv.org/abs/2005.14178} {arXiv:2005.14178} \BibitemShut
  {NoStop}%
\bibitem [{\citenamefont {Wen}(1989)}]{W8987}%
  \BibitemOpen
  \bibfield  {author} {\bibinfo {author} {\bibfnamefont {X.-G.}\ \bibnamefont
  {Wen}},\ }\bibfield  {title} {\bibinfo {title} {Vacuum degeneracy of chiral
  spin states in compactified space},\ }\href
  {https://doi.org/10.1103/physrevb.40.7387} {\bibfield  {journal} {\bibinfo
  {journal} {Phys. Rev. B}\ }\textbf {\bibinfo {volume} {40}},\ \bibinfo
  {pages} {7387} (\bibinfo {year} {1989})}\BibitemShut {NoStop}%
\bibitem [{\citenamefont {Wen}(1990)}]{W9039}%
  \BibitemOpen
  \bibfield  {author} {\bibinfo {author} {\bibfnamefont {X.-G.}\ \bibnamefont
  {Wen}},\ }\bibfield  {title} {\bibinfo {title} {Topological orders in rigid
  states},\ }\href {https://doi.org/10.1142/s0217979290000139} {\bibfield
  {journal} {\bibinfo  {journal} {Int. J. Mod. Phys. B}\ }\textbf {\bibinfo
  {volume} {04}},\ \bibinfo {pages} {239} (\bibinfo {year} {1990})}\BibitemShut
  {NoStop}%
\bibitem [{\citenamefont {Freed}(2014)}]{F14047224}%
  \BibitemOpen
  \bibfield  {author} {\bibinfo {author} {\bibfnamefont {D.~S.}\ \bibnamefont
  {Freed}},\ }\bibfield  {title} {\bibinfo {title} {Anomalies and invertible
  field theories},\ }in\ \href@noop {} {\emph {\bibinfo {booktitle} {Proc.
  Symp. Pure Math}}},\ Vol.~\bibinfo {volume} {88}\ (\bibinfo {year} {2014})\
  pp.\ \bibinfo {pages} {25--46}\BibitemShut {NoStop}%
\bibitem [{\citenamefont {Kong}\ \emph {et~al.}(2017)\citenamefont {Kong},
  \citenamefont {Wen},\ and\ \citenamefont {Zheng}}]{KZ170200673}%
  \BibitemOpen
  \bibfield  {author} {\bibinfo {author} {\bibfnamefont {L.}~\bibnamefont
  {Kong}}, \bibinfo {author} {\bibfnamefont {X.-G.}\ \bibnamefont {Wen}},\ and\
  \bibinfo {author} {\bibfnamefont {H.}~\bibnamefont {Zheng}},\ }\bibfield
  {title} {\bibinfo {title} {Boundary-bulk relation in topological orders},\
  }\href {https://doi.org/10.1016/j.nuclphysb.2017.06.023} {\bibfield
  {journal} {\bibinfo  {journal} {Nucl. Phys. B}\ }\textbf {\bibinfo {volume}
  {922}},\ \bibinfo {pages} {62} (\bibinfo {year} {2017})},\ \Eprint
  {https://arxiv.org/abs/1702.00673} {arXiv:1702.00673} \BibitemShut {NoStop}%
\bibitem [{\citenamefont {Freed}\ and\ \citenamefont
  {Teleman}(2022)}]{FT180600008}%
  \BibitemOpen
  \bibfield  {author} {\bibinfo {author} {\bibfnamefont {D.~S.}\ \bibnamefont
  {Freed}}\ and\ \bibinfo {author} {\bibfnamefont {C.}~\bibnamefont
  {Teleman}},\ }\bibfield  {title} {\bibinfo {title} {Topological dualities in
  the ising model},\ }\href@noop {} {\bibfield  {journal} {\bibinfo  {journal}
  {Geometry \& Topology}\ }\textbf {\bibinfo {volume} {26}},\ \bibinfo {pages}
  {1907} (\bibinfo {year} {2022})}\BibitemShut {NoStop}%
\bibitem [{\citenamefont {{Thorngren}}\ and\ \citenamefont
  {{Wang}}(2019)}]{TW191202817}%
  \BibitemOpen
  \bibfield  {author} {\bibinfo {author} {\bibfnamefont {R.}~\bibnamefont
  {{Thorngren}}}\ and\ \bibinfo {author} {\bibfnamefont {Y.}~\bibnamefont
  {{Wang}}},\ }\bibfield  {title} {\bibinfo {title} {{Fusion Category Symmetry
  I: Anomaly In-Flow and Gapped Phases}},\ }\href@noop {} {\  (\bibinfo {year}
  {2019})},\ \Eprint {https://arxiv.org/abs/1912.02817} {arXiv:1912.02817}
  \BibitemShut {NoStop}%
\bibitem [{\citenamefont {Ji}\ and\ \citenamefont
  {Wen}(2020{\natexlab{a}})}]{JW191209391}%
  \BibitemOpen
  \bibfield  {author} {\bibinfo {author} {\bibfnamefont {W.}~\bibnamefont
  {Ji}}\ and\ \bibinfo {author} {\bibfnamefont {X.-G.}\ \bibnamefont {Wen}},\
  }\bibfield  {title} {\bibinfo {title} {Metallic states beyond the
  tomonaga-luttinger liquid in one dimension},\ }\bibfield  {journal} {\bibinfo
   {journal} {Phys. Rev. B}\ }\textbf {\bibinfo {volume} {102}},\ \href
  {https://doi.org/10.1103/physrevb.102.195107} {10.1103/physrevb.102.195107}
  (\bibinfo {year} {2020}{\natexlab{a}}),\ \Eprint
  {https://arxiv.org/abs/1912.09391} {arXiv:1912.09391} \BibitemShut {NoStop}%
\bibitem [{\citenamefont {Ji}\ and\ \citenamefont
  {Wen}(2020{\natexlab{b}})}]{JW191213492}%
  \BibitemOpen
  \bibfield  {author} {\bibinfo {author} {\bibfnamefont {W.}~\bibnamefont
  {Ji}}\ and\ \bibinfo {author} {\bibfnamefont {X.-G.}\ \bibnamefont {Wen}},\
  }\bibfield  {title} {\bibinfo {title} {Categorical symmetry and noninvertible
  anomaly in symmetry-breaking and topological phase transitions},\ }\href
  {https://doi.org/10.1103/PhysRevResearch.2.033417} {\bibfield  {journal}
  {\bibinfo  {journal} {Phys. Rev. Research}\ }\textbf {\bibinfo {volume}
  {2}},\ \bibinfo {pages} {033417} (\bibinfo {year} {2020}{\natexlab{b}})},\
  \Eprint {https://arxiv.org/abs/1912.13492} {arXiv:1912.13492} \BibitemShut
  {NoStop}%
\bibitem [{\citenamefont {Moudgalya}\ and\ \citenamefont
  {Motrunich}(2022{\natexlab{a}})}]{MM210810324}%
  \BibitemOpen
  \bibfield  {author} {\bibinfo {author} {\bibfnamefont {S.}~\bibnamefont
  {Moudgalya}}\ and\ \bibinfo {author} {\bibfnamefont {O.~I.}\ \bibnamefont
  {Motrunich}},\ }\bibfield  {title} {\bibinfo {title} {Hilbert space
  fragmentation and commutant algebras},\ }\href@noop {} {\bibfield  {journal}
  {\bibinfo  {journal} {Physical Review X}\ }\textbf {\bibinfo {volume} {12}},\
  \bibinfo {pages} {011050} (\bibinfo {year} {2022}{\natexlab{a}})}\BibitemShut
  {NoStop}%
\bibitem [{\citenamefont {Moudgalya}\ and\ \citenamefont
  {Motrunich}(2022{\natexlab{b}})}]{MM220903370}%
  \BibitemOpen
  \bibfield  {author} {\bibinfo {author} {\bibfnamefont {S.}~\bibnamefont
  {Moudgalya}}\ and\ \bibinfo {author} {\bibfnamefont {O.~I.}\ \bibnamefont
  {Motrunich}},\ }\bibfield  {title} {\bibinfo {title} {From symmetries to
  commutant algebras in standard hamiltonians},\ }\href@noop {} {\bibfield
  {journal} {\bibinfo  {journal} {arXiv preprint arXiv:2209.03370}\ } (\bibinfo
  {year} {2022}{\natexlab{b}})}\BibitemShut {NoStop}%
\bibitem [{\citenamefont {Kong}\ \emph {et~al.}(2022)\citenamefont {Kong},
  \citenamefont {Wen},\ and\ \citenamefont {Zheng}}]{KZ210808835}%
  \BibitemOpen
  \bibfield  {author} {\bibinfo {author} {\bibfnamefont {L.}~\bibnamefont
  {Kong}}, \bibinfo {author} {\bibfnamefont {X.-G.}\ \bibnamefont {Wen}},\ and\
  \bibinfo {author} {\bibfnamefont {H.}~\bibnamefont {Zheng}},\ }\bibfield
  {title} {\bibinfo {title} {One dimensional gapped quantum phases and enriched
  fusion categories},\ }\href@noop {} {\bibfield  {journal} {\bibinfo
  {journal} {Journal of High Energy Physics}\ }\textbf {\bibinfo {volume}
  {2022}},\ \bibinfo {pages} {1} (\bibinfo {year} {2022})}\BibitemShut
  {NoStop}%
\bibitem [{\citenamefont {Haag}\ and\ \citenamefont {Kastler}(1964)}]{HK6448}%
  \BibitemOpen
  \bibfield  {author} {\bibinfo {author} {\bibfnamefont {R.}~\bibnamefont
  {Haag}}\ and\ \bibinfo {author} {\bibfnamefont {D.}~\bibnamefont {Kastler}},\
  }\bibfield  {title} {\bibinfo {title} {{An algebraic approach to quantum
  field theory}},\ }\href@noop {} {\bibfield  {journal} {\bibinfo  {journal}
  {J. Math. Phys.}\ }\textbf {\bibinfo {volume} {5}},\ \bibinfo {pages}
  {848–861} (\bibinfo {year} {1964})}\BibitemShut {NoStop}%
\bibitem [{\citenamefont {{Kong}}\ and\ \citenamefont
  {{Zheng}}(2022)}]{KZ220105726}%
  \BibitemOpen
  \bibfield  {author} {\bibinfo {author} {\bibfnamefont {L.}~\bibnamefont
  {{Kong}}}\ and\ \bibinfo {author} {\bibfnamefont {H.}~\bibnamefont
  {{Zheng}}},\ }\bibfield  {title} {\bibinfo {title} {{Categories of quantum
  liquids III}},\ }\href@noop {} {\  (\bibinfo {year} {2022})},\ \Eprint
  {https://arxiv.org/abs/2201.05726} {arXiv:2201.05726} \BibitemShut {NoStop}%
\bibitem [{\citenamefont {Buerschaper}\ \emph {et~al.}(2013)\citenamefont
  {Buerschaper}, \citenamefont {Christandl}, \citenamefont {Kong},\ and\
  \citenamefont {Aguado}}]{BA10065823}%
  \BibitemOpen
  \bibfield  {author} {\bibinfo {author} {\bibfnamefont {O.}~\bibnamefont
  {Buerschaper}}, \bibinfo {author} {\bibfnamefont {M.}~\bibnamefont
  {Christandl}}, \bibinfo {author} {\bibfnamefont {L.}~\bibnamefont {Kong}},\
  and\ \bibinfo {author} {\bibfnamefont {M.}~\bibnamefont {Aguado}},\
  }\bibfield  {title} {\bibinfo {title} {Electric-magnetic duality of lattice
  systems with topological order},\ }\href
  {https://doi.org/10.1016/j.nuclphysb.2013.08.014} {\bibfield  {journal}
  {\bibinfo  {journal} {Nuclear Physics B}\ }\textbf {\bibinfo {volume}
  {876}},\ \bibinfo {pages} {619} (\bibinfo {year} {2013})},\ \Eprint
  {https://arxiv.org/abs/1006.5823} {arXiv:1006.5823} \BibitemShut {NoStop}%
\bibitem [{\citenamefont {{Lootens}}\ \emph {et~al.}(2021)\citenamefont
  {{Lootens}}, \citenamefont {{Delcamp}}, \citenamefont {{Ortiz}},\ and\
  \citenamefont {{Verstraete}}}]{LV211209091}%
  \BibitemOpen
  \bibfield  {author} {\bibinfo {author} {\bibfnamefont {L.}~\bibnamefont
  {{Lootens}}}, \bibinfo {author} {\bibfnamefont {C.}~\bibnamefont
  {{Delcamp}}}, \bibinfo {author} {\bibfnamefont {G.}~\bibnamefont {{Ortiz}}},\
  and\ \bibinfo {author} {\bibfnamefont {F.}~\bibnamefont {{Verstraete}}},\
  }\bibfield  {title} {\bibinfo {title} {{Category-theoretic recipe for
  dualities in one-dimensional quantum lattice models}},\ }\href@noop {} {\
  (\bibinfo {year} {2021})},\ \Eprint {https://arxiv.org/abs/2112.09091}
  {arXiv:2112.09091} \BibitemShut {NoStop}%
\bibitem [{\citenamefont {Fr\"ohlich}\ \emph {et~al.}(2007)\citenamefont
  {Fr\"ohlich}, \citenamefont {Fuchs}, \citenamefont {Runkel},\ and\
  \citenamefont {Schweigert}}]{FSh0607247}%
  \BibitemOpen
  \bibfield  {author} {\bibinfo {author} {\bibfnamefont {J.}~\bibnamefont
  {Fr\"ohlich}}, \bibinfo {author} {\bibfnamefont {J.}~\bibnamefont {Fuchs}},
  \bibinfo {author} {\bibfnamefont {I.}~\bibnamefont {Runkel}},\ and\ \bibinfo
  {author} {\bibfnamefont {C.}~\bibnamefont {Schweigert}},\ }\bibfield  {title}
  {\bibinfo {title} {Duality and defects in rational conformal field theory},\
  }\href@noop {} {\bibfield  {journal} {\bibinfo  {journal} {Nucl. Phys. B}\
  }\textbf {\bibinfo {volume} {763}},\ \bibinfo {pages} {354} (\bibinfo {year}
  {2007})},\ \Eprint {https://arxiv.org/abs/hep-th/0607247}
  {arXiv:hep-th/0607247} \BibitemShut {NoStop}%
\bibitem [{\citenamefont {Davydov}\ \emph
  {et~al.}(2011{\natexlab{a}})\citenamefont {Davydov}, \citenamefont {Kong},\
  and\ \citenamefont {Runkel}}]{DR10044725}%
  \BibitemOpen
  \bibfield  {author} {\bibinfo {author} {\bibfnamefont {A.}~\bibnamefont
  {Davydov}}, \bibinfo {author} {\bibfnamefont {L.}~\bibnamefont {Kong}},\ and\
  \bibinfo {author} {\bibfnamefont {I.}~\bibnamefont {Runkel}},\ }\bibfield
  {title} {\bibinfo {title} {Invertible defects and isomorphisms of rational
  cfts},\ }\href@noop {} {\bibfield  {journal} {\bibinfo  {journal} {Adv.
  Theor. Math. Phys.}\ }\textbf {\bibinfo {volume} {15}},\ \bibinfo {pages}
  {43} (\bibinfo {year} {2011}{\natexlab{a}})},\ \Eprint
  {https://arxiv.org/abs/1004.4725} {arXiv:1004.4725} \BibitemShut {NoStop}%
\bibitem [{\citenamefont {Davydov}\ \emph
  {et~al.}(2011{\natexlab{b}})\citenamefont {Davydov}, \citenamefont {Kong},\
  and\ \citenamefont {Runkel}}]{DR11070495}%
  \BibitemOpen
  \bibfield  {author} {\bibinfo {author} {\bibfnamefont {A.}~\bibnamefont
  {Davydov}}, \bibinfo {author} {\bibfnamefont {L.}~\bibnamefont {Kong}},\ and\
  \bibinfo {author} {\bibfnamefont {I.}~\bibnamefont {Runkel}},\ }\bibfield
  {title} {\bibinfo {title} {Field theories with defects and the centre
  functor},\ }\href@noop {} {\bibfield  {journal} {\bibinfo  {journal} {Nuclear
  Physics B}\ } (\bibinfo {year} {2011}{\natexlab{b}})},\ \Eprint
  {https://arxiv.org/abs/1107.0495} {arXiv:1107.0495} \BibitemShut {NoStop}%
\bibitem [{\citenamefont {Chang}\ \emph {et~al.}(2019)\citenamefont {Chang},
  \citenamefont {Lin}, \citenamefont {Shao}, \citenamefont {Wang},\ and\
  \citenamefont {Yin}}]{CY180204445}%
  \BibitemOpen
  \bibfield  {author} {\bibinfo {author} {\bibfnamefont {C.-M.}\ \bibnamefont
  {Chang}}, \bibinfo {author} {\bibfnamefont {Y.-H.}\ \bibnamefont {Lin}},
  \bibinfo {author} {\bibfnamefont {S.-H.}\ \bibnamefont {Shao}}, \bibinfo
  {author} {\bibfnamefont {Y.}~\bibnamefont {Wang}},\ and\ \bibinfo {author}
  {\bibfnamefont {X.}~\bibnamefont {Yin}},\ }\bibfield  {title} {\bibinfo
  {title} {Topological defect lines and renormalization group flows in two
  dimensions},\ }\href {https://doi.org/10.1007/jhep01(2019)026} {\bibfield
  {journal} {\bibinfo  {journal} {J. High Energ. Phys.}\ }\textbf {\bibinfo
  {volume} {2019}}\bibfield  {number} {\bibinfo  {number} { (1)},\ \bibinfo
  {pages} {26}},\ }\Eprint {https://arxiv.org/abs/1802.04445}
  {arXiv:1802.04445} \BibitemShut {NoStop}%
\bibitem [{\citenamefont {Metlitski}\ and\ \citenamefont
  {Thorngren}(2018)}]{MT170707686}%
  \BibitemOpen
  \bibfield  {author} {\bibinfo {author} {\bibfnamefont {M.~A.}\ \bibnamefont
  {Metlitski}}\ and\ \bibinfo {author} {\bibfnamefont {R.}~\bibnamefont
  {Thorngren}},\ }\bibfield  {title} {\bibinfo {title} {Intrinsic and emergent
  anomalies at deconfined critical points},\ }\href
  {https://doi.org/10.1103/physrevb.98.085140} {\bibfield  {journal} {\bibinfo
  {journal} {Phys. Rev. B}\ }\textbf {\bibinfo {volume} {98}},\ \bibinfo
  {pages} {085140} (\bibinfo {year} {2018})},\ \Eprint
  {https://arxiv.org/abs/1707.07686} {arXiv:1707.07686} \BibitemShut {NoStop}%
\bibitem [{\citenamefont {Kong}\ and\ \citenamefont
  {Zheng}(2018)}]{KZ170501087}%
  \BibitemOpen
  \bibfield  {author} {\bibinfo {author} {\bibfnamefont {L.}~\bibnamefont
  {Kong}}\ and\ \bibinfo {author} {\bibfnamefont {H.}~\bibnamefont {Zheng}},\
  }\bibfield  {title} {\bibinfo {title} {Gapless edges of 2d topological orders
  and enriched monoidal categories},\ }\href
  {https://doi.org/10.1016/j.nuclphysb.2017.12.007} {\bibfield  {journal}
  {\bibinfo  {journal} {Nucl. Phys. B}\ }\textbf {\bibinfo {volume} {927}},\
  \bibinfo {pages} {140} (\bibinfo {year} {2018})},\ \Eprint
  {https://arxiv.org/abs/1705.01087} {arXiv:1705.01087} \BibitemShut {NoStop}%
\bibitem [{\citenamefont {{Kong}}\ and\ \citenamefont
  {{Zheng}}(2020)}]{KZ190504924}%
  \BibitemOpen
  \bibfield  {author} {\bibinfo {author} {\bibfnamefont {L.}~\bibnamefont
  {{Kong}}}\ and\ \bibinfo {author} {\bibfnamefont {H.}~\bibnamefont
  {{Zheng}}},\ }\bibfield  {title} {\bibinfo {title} {{A mathematical theory of
  gapless edges of 2d topological orders I}},\ }\href
  {https://doi.org/10.1007/JHEP02(2020)150} {\bibfield  {journal} {\bibinfo
  {journal} {J. High Energ. Phys.}\ }\textbf {\bibinfo {volume} {2020}},\
  \bibinfo {pages} {150}},\ \Eprint {https://arxiv.org/abs/1905.04924}
  {arXiv:1905.04924} \BibitemShut {NoStop}%
\bibitem [{\citenamefont {{Kong}}\ and\ \citenamefont
  {{Zheng}}(2021)}]{KZ191201760}%
  \BibitemOpen
  \bibfield  {author} {\bibinfo {author} {\bibfnamefont {L.}~\bibnamefont
  {{Kong}}}\ and\ \bibinfo {author} {\bibfnamefont {H.}~\bibnamefont
  {{Zheng}}},\ }\bibfield  {title} {\bibinfo {title} {{A mathematical theory of
  gapless edges of 2d topological orders. Part {II}}},\ }\href@noop {}
  {\bibfield  {journal} {\bibinfo  {journal} {Nuclear Physics B}\ }\textbf
  {\bibinfo {volume} {966}},\ \bibinfo {pages} {115384} (\bibinfo {year}
  {2021})},\ \Eprint {https://arxiv.org/abs/1912.01760} {arXiv:1912.01760}
  \BibitemShut {NoStop}%
\bibitem [{\citenamefont {Chen}\ \emph {et~al.}(2010)\citenamefont {Chen},
  \citenamefont {Gu},\ and\ \citenamefont {Wen}}]{CGW1038}%
  \BibitemOpen
  \bibfield  {author} {\bibinfo {author} {\bibfnamefont {X.}~\bibnamefont
  {Chen}}, \bibinfo {author} {\bibfnamefont {Z.-C.}\ \bibnamefont {Gu}},\ and\
  \bibinfo {author} {\bibfnamefont {X.-G.}\ \bibnamefont {Wen}},\ }\bibfield
  {title} {\bibinfo {title} {Local unitary transformation, long-range quantum
  entanglement, wave function renormalization, and topological order},\ }\href
  {https://doi.org/10.1103/physrevb.82.155138} {\bibfield  {journal} {\bibinfo
  {journal} {Phys. Rev. B}\ }\textbf {\bibinfo {volume} {82}},\ \bibinfo
  {pages} {155138} (\bibinfo {year} {2010})},\ \Eprint
  {https://arxiv.org/abs/1004.3835} {arXiv:1004.3835} \BibitemShut {NoStop}%
\bibitem [{\citenamefont {Zeng}\ and\ \citenamefont {Wen}(2015)}]{ZW1490}%
  \BibitemOpen
  \bibfield  {author} {\bibinfo {author} {\bibfnamefont {B.}~\bibnamefont
  {Zeng}}\ and\ \bibinfo {author} {\bibfnamefont {X.-G.}\ \bibnamefont {Wen}},\
  }\bibfield  {title} {\bibinfo {title} {Gapped quantum liquids and topological
  order, stochastic local transformations and emergence of unitarity},\ }\href
  {https://doi.org/10.1103/physrevb.91.125121} {\bibfield  {journal} {\bibinfo
  {journal} {Phys. Rev. B}\ }\textbf {\bibinfo {volume} {91}},\ \bibinfo
  {pages} {125121} (\bibinfo {year} {2015})},\ \Eprint
  {https://arxiv.org/abs/1406.5090} {arXiv:1406.5090} \BibitemShut {NoStop}%
\bibitem [{\citenamefont {{Apruzzi}}\ \emph {et~al.}(2021)\citenamefont
  {{Apruzzi}}, \citenamefont {{Bonetti}}, \citenamefont {{Garc{\'\i}a
  Etxebarria}}, \citenamefont {{Hosseini}},\ and\ \citenamefont
  {{Schafer-Nameki}}}]{AS211202092}%
  \BibitemOpen
  \bibfield  {author} {\bibinfo {author} {\bibfnamefont {F.}~\bibnamefont
  {{Apruzzi}}}, \bibinfo {author} {\bibfnamefont {F.}~\bibnamefont
  {{Bonetti}}}, \bibinfo {author} {\bibfnamefont {I.}~\bibnamefont
  {{Garc{\'\i}a Etxebarria}}}, \bibinfo {author} {\bibfnamefont {S.~S.}\
  \bibnamefont {{Hosseini}}},\ and\ \bibinfo {author} {\bibfnamefont
  {S.}~\bibnamefont {{Schafer-Nameki}}},\ }\bibfield  {title} {\bibinfo {title}
  {{Symmetry TFTs from String Theory}},\ }\href@noop {} {\bibfield  {journal}
  {\bibinfo  {journal} {arXiv e-prints}\ ,\ \bibinfo {eid} {arXiv:2112.02092}}
  (\bibinfo {year} {2021})},\ \Eprint {https://arxiv.org/abs/2112.02092}
  {arXiv:2112.02092} \BibitemShut {NoStop}%
\bibitem [{\citenamefont {{Freed}}\ \emph {et~al.}(2022)\citenamefont
  {{Freed}}, \citenamefont {{Moore}},\ and\ \citenamefont
  {{Teleman}}}]{FT220907471}%
  \BibitemOpen
  \bibfield  {author} {\bibinfo {author} {\bibfnamefont {D.~S.}\ \bibnamefont
  {{Freed}}}, \bibinfo {author} {\bibfnamefont {G.~W.}\ \bibnamefont
  {{Moore}}},\ and\ \bibinfo {author} {\bibfnamefont {C.}~\bibnamefont
  {{Teleman}}},\ }\bibfield  {title} {\bibinfo {title} {{Topological symmetry
  in quantum field theory}},\ }\href@noop {} {\  (\bibinfo {year} {2022})},\
  \Eprint {https://arxiv.org/abs/2209.07471} {arXiv:2209.07471} \BibitemShut
  {NoStop}%
\bibitem [{\citenamefont {Kawagoe}\ and\ \citenamefont {Levin}()}]{Kawagoe}%
  \BibitemOpen
  \bibfield  {author} {\bibinfo {author} {\bibfnamefont {K.}~\bibnamefont
  {Kawagoe}}\ and\ \bibinfo {author} {\bibfnamefont {M.}~\bibnamefont
  {Levin}},\ }\bibfield  {title} {\bibinfo {title} {{Microscopic definitions of
  anyon data}},\ }\href@noop {} {\ }\Eprint
  {https://arxiv.org/abs/1910.11353v1} {arXiv:1910.11353v1} \BibitemShut
  {NoStop}%
\bibitem [{\citenamefont {Levin}\ and\ \citenamefont {Wen}(2003)}]{LW0316}%
  \BibitemOpen
  \bibfield  {author} {\bibinfo {author} {\bibfnamefont {M.}~\bibnamefont
  {Levin}}\ and\ \bibinfo {author} {\bibfnamefont {X.-G.}\ \bibnamefont
  {Wen}},\ }\bibfield  {title} {\bibinfo {title} {Fermions, strings, and gauge
  fields in lattice spin models},\ }\href
  {https://doi.org/10.1103/physrevb.67.245316} {\bibfield  {journal} {\bibinfo
  {journal} {Phys. Rev. B}\ }\textbf {\bibinfo {volume} {67}},\ \bibinfo
  {pages} {245316} (\bibinfo {year} {2003})},\ \Eprint
  {https://arxiv.org/abs/cond-mat/0302460} {arXiv:cond-mat/0302460}
  \BibitemShut {NoStop}%
\bibitem [{\citenamefont {Lan}\ \emph {et~al.}(2016)\citenamefont {Lan},
  \citenamefont {Kong},\ and\ \citenamefont {Wen}}]{LW160205936}%
  \BibitemOpen
  \bibfield  {author} {\bibinfo {author} {\bibfnamefont {T.}~\bibnamefont
  {Lan}}, \bibinfo {author} {\bibfnamefont {L.}~\bibnamefont {Kong}},\ and\
  \bibinfo {author} {\bibfnamefont {X.-G.}\ \bibnamefont {Wen}},\ }\bibfield
  {title} {\bibinfo {title} {Modular extensions of unitary braided fusion
  categories and {2+1D} {Topological/SPT} orders with symmetries},\ }\href
  {https://doi.org/10.1007/s00220-016-2748-y} {\bibfield  {journal} {\bibinfo
  {journal} {Commun. Math. Phys.}\ }\textbf {\bibinfo {volume} {351}},\
  \bibinfo {pages} {709} (\bibinfo {year} {2016})},\ \Eprint
  {https://arxiv.org/abs/1602.05936} {arXiv:1602.05936} \BibitemShut {NoStop}%
\bibitem [{\citenamefont {Albert}\ \emph {et~al.}(2021)\citenamefont {Albert},
  \citenamefont {Aasen}, \citenamefont {Xu}, \citenamefont {Ji}, \citenamefont
  {Alicea},\ and\ \citenamefont {Preskill}}]{AAX211112096}%
  \BibitemOpen
  \bibfield  {author} {\bibinfo {author} {\bibfnamefont {V.~V.}\ \bibnamefont
  {Albert}}, \bibinfo {author} {\bibfnamefont {D.}~\bibnamefont {Aasen}},
  \bibinfo {author} {\bibfnamefont {W.}~\bibnamefont {Xu}}, \bibinfo {author}
  {\bibfnamefont {W.}~\bibnamefont {Ji}}, \bibinfo {author} {\bibfnamefont
  {J.}~\bibnamefont {Alicea}},\ and\ \bibinfo {author} {\bibfnamefont
  {J.}~\bibnamefont {Preskill}},\ }\bibfield  {title} {\bibinfo {title} {Spin
  chains, defects, and quantum wires for the quantum-double edge},\ }\href@noop
  {} {\bibfield  {journal} {\bibinfo  {journal} {arXiv preprint
  arXiv:2111.12096}\ } (\bibinfo {year} {2021})}\BibitemShut {NoStop}%
\bibitem [{\citenamefont {Chen}\ \emph {et~al.}(2011)\citenamefont {Chen},
  \citenamefont {Liu},\ and\ \citenamefont {Wen}}]{CLW1141}%
  \BibitemOpen
  \bibfield  {author} {\bibinfo {author} {\bibfnamefont {X.}~\bibnamefont
  {Chen}}, \bibinfo {author} {\bibfnamefont {Z.-X.}\ \bibnamefont {Liu}},\ and\
  \bibinfo {author} {\bibfnamefont {X.-G.}\ \bibnamefont {Wen}},\ }\bibfield
  {title} {\bibinfo {title} {Two-dimensional symmetry-protected topological
  orders and their protected gapless edge excitations},\ }\href
  {https://doi.org/10.1103/physrevb.84.235141} {\bibfield  {journal} {\bibinfo
  {journal} {Phys. Rev. B}\ }\textbf {\bibinfo {volume} {84}},\ \bibinfo
  {pages} {235141} (\bibinfo {year} {2011})},\ \Eprint
  {https://arxiv.org/abs/1106.4752} {arXiv:1106.4752} \BibitemShut {NoStop}%
\bibitem [{\citenamefont {Chen}\ \emph {et~al.}(2013)\citenamefont {Chen},
  \citenamefont {Gu}, \citenamefont {Liu},\ and\ \citenamefont
  {Wen}}]{CGL1314}%
  \BibitemOpen
  \bibfield  {author} {\bibinfo {author} {\bibfnamefont {X.}~\bibnamefont
  {Chen}}, \bibinfo {author} {\bibfnamefont {Z.-C.}\ \bibnamefont {Gu}},
  \bibinfo {author} {\bibfnamefont {Z.-X.}\ \bibnamefont {Liu}},\ and\ \bibinfo
  {author} {\bibfnamefont {X.-G.}\ \bibnamefont {Wen}},\ }\bibfield  {title}
  {\bibinfo {title} {Symmetry protected topological orders and the group
  cohomology of their symmetry group},\ }\href
  {https://doi.org/10.1103/physrevb.87.155114} {\bibfield  {journal} {\bibinfo
  {journal} {Phys. Rev. B}\ }\textbf {\bibinfo {volume} {87}},\ \bibinfo
  {pages} {155114} (\bibinfo {year} {2013})},\ \Eprint
  {https://arxiv.org/abs/1106.4772} {arXiv:1106.4772} \BibitemShut {NoStop}%
\bibitem [{\citenamefont {Wang}\ \emph {et~al.}(2018)\citenamefont {Wang},
  \citenamefont {Wen},\ and\ \citenamefont {Witten}}]{WW170506728}%
  \BibitemOpen
  \bibfield  {author} {\bibinfo {author} {\bibfnamefont {J.}~\bibnamefont
  {Wang}}, \bibinfo {author} {\bibfnamefont {X.-G.}\ \bibnamefont {Wen}},\ and\
  \bibinfo {author} {\bibfnamefont {E.}~\bibnamefont {Witten}},\ }\bibfield
  {title} {\bibinfo {title} {Symmetric gapped interfaces of {SPT} and {SET}
  states: systematic constructions},\ }\href
  {https://doi.org/10.1103/physrevx.8.031048} {\bibfield  {journal} {\bibinfo
  {journal} {Phys. Rev. X}\ }\textbf {\bibinfo {volume} {8}},\ \bibinfo {pages}
  {031048} (\bibinfo {year} {2018})},\ \Eprint
  {https://arxiv.org/abs/1705.06728} {arXiv:1705.06728} \BibitemShut {NoStop}%
\bibitem [{\citenamefont {Wen}\ and\ \citenamefont {Zee}(1992)}]{WZ9290}%
  \BibitemOpen
  \bibfield  {author} {\bibinfo {author} {\bibfnamefont {X.-G.}\ \bibnamefont
  {Wen}}\ and\ \bibinfo {author} {\bibfnamefont {A.}~\bibnamefont {Zee}},\
  }\bibfield  {title} {\bibinfo {title} {Classification of {A}belian quantum
  {H}all states and matrix formulation of topological fluids},\ }\href
  {https://doi.org/10.1103/physrevb.46.2290} {\bibfield  {journal} {\bibinfo
  {journal} {Phys. Rev. B}\ }\textbf {\bibinfo {volume} {46}},\ \bibinfo
  {pages} {2290} (\bibinfo {year} {1992})}\BibitemShut {NoStop}%
\bibitem [{\citenamefont {Fr\"ohlich}\ and\ \citenamefont
  {Studer}(1993)}]{FS9333}%
  \BibitemOpen
  \bibfield  {author} {\bibinfo {author} {\bibfnamefont {J.}~\bibnamefont
  {Fr\"ohlich}}\ and\ \bibinfo {author} {\bibfnamefont {U.~M.}\ \bibnamefont
  {Studer}},\ }\bibfield  {title} {\bibinfo {title} {Gauge invariance and
  current algebra in nonrelativistic many-body theory},\ }\href
  {https://doi.org/10.1103/revmodphys.65.733} {\bibfield  {journal} {\bibinfo
  {journal} {Rev. Mod. Phys.}\ }\textbf {\bibinfo {volume} {65}},\ \bibinfo
  {pages} {733} (\bibinfo {year} {1993})}\BibitemShut {NoStop}%
\bibitem [{\citenamefont {Levin}\ and\ \citenamefont {Gu}(2012)}]{LG1209}%
  \BibitemOpen
  \bibfield  {author} {\bibinfo {author} {\bibfnamefont {M.}~\bibnamefont
  {Levin}}\ and\ \bibinfo {author} {\bibfnamefont {Z.-C.}\ \bibnamefont {Gu}},\
  }\bibfield  {title} {\bibinfo {title} {Braiding statistics approach to
  symmetry-protected topological phases},\ }\href
  {https://doi.org/10.1103/physrevb.86.115109} {\bibfield  {journal} {\bibinfo
  {journal} {Phys. Rev. B}\ }\textbf {\bibinfo {volume} {86}},\ \bibinfo
  {pages} {115109} (\bibinfo {year} {2012})},\ \Eprint
  {https://arxiv.org/abs/1202.3120} {arXiv:1202.3120} \BibitemShut {NoStop}%
\bibitem [{\citenamefont {Wen}(2014)}]{W1447}%
  \BibitemOpen
  \bibfield  {author} {\bibinfo {author} {\bibfnamefont {X.-G.}\ \bibnamefont
  {Wen}},\ }\bibfield  {title} {\bibinfo {title} {Symmetry-protected
  topological invariants of symmetry-protected topological phases of
  interacting bosons and fermions},\ }\href
  {https://doi.org/10.1103/physrevb.89.035147} {\bibfield  {journal} {\bibinfo
  {journal} {Phys. Rev. B}\ }\textbf {\bibinfo {volume} {89}},\ \bibinfo
  {pages} {035147} (\bibinfo {year} {2014})},\ \Eprint
  {https://arxiv.org/abs/1301.7675} {arXiv:1301.7675} \BibitemShut {NoStop}%
\bibitem [{\citenamefont {{Ellison}}\ \emph {et~al.}(2022)\citenamefont
  {{Ellison}}, \citenamefont {{Chen}}, \citenamefont {{Dua}}, \citenamefont
  {{Shirley}}, \citenamefont {{Tantivasadakarn}},\ and\ \citenamefont
  {{Williamson}}}]{EW211211394}%
  \BibitemOpen
  \bibfield  {author} {\bibinfo {author} {\bibfnamefont {T.~D.}\ \bibnamefont
  {{Ellison}}}, \bibinfo {author} {\bibfnamefont {Y.-A.}\ \bibnamefont
  {{Chen}}}, \bibinfo {author} {\bibfnamefont {A.}~\bibnamefont {{Dua}}},
  \bibinfo {author} {\bibfnamefont {W.}~\bibnamefont {{Shirley}}}, \bibinfo
  {author} {\bibfnamefont {N.}~\bibnamefont {{Tantivasadakarn}}},\ and\
  \bibinfo {author} {\bibfnamefont {D.~J.}\ \bibnamefont {{Williamson}}},\
  }\bibfield  {title} {\bibinfo {title} {{Pauli Stabilizer Models of Twisted
  Quantum Doubles}},\ }\href {https://doi.org/10.1103/PRXQuantum.3.010353}
  {\bibfield  {journal} {\bibinfo  {journal} {Phys. Rev. X Quantum}\ }\textbf
  {\bibinfo {volume} {3}},\ \bibinfo {pages} {010353} (\bibinfo {year}
  {2022})},\ \Eprint {https://arxiv.org/abs/2112.11394} {arXiv:2112.11394}
  \BibitemShut {NoStop}%
\bibitem [{\citenamefont {{Lan}}\ \emph {et~al.}(2019)\citenamefont {{Lan}},
  \citenamefont {{Zhu}},\ and\ \citenamefont {{Wen}}}]{LW180901112}%
  \BibitemOpen
  \bibfield  {author} {\bibinfo {author} {\bibfnamefont {T.}~\bibnamefont
  {{Lan}}}, \bibinfo {author} {\bibfnamefont {C.}~\bibnamefont {{Zhu}}},\ and\
  \bibinfo {author} {\bibfnamefont {X.-G.}\ \bibnamefont {{Wen}}},\ }\bibfield
  {title} {\bibinfo {title} {Fermion decoration construction of symmetry
  protected trivial orders for fermion systems with any symmetries $g_f$ and in
  any dimensions},\ }\href {https://doi.org/10.1103/PhysRevB.100.235141}
  {\bibfield  {journal} {\bibinfo  {journal} {Phys. Rev. B}\ }\textbf {\bibinfo
  {volume} {100}},\ \bibinfo {pages} {235141} (\bibinfo {year} {2019})},\
  \Eprint {https://arxiv.org/abs/1809.01112} {arXiv:1809.01112} \BibitemShut
  {NoStop}%
\bibitem [{\citenamefont {Levin}(2020)}]{L190309028}%
  \BibitemOpen
  \bibfield  {author} {\bibinfo {author} {\bibfnamefont {M.}~\bibnamefont
  {Levin}},\ }\bibfield  {title} {\bibinfo {title} {Constraints on order and
  disorder parameters in quantum spin chains},\ }\href
  {https://doi.org/10.1007/s00220-020-03802-4} {\bibfield  {journal} {\bibinfo
  {journal} {Commun. Math. Phys.}\ }\textbf {\bibinfo {volume} {378}},\
  \bibinfo {pages} {1081} (\bibinfo {year} {2020})},\ \Eprint
  {https://arxiv.org/abs/1903.09028} {arXiv:1903.09028} \BibitemShut {NoStop}%
\bibitem [{\citenamefont {Lan}\ \emph {et~al.}(2018)\citenamefont {Lan},
  \citenamefont {Kong},\ and\ \citenamefont {Wen}}]{LW170404221}%
  \BibitemOpen
  \bibfield  {author} {\bibinfo {author} {\bibfnamefont {T.}~\bibnamefont
  {Lan}}, \bibinfo {author} {\bibfnamefont {L.}~\bibnamefont {Kong}},\ and\
  \bibinfo {author} {\bibfnamefont {X.-G.}\ \bibnamefont {Wen}},\ }\bibfield
  {title} {\bibinfo {title} {Classification of {(3+1)D} bosonic topological
  orders: the case when pointlike excitations are all bosons},\ }\href
  {https://doi.org/10.1103/physrevx.8.021074} {\bibfield  {journal} {\bibinfo
  {journal} {Phys. Rev. X}\ }\textbf {\bibinfo {volume} {8}},\ \bibinfo {pages}
  {021074} (\bibinfo {year} {2018})},\ \Eprint
  {https://arxiv.org/abs/1704.04221} {arXiv:1704.04221} \BibitemShut {NoStop}%
\bibitem [{\citenamefont {{Gaiotto}}\ and\ \citenamefont
  {{Johnson-Freyd}}(2019)}]{GJ190509566}%
  \BibitemOpen
  \bibfield  {author} {\bibinfo {author} {\bibfnamefont {D.}~\bibnamefont
  {{Gaiotto}}}\ and\ \bibinfo {author} {\bibfnamefont {T.}~\bibnamefont
  {{Johnson-Freyd}}},\ }\bibfield  {title} {\bibinfo {title} {{Condensations in
  higher categories}},\ }\href@noop {} {\  (\bibinfo {year} {2019})},\ \Eprint
  {https://arxiv.org/abs/1905.09566} {arXiv:1905.09566} \BibitemShut {NoStop}%
\bibitem [{\citenamefont {Delcamp}(2022)}]{D2239}%
  \BibitemOpen
  \bibfield  {author} {\bibinfo {author} {\bibfnamefont {C.}~\bibnamefont
  {Delcamp}},\ }\bibfield  {title} {\bibinfo {title} {Tensor network approach
  to electromagnetic duality in (3+ 1) d topological gauge models},\
  }\href@noop {} {\bibfield  {journal} {\bibinfo  {journal} {Journal of High
  Energy Physics}\ }\textbf {\bibinfo {volume} {2022}},\ \bibinfo {pages} {1}
  (\bibinfo {year} {2022})}\BibitemShut {NoStop}%
\bibitem [{\citenamefont {Joyal}\ and\ \citenamefont {Street}(1991)}]{JS9151}%
  \BibitemOpen
  \bibfield  {author} {\bibinfo {author} {\bibfnamefont {A.}~\bibnamefont
  {Joyal}}\ and\ \bibinfo {author} {\bibfnamefont {R.}~\bibnamefont {Street}},\
  }\bibfield  {title} {\bibinfo {title} {Tortile yang-baxter operators in
  tensor categories},\ }\href@noop {} {\bibfield  {journal} {\bibinfo
  {journal} {Journal of Pure and Applied Algebra}\ }\textbf {\bibinfo {volume}
  {71}},\ \bibinfo {pages} {43} (\bibinfo {year} {1991})}\BibitemShut {NoStop}%
\bibitem [{\citenamefont {M{\"u}ger}(2003)}]{M0319}%
  \BibitemOpen
  \bibfield  {author} {\bibinfo {author} {\bibfnamefont {M.}~\bibnamefont
  {M{\"u}ger}},\ }\bibfield  {title} {\bibinfo {title} {From subfactors to
  categories and topology ii: The quantum double of tensor categories and
  subfactors},\ }\href@noop {} {\bibfield  {journal} {\bibinfo  {journal}
  {Journal of Pure and Applied Algebra}\ }\textbf {\bibinfo {volume} {180}},\
  \bibinfo {pages} {159} (\bibinfo {year} {2003})}\BibitemShut {NoStop}%
\bibitem [{\citenamefont {Goff}\ \emph {et~al.}(2007)\citenamefont {Goff},
  \citenamefont {Mason},\ and\ \citenamefont {Ng}}]{GNm0603191}%
  \BibitemOpen
  \bibfield  {author} {\bibinfo {author} {\bibfnamefont {C.}~\bibnamefont
  {Goff}}, \bibinfo {author} {\bibfnamefont {G.}~\bibnamefont {Mason}},\ and\
  \bibinfo {author} {\bibfnamefont {S.-H.}\ \bibnamefont {Ng}},\ }\bibfield
  {title} {\bibinfo {title} {On the gauge equivalence of twisted quantum
  doubles of elementary abelian and extra-special 2-groups},\ }\href
  {https://doi.org/10.1016/j.jalgebra.2006.10.022} {\bibfield  {journal}
  {\bibinfo  {journal} {Journal of Algebra}\ }\textbf {\bibinfo {volume}
  {312}},\ \bibinfo {pages} {849} (\bibinfo {year} {2007})},\ \Eprint
  {https://arxiv.org/abs/math/0603191} {arXiv:math/0603191} \BibitemShut
  {NoStop}%
\bibitem [{\citenamefont {Wang}\ and\ \citenamefont {Wen}(2015)}]{WW1454}%
  \BibitemOpen
  \bibfield  {author} {\bibinfo {author} {\bibfnamefont {J.~C.}\ \bibnamefont
  {Wang}}\ and\ \bibinfo {author} {\bibfnamefont {X.-G.}\ \bibnamefont {Wen}},\
  }\bibfield  {title} {\bibinfo {title} {Non-{A}belian string and particle
  braiding in topological order: {Modular SL(3,Z) representation}
  and(3+1)-dimensional twisted gauge theory},\ }\href
  {https://doi.org/10.1103/physrevb.91.035134} {\bibfield  {journal} {\bibinfo
  {journal} {Phys. Rev. B}\ }\textbf {\bibinfo {volume} {91}},\ \bibinfo
  {pages} {035134} (\bibinfo {year} {2015})},\ \Eprint
  {https://arxiv.org/abs/1404.7854} {arXiv:1404.7854} \BibitemShut {NoStop}%
\bibitem [{\citenamefont {Wen}(2015)}]{W150605768}%
  \BibitemOpen
  \bibfield  {author} {\bibinfo {author} {\bibfnamefont {X.-G.}\ \bibnamefont
  {Wen}},\ }\bibfield  {title} {\bibinfo {title} {A theory of {2+1D} bosonic
  topological orders},\ }\href {https://doi.org/10.1093/nsr/nwv077} {\bibfield
  {journal} {\bibinfo  {journal} {Nat. Sci. Rev.}\ }\textbf {\bibinfo {volume}
  {3}},\ \bibinfo {pages} {68} (\bibinfo {year} {2015})},\ \Eprint
  {https://arxiv.org/abs/1506.05768} {arXiv:1506.05768} \BibitemShut {NoStop}%
\bibitem [{\citenamefont {Wen}(2017)}]{W161201418}%
  \BibitemOpen
  \bibfield  {author} {\bibinfo {author} {\bibfnamefont {X.-G.}\ \bibnamefont
  {Wen}},\ }\bibfield  {title} {\bibinfo {title} {Exactly soluble local bosonic
  cocycle models, statistical transmutation, and simplest time-reversal
  symmetric topological orders in 3+1 dimensions},\ }\href
  {https://doi.org/10.1103/physrevb.95.205142} {\bibfield  {journal} {\bibinfo
  {journal} {Phys. Rev. B}\ }\textbf {\bibinfo {volume} {95}},\ \bibinfo
  {pages} {205142} (\bibinfo {year} {2017})},\ \Eprint
  {https://arxiv.org/abs/1612.01418} {arXiv:1612.01418} \BibitemShut {NoStop}%
\bibitem [{\citenamefont {{Petkova}}\ and\ \citenamefont
  {{Zuber}}(2001)}]{PZh0011021}%
  \BibitemOpen
  \bibfield  {author} {\bibinfo {author} {\bibfnamefont {V.~B.}\ \bibnamefont
  {{Petkova}}}\ and\ \bibinfo {author} {\bibfnamefont {J.~B.}\ \bibnamefont
  {{Zuber}}},\ }\bibfield  {title} {\bibinfo {title} {{Generalised twisted
  partition functions}},\ }\href
  {https://doi.org/10.1016/S0370-2693(01)00276-3} {\bibfield  {journal}
  {\bibinfo  {journal} {Physics Letters B}\ }\textbf {\bibinfo {volume}
  {504}},\ \bibinfo {pages} {157} (\bibinfo {year} {2001})},\ \Eprint
  {https://arxiv.org/abs/hep-th/0011021} {arXiv:hep-th/0011021} \BibitemShut
  {NoStop}%
\bibitem [{\citenamefont {{Coquereaux}}\ and\ \citenamefont
  {{Schieber}}(2002)}]{CSh0107001}%
  \BibitemOpen
  \bibfield  {author} {\bibinfo {author} {\bibfnamefont {R.}~\bibnamefont
  {{Coquereaux}}}\ and\ \bibinfo {author} {\bibfnamefont {G.}~\bibnamefont
  {{Schieber}}},\ }\bibfield  {title} {\bibinfo {title} {{Twisted partition
  functions for ADE boundary conformal field theories and Ocneanu algebras of
  quantum symmetries}},\ }\href {https://doi.org/10.1016/S0393-0440(01)00090-0}
  {\bibfield  {journal} {\bibinfo  {journal} {Journal of Geometry and Physics}\
  }\textbf {\bibinfo {volume} {42}},\ \bibinfo {pages} {216} (\bibinfo {year}
  {2002})},\ \Eprint {https://arxiv.org/abs/hep-th/0107001}
  {arXiv:hep-th/0107001} \BibitemShut {NoStop}%
\bibitem [{\citenamefont {{Fuchs}}\ \emph {et~al.}(2002)\citenamefont
  {{Fuchs}}, \citenamefont {{Runkel}},\ and\ \citenamefont
  {{Schweigert}}}]{FSh0204148}%
  \BibitemOpen
  \bibfield  {author} {\bibinfo {author} {\bibfnamefont {J.}~\bibnamefont
  {{Fuchs}}}, \bibinfo {author} {\bibfnamefont {I.}~\bibnamefont {{Runkel}}},\
  and\ \bibinfo {author} {\bibfnamefont {C.}~\bibnamefont {{Schweigert}}},\
  }\bibfield  {title} {\bibinfo {title} {{TFT construction of RCFT correlators
  I: partition functions}},\ }\href
  {https://doi.org/10.1016/S0550-3213(02)00744-7} {\bibfield  {journal}
  {\bibinfo  {journal} {Nuclear Physics B}\ }\textbf {\bibinfo {volume}
  {646}},\ \bibinfo {pages} {353} (\bibinfo {year} {2002})},\ \Eprint
  {https://arxiv.org/abs/hep-th/0204148} {arXiv:hep-th/0204148} \BibitemShut
  {NoStop}%
\bibitem [{\citenamefont {{Inamura}}(2021)}]{I210315588}%
  \BibitemOpen
  \bibfield  {author} {\bibinfo {author} {\bibfnamefont {K.}~\bibnamefont
  {{Inamura}}},\ }\bibfield  {title} {\bibinfo {title} {{Topological field
  theories and symmetry protected topological phases with fusion category
  symmetries}},\ }\href {https://doi.org/10.1007/JHEP05(2021)204} {\bibfield
  {journal} {\bibinfo  {journal} {Journal of High Energy Physics}\ }\textbf
  {\bibinfo {volume} {2021}},\ \bibinfo {eid} {204} (\bibinfo {year} {2021})},\
  \Eprint {https://arxiv.org/abs/2103.15588} {arXiv:2103.15588} \BibitemShut
  {NoStop}%
\bibitem [{\citenamefont {{Quella}}(2020)}]{Q200509072}%
  \BibitemOpen
  \bibfield  {author} {\bibinfo {author} {\bibfnamefont {T.}~\bibnamefont
  {{Quella}}},\ }\bibfield  {title} {\bibinfo {title} {{Symmetry-protected
  topological phases beyond groups: The q-deformed
  {Affleck-Kennedy-Lieb-Tasaki} model}},\ }\href
  {https://doi.org/10.1103/PhysRevB.102.081120} {\bibfield  {journal} {\bibinfo
   {journal} {\prb}\ }\textbf {\bibinfo {volume} {102}},\ \bibinfo {pages}
  {081120} (\bibinfo {year} {2020})},\ \Eprint
  {https://arxiv.org/abs/2005.09072} {arXiv:2005.09072} \BibitemShut {NoStop}%
\end{thebibliography}%

\end{document}